\documentclass[chaos,superscriptaddress,amsmath,amssymb,reprint,author-numerical,showpacs,showkeys]{revtex4-1} % una columna, doble espaciado
\usepackage{graphicx}% Include figure files
\usepackage{dcolumn}% Align table columns on decimal point
\usepackage{bm}% bold math
\usepackage{latexsym}
\usepackage{color}
\begin{document}

\preprint{AIP/123-QED}

\title[]{Assessing the  direction of climate interactions by means of complex networks and information theoretic tools}% Force line breaks with \\
%\thanks{Footnote to title of article. }

\author{J.  I.  Deza}\email{juan.ignacio.deza@upc.edu.}
\affiliation{Departament de F\'isica i Enginyeria Nuclear, Universitat Polit\`ecnica de Catalunya, Colom 11, E-08222, Terrassa, Barcelona, Spain. }%Lines break automatically or can be forced with \\

\author{M.  Barreiro}
% \homepage{http://www. Second. institution. edu/~Charlie. Author. }
\affiliation{
Instituto de F\'isica, Facultad de Ciencias, Universidad de la Rep\'ublica, Igu\'a 4225, Montevideo, Uruguay. %\\This line break forced with \textbackslash\textbackslash
}%

\author{C.  Masoller}
 \affiliation{Departament de F\'isica i Enginyeria Nuclear, Universitat Polit\`ecnica de Catalunya, Colom 11, E-08222, Terrassa, Barcelona, Spain. }%Lines break automatically or can be forced with \\

\date{\today}% It is always \today, today,

\begin{abstract}
An estimate of the net direction of climate interactions in different geographical regions is made by constructing a directed climate network from a regular latitude-longitude grid of nodes, using a directionality index ($\mbox{DI}$) based on  conditional mutual information.  Two datasets of surface air temperature anomalies---one monthly-averaged and another daily-averaged---are analyzed and compared. The network links are interpreted in terms of known atmospheric tropical and extra-tropical variability patterns.  Specific and relevant geographical regions are selected, the net direction of propagation of the atmospheric patterns is analyzed and the direction of the inferred links is validated by recovering some well-known climate variability structures. These patterns are found to be acting at various time-scales, such as atmospheric waves in the extratropics or longer range events in the tropics. This analysis demonstrates the capability of the $\mbox{DI}$ measure to infer the net direction of climate interactions and  may contribute to improve the present understanding of climate phenomena and climate predictability. The work presented here also stands out as an application of advanced tools to the analysis of empirical, real-world data.
%
%Valid PACS numbers may be entered using the \verb+\pacs{#1}+ command.
\end{abstract}

%Use showkeys class option if keyword
\maketitle

\begin{quotation}
{\bf
Information-theoretic tools are used to construct {\em directed} climate networks from time-series analysis of observed climatological data. Specifically, surface air temperature (SAT) anomalies are considered. Two datasets---one monthly-averaged and another daily-averaged--- are used. Directed links among the network nodes are defined via an analysis of the net direction of information transfer. A predictability measure---based on  conditional mutual information---quantifying the amount of information  in a time-series $x(t)$, contained in $\tau$ time units in the past of another time series, $y(t)$ is used.
The resulting directed network is then studied and a full agreement with state-of-the-art knowledge in climate phenomena has been found, validating this methodology for inferring the net directionality of climate interactions, directly from the data.  No weather assumptions or models are made, except for the appropriate setting of the parameter $\tau$ which is sensible to the shorter or longer auto-correlation of the time series.
}
\end{quotation}

\section{\label{sec:introduction} Introduction}

Network theory is a well-known framework for describing complex systems composed of many interacting components \cite{Watts1998,Albert2002,Boccaletti2006a,Arenas2008}.   Many systems can be straightforwardly represented in terms of a well-defined set of nodes coupled among them via links that have a clear physical interpretation. This is the case, for example, of airport networks \cite{Colizza2006}, of social interactions \cite{Barabasi2002} or the Internet \cite{Albert1999}, just to name some.  In other systems it is not clear how to define the relevant nodes, and/or there is no obvious interaction that can be used to define links.  An example of this situation is the Earth climate system, in which the lattice of  grid points from measurements or models, is defined to be the network set of nodes, and it depends on the resolution of the dataset analyzed. Many climatological fields---such as surface air temperature (SAT) or the geopotential height (GH) at a certain pressure level---can be used to define links via an analysis of significant correlations \cite{Tsonis2004,Yamasaki2008a,Steinhaeuser2009}.

Climate networks (CNs), which represent the statistical similarity structure of  spatio-temporal
resolved climatological variables, depend on the definition of nodes and links \cite{Zerenner2014}. Their regular spatial sampling results in a small-world topology \cite{Bialonski2010}, and thus, a careful interpretation of the inferred network is required. Nevertheless, CNs have been successfully employed to analyze climate features including the global connectivity
\cite{Donges2009,Donges2009a,Donges2011}, the identification of community structures \cite{Tsonis2008,Tantet2014} and the study of the possible collapse of the meridional overturning circulation (MOC)  \cite{VanderMheen2013} on the north Atlantic, among others.

A relevant drawback of this correlation analysis, which uses symmetric similarity measures (such as cross-correlation or mutual information), is that it yields non-directed networks where the presence of a link reveals inter-dependency but the direction associated (if any) of the underlying interaction is not established. For improving the understanding of climate phenomena and its predictability, it is of foremost importance not only to be able to infer the presence of a link between two nodes, but also, to infer the direction of this interaction.

A path to overcome this limitation is by constructing weighted climate networks, where the weight of each link is composed of two numbers: the correlation strength between the two nodes and the time delay  which maximizes this correlation strength. The sign of the time delay gives information about the direction of the link.  Using this approach, the high sensitivity of the network links to El-Ni\~no events was demonstrated, even in geographical regions far from the Pacific ocean \cite{Gozolchiani2008,Yamasaki2008b}.

An alternative approach for assessing the directionality of climate interactions involves the use of Granger Causality. In \cite{Wang2012} it was argued that the inclusion of several variability patterns like  NAO (North Atlantic Oscillation),  PDO (Pacific Decadal Oscillation),  ENSO (El Ni\~no--Southern Oscillation), and NPI (North Pacific Index)---which occur naturally and explain an important part of the global atmospheric variability---into a nonlinear network-like prediction method, largely improves the predictability of global temperature over seasonal time scales. This is so even when the time scales of the patterns used of in the order of decades. The study  suggested a causal directional influence among these major oceanic and atmospheric modes and global temperature variability, not only over their own time-scales but also over much shorter ones. Granger causality has also been used to test interdependence among ENSO and the Indian monsoon \cite{Mokhov2011}. A non-symmetric bidirectional and even alternating character of coupling was found that extends previous knowledge about the presence of negative correlation and intervals of phase synchronicity between the processes.

{ A third approach for directionality detection is based on information-theoretic measures \cite{Hlavackovaschindler2007}. For example, in \cite{Palus2014} information transfer from larger to smaller time scales was detected in daily-averaged SAT time series as the causal influence of the phase of slow oscillatory phenomena with periods of about 6-11 years in the amplitudes of the variability, characterized by smaller timescales, from a few months to 4-5 years.}

The reliability of directed climate networks detected by bivariate nonlinear methods based on information theory, compared to those generated by linear Granger causality analysis, was studied in \cite{Hlinka2013}. Several algorithms for estimating  transfer entropy with a wide range of parameter choices were considered.  As transfer entropy is a special case of the conditional mutual information, it reduces to Granger causality for linear Gaussian processes, and it usually requires longer time-series for accurate estimation. It was shown that all the causality methods considered provided reproducible estimates of climate directed interactions, with the linear method outperforming the nonlinear ones in terms of reliability.

In the present paper,  directed climate networks are constructed by using a predictability measure which, to the best of our knowledge, has not yet been applied to climatological data.  This bivariate analysis quantifies the amount of information contained in a time-series, $x(t)$, about the past of another time series, $y(t)$,  $\tau$ time units before. {More specifically,  the {\it directionality index} (DI) is used, which
quantifies the net direction of information flow between two time series, and has been successfully employed for the analysis of numerical data generated from coupled oscillators, and empirical data from cardiorespiratory recordings and electroencephalographic recordings, to name just a few \cite{Bahraminasab2008,Palus2003,Lehnertz2014}.}

The objective of this work is to demonstrate that the $\mbox{DI}$ methodology for inferring the net direction of information transfer indeed unveils climatologically relevant phenomena. Clear patterns of variability are uncovered, both in the tropics and in the extra-tropical regions, and their direction of propagation is shown to be in good agreement with the current understanding of main climate phenomena. The method is shown to work with both  monthly- and  daily-averaged data, and furthermore,  well-defined atmospheric patterns are uncovered on daily-averaged data, which are not seen with monthly-averaged data.

The paper is organized as follows. Section \ref{sec:methods} presents the data analyzed and the method used for constructing directed climate networks.  The statistical significance test is also discussed.  Section \ref{sec:results} presents the results and provide a comparison between directed and non-directed climate networks. Relevant patterns of global atmospheric variability  are interpreted from the maps obtained through the $\mbox{DI}$ analysis.  Finally Sec. \ref{sec:conclusions} presents the conclusions.

\section{\label{sec:methods} Data and Methods}
\subsection{Data}

We consider two datasets, both corresponding to SAT anomalies from the reanalysis of the National Center for Environmental Prediction/National Center for Atmospheric Research, (NCEP/NCAR)\cite{Kalnay1996}.  The data covers a regular grid over the Earth's surface with latitudinal and longitudinal resolution of $2. 5^\circ$, resulting in $N=10226$ grid points or network nodes.

The first dataset corresponds to monthly-averaged SAT data.  Since the data cover the period from January 1949 to December 2013, in each node there is a time series of $780$ data points.  The SAT anomalies are calculated as the actual temperature values minus the monthly average, and they are normalized by the standard deviation.  Each time-series is linearly detrended.

The second dataset consists of daily-averaged SAT from the same source, with the same spatial resolution, and covering the same period of time (the length of each time series is $23725$ data points).  In this case, for calculating the anomalies, the daily average has been subtracted from the actual temperature value, and the leap days have been discarded.  The time series are also detrended and normalized by the standard deviation.

\subsection{Directionality measure}

We consider the SAT anomalies time series in two nodes, $X(t)$ and $Y(t)$, which are characterized by probability distribution functions (PDFs) $p_X$, $p_Y$, and by their joint PDF, $p_{XY}$. To assess the directionality of the link between these two nodes  the \emph{directionality index} ($\mbox{DI}$) as defined in \cite{Bahraminasab2008,Palus2007a} is used:

\begin{equation}
\mbox{DI}_{XY}(\tau)=\frac{I_{XY}(\tau)-I_{YX}(\tau)}{I_{XY}(\tau)+I_{YX}(\tau)},
\label{eq:di}
\end{equation}
where $I_{XY}(\tau)$, $I_{YX}(\tau)$ (conditional mutual information) are defined as:
\begin{eqnarray}
I_{XY}(\tau) &=& I(X;Y|X_\tau) \nonumber \\ &=&  H(X|X_\tau) + H(Y|X_\tau) - H(X,Y | X_\tau);\\
I_{YX}(\tau) &=& I(Y;X|Y_\tau) \nonumber \\ &=&  H(Y|Y_\tau) + H(X|Y_\tau) - H(Y,X | Y_\tau),
\end{eqnarray}
with $X_\tau=X(t-\tau)$, $Y_\tau=Y(t-\tau)$ and $H(X|Y)$ being the conditional entropy \cite{Bahraminasab2008,Palus2007a}.

In terms of three time series $X(t)$, $Y(t)$ and $Z(t)$, $I(X;Y|Z)$ indicates the amount of information shared between $X(t)$ and $Y(t)$,  given the effect of $Z(t)$ over $Y(t)$. In this way, to assess the information transfer from $X(t)$ to $Y(t)$, $Z(t)$ is replaced by the past of time-series $X(t)$. Thus,  $I(X;Y|X_{\tau})$ quantifies the amount of information shared between $X(t)$ and $Y(t)$, given the influence of $X(t-\tau)$ over $Y(t)$. Analogously, to assess the information transfer from $Y$ to $X$, $Z(t)$ is replaced by the past of $Y(t)$. This special case of conditional mutual information is also referred as {\em transfer entropy}.

The directionality index, $\mbox{DI}_{XY}$, then quantifies the \emph{net} information flow. From the definition of $\mbox{DI}_{XY}$, Eq.(\ref{eq:di}), it is clear that  $\mbox{DI}_{XY}=-\mbox{DI}_{YX}$. Also, $-1 \leq \mbox{DI}_{XY} \leq 1$: $\mbox{DI}_{XY}=1$ if and only if $I_{XY}\ne 0$, $I_{YX}=0$ (\textit{i.e.}, the information flow is $X\rightarrow Y$ and there is no back coupling $Y\rightarrow X$) and $\mbox{DI}_{XY}=-1$ if and only if $I_{XY}= 0$, $I_{YX}\ne0$ (\textit{i.e.}, the information flow is $Y\rightarrow X$ and there is no back coupling $X\rightarrow Y$).

Naturally, $\tau > 0$ is a parameter that has to be tuned appropriately to the time-scales relevant to the particular dataset.
If $\tau$ is too small $\mbox{DI}_{XY}(\tau)$ will capture short time scale directionality, and may fail if the time series behave too similarly on those time scales as they do if they are subjected to the same external forcing. On the other hand, if $\tau$ is too large, larger than the decorrelation time of the time-series, the effect of the past $X$ over $Y$ (and of $Y$ over $X$) will be negligible and $\mbox{DI}_{XY}(\tau)$ will be a small and in principle random value. In the next section the criterion employed for accessing the statistical significance of the $\mbox{DI}$ values will be discussed, and in section \ref{sec:results},  a thorough study of the effect of varying the parameter $\tau$ will be performed. It is worth noticing that in the two datasets considered, $\tau = 1$ has a different meaning: in the monthly-averaged SAT time-series, the minimum value of $\tau$ is one month, while in the daily-averaged data, the minimum value of $\tau$ is one day.

To compute the PDFs associated to each time series, and the joint PDFs a 10-bin histograms of values is used. An alternative approach is to use a symbolic transformation known as \emph{ordinal analysis} \citep{Amigo2010,Barreiro2011,Bahraminasab2008, Deza2013}. A preliminary study suggests that using ordinal analysis offers as an additional advantage the possibility of finding the directionality of the links at different time-scales; this study is still in progress.

It is important to note a drawback of using the directionality index for network construction: it does not distinguish indirect from direct information transfer. Given three time-series, $X$, $Y$ and $Z$,  positive and significant $\mbox{DI}$ values  $I_{XY}>0$ and $I_{XZ}>0$ might also give  a significant value of $I_{YZ}$---either positive or negative---even if there is no ``direct'' information in $Y$ about the future or the past of $Z$, as the information will have been ``indirectly'' contained in $X$.

\section{\label{sec:results} Results}

{Firstly, the results of the statistical significance analysis of link directionality, when using monthly-averaged SAT datasets, are presented. Monthly datasets are used in order to to compare with the undirected networks reported in \cite{Deza2013}. Afterwards, the results of the analysis of the daily-averaged datasets are presented. They are consistent with the monthly datasets results, furthermore uncovering additional patterns of atmospheric variability, not observed in monthly data as the averaging procedure filters high frequency and variability.}

\subsection{Statistical significance analysis} \label{statistAnalysis}

\begin{figure*}[htb]
\begin{center}
\includegraphics[width=0.32\textwidth]{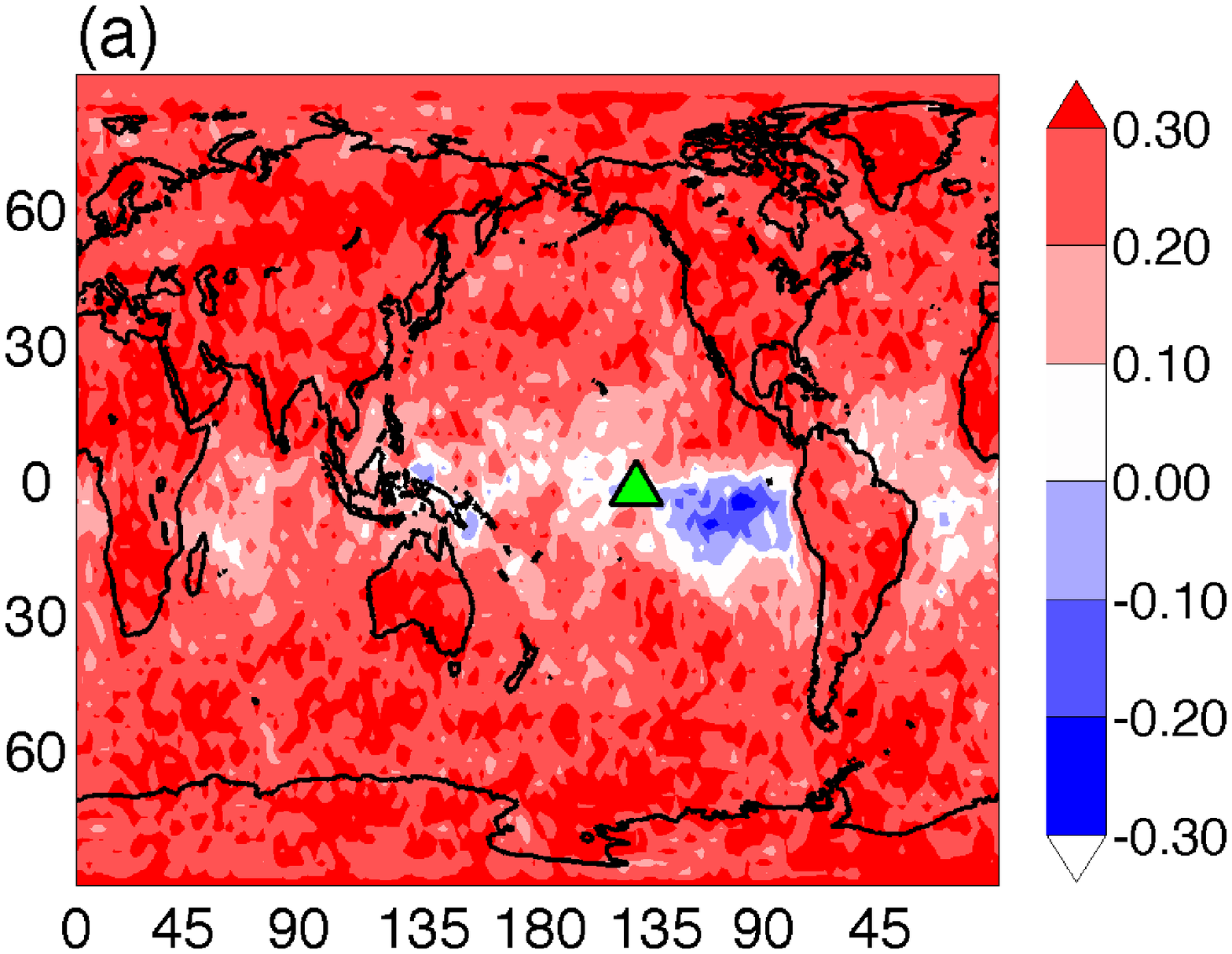}
\includegraphics[width=0.32\textwidth]{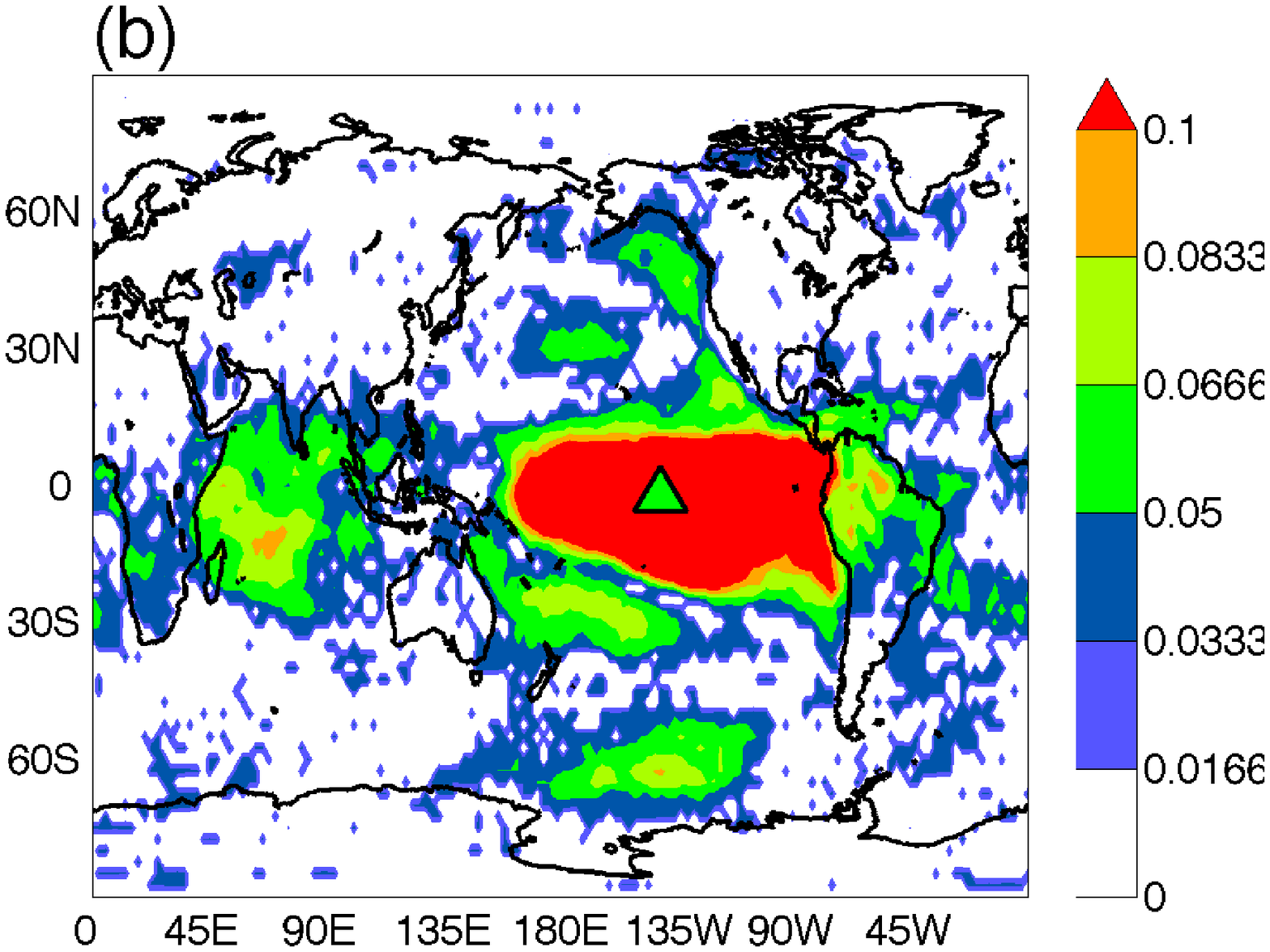}
\includegraphics[width=0.32\textwidth]{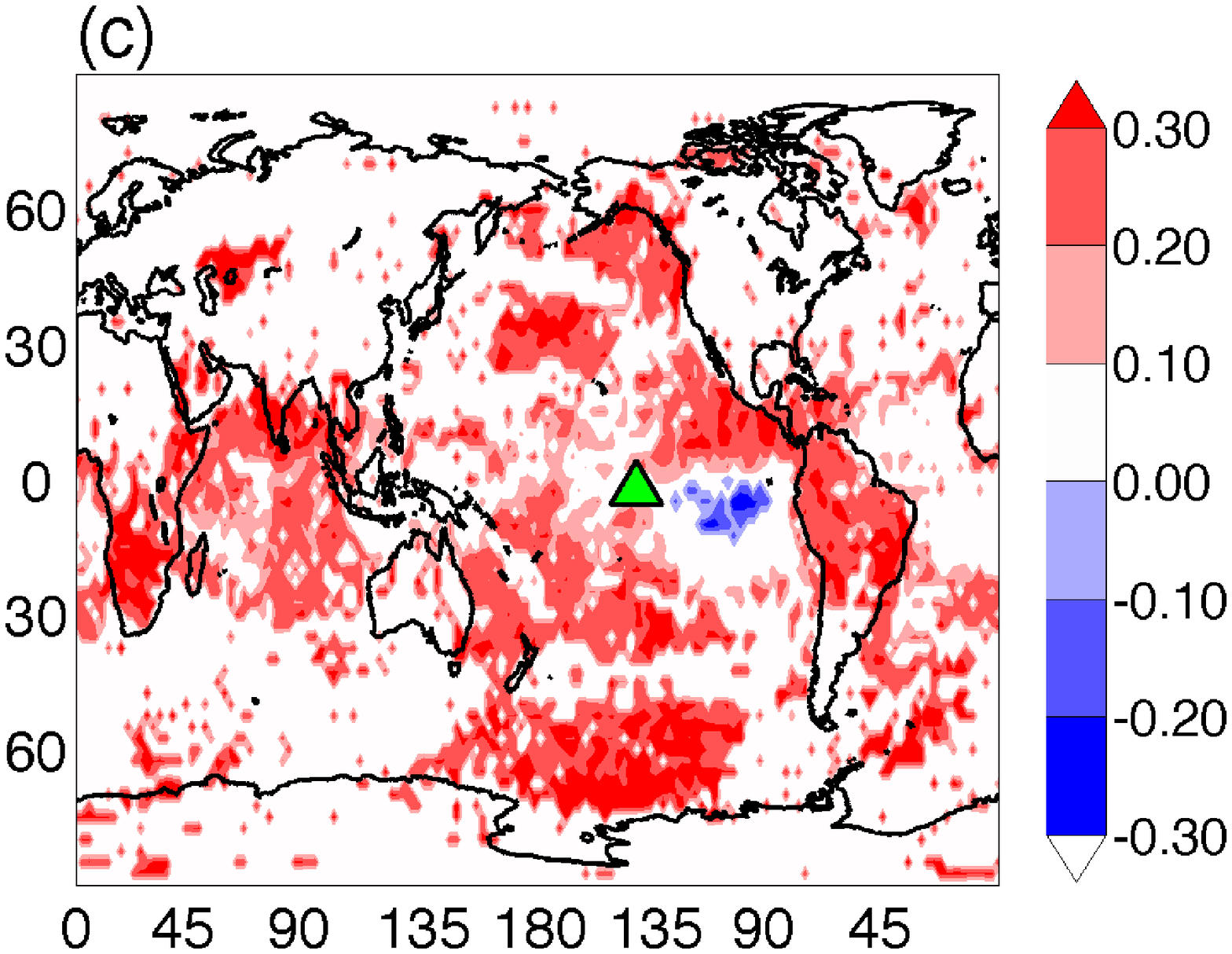}
\includegraphics[width=0.32\textwidth]{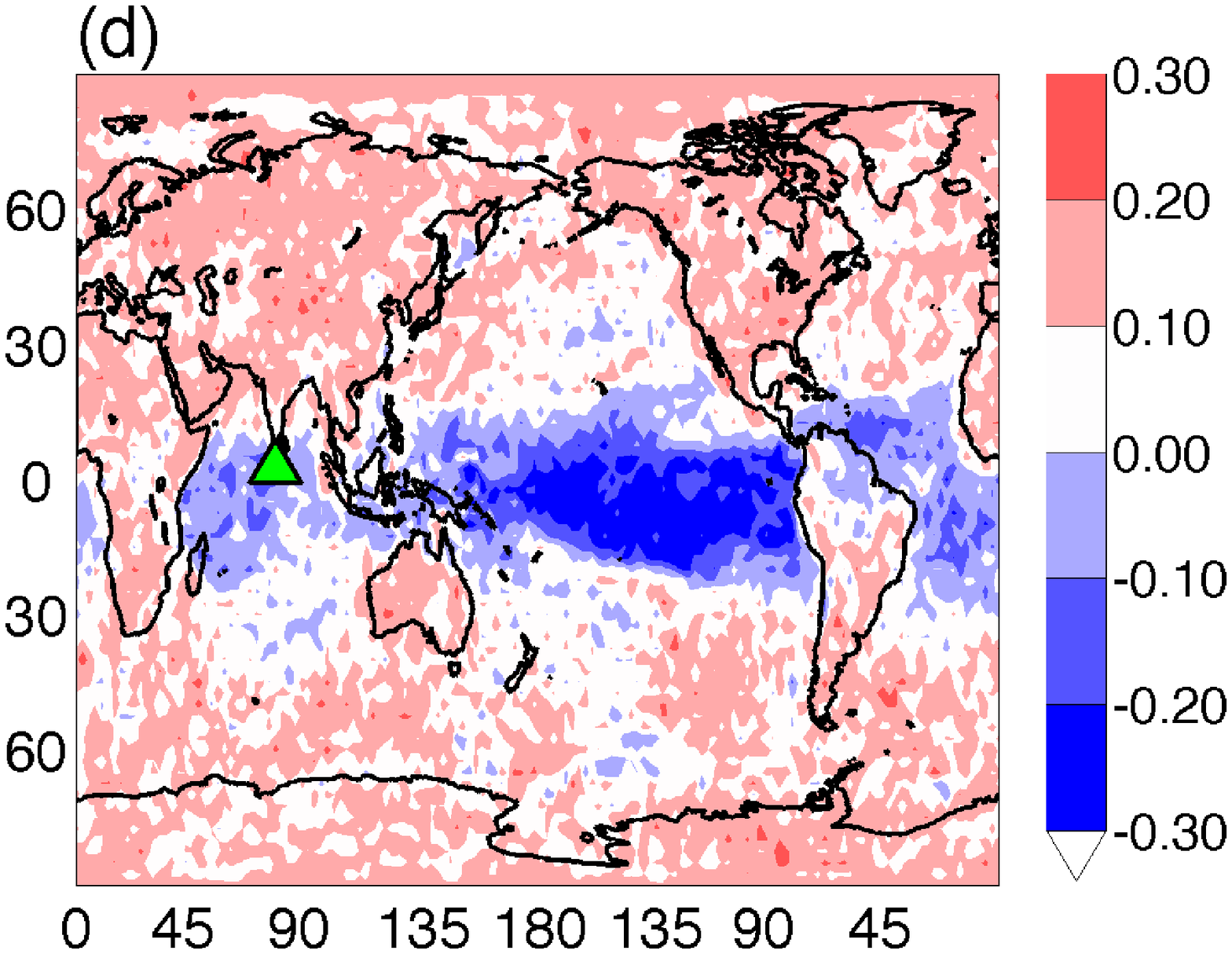}
\includegraphics[width=0.32\textwidth]{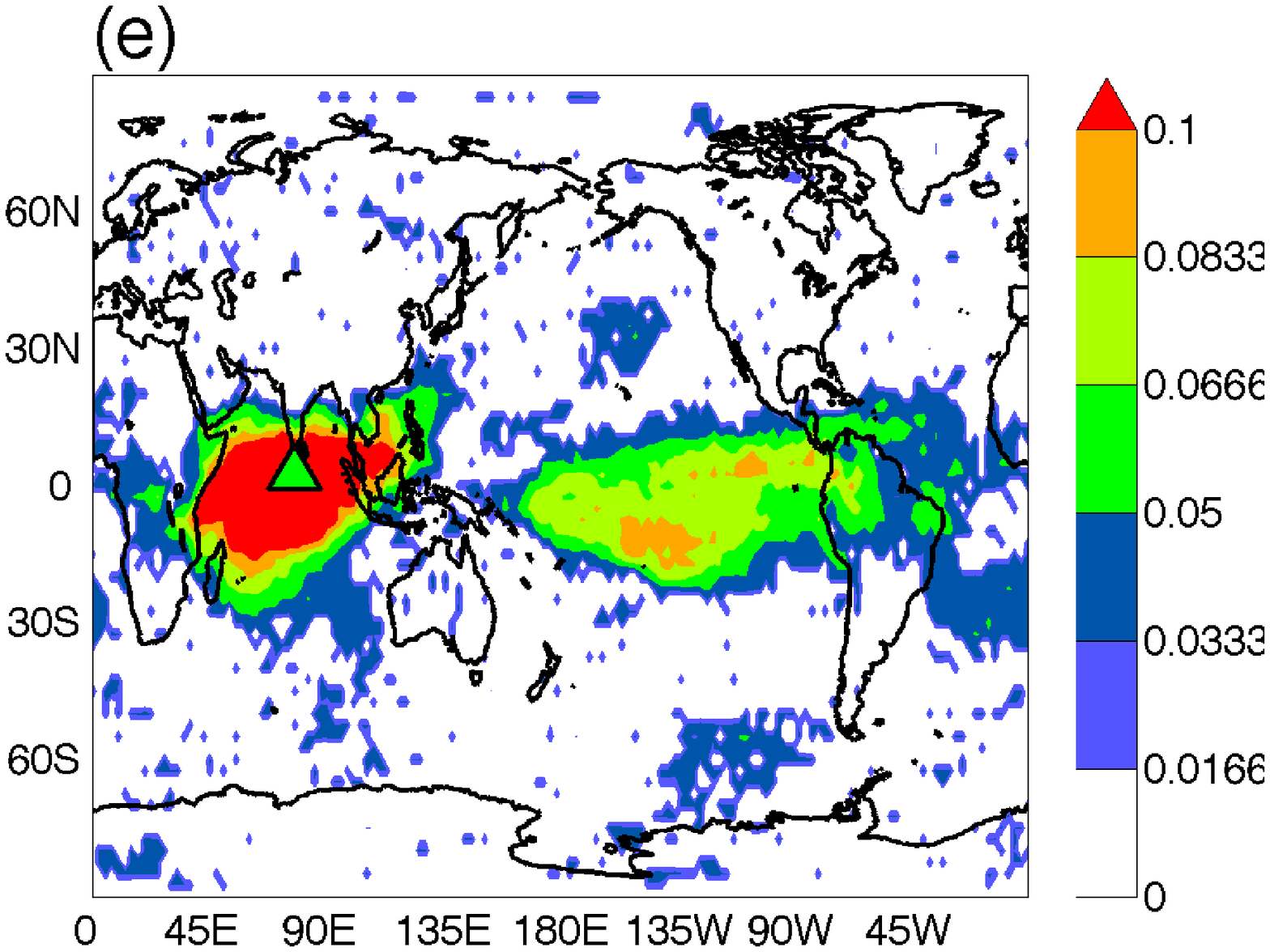}
\includegraphics[width=0.32\textwidth]{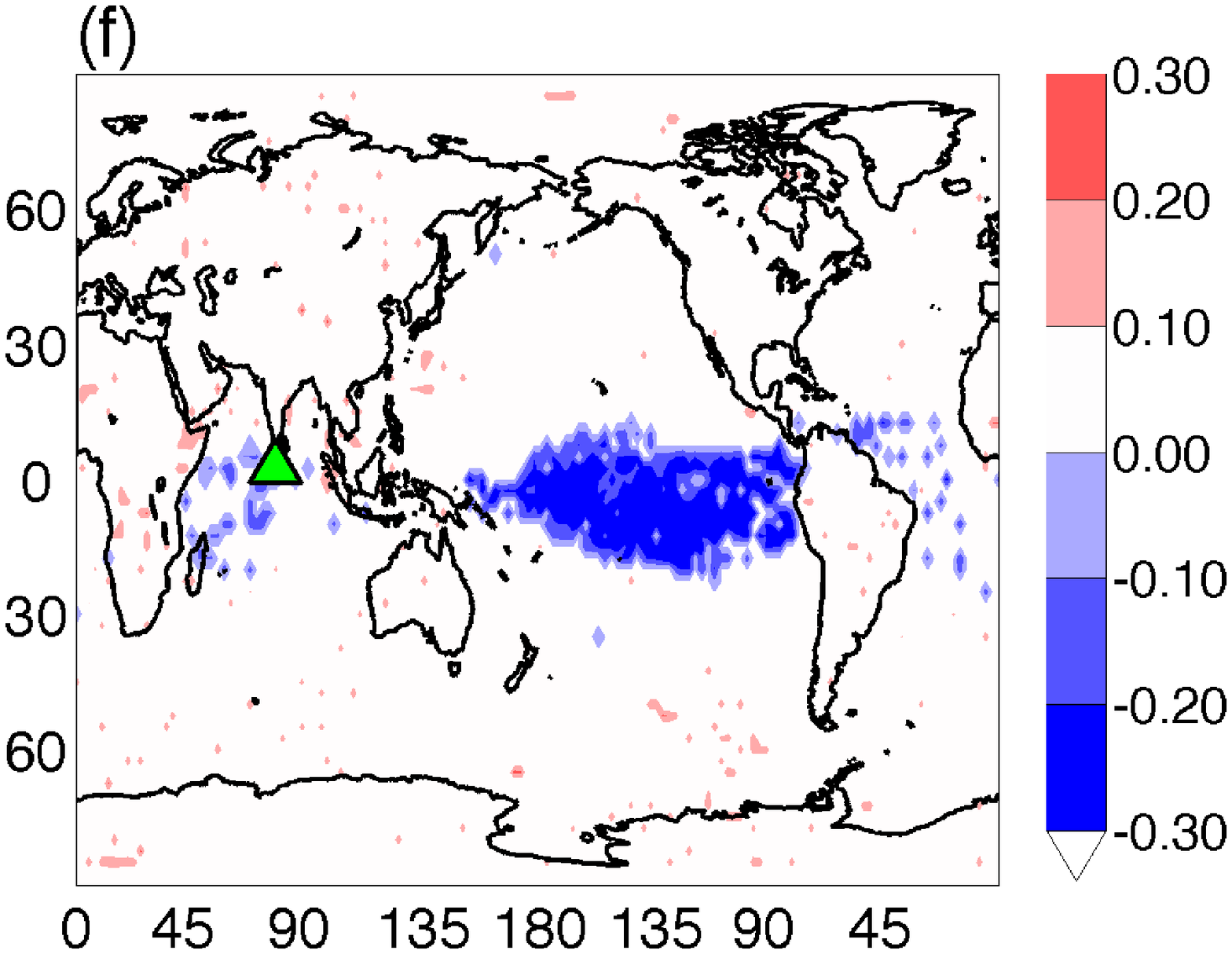}
\caption{(Color Online).
{\color{black}{Procedure of constructing significant directionality maps from raw $\mbox{DI}$ calculations.  In panels (a) and (d) the unfiltered $\mbox{DI}$ maps are shown for two nodes, one in the central Pacific ocean and one in the Indian ocean, indicated with triangles. In panels (b) and (e) only the statistically significant $\mbox{MI}$ values are shown. These results are combined in (c) and (f) where only the links that have both, statistically significant $\mbox{MI}$ and $\mbox{DI}$ values are shown.}}
} \label{fig:construction}
\end{center}
\end{figure*}

\begin{figure}[htb]
\begin{center}
\includegraphics[width=0.49\textwidth]{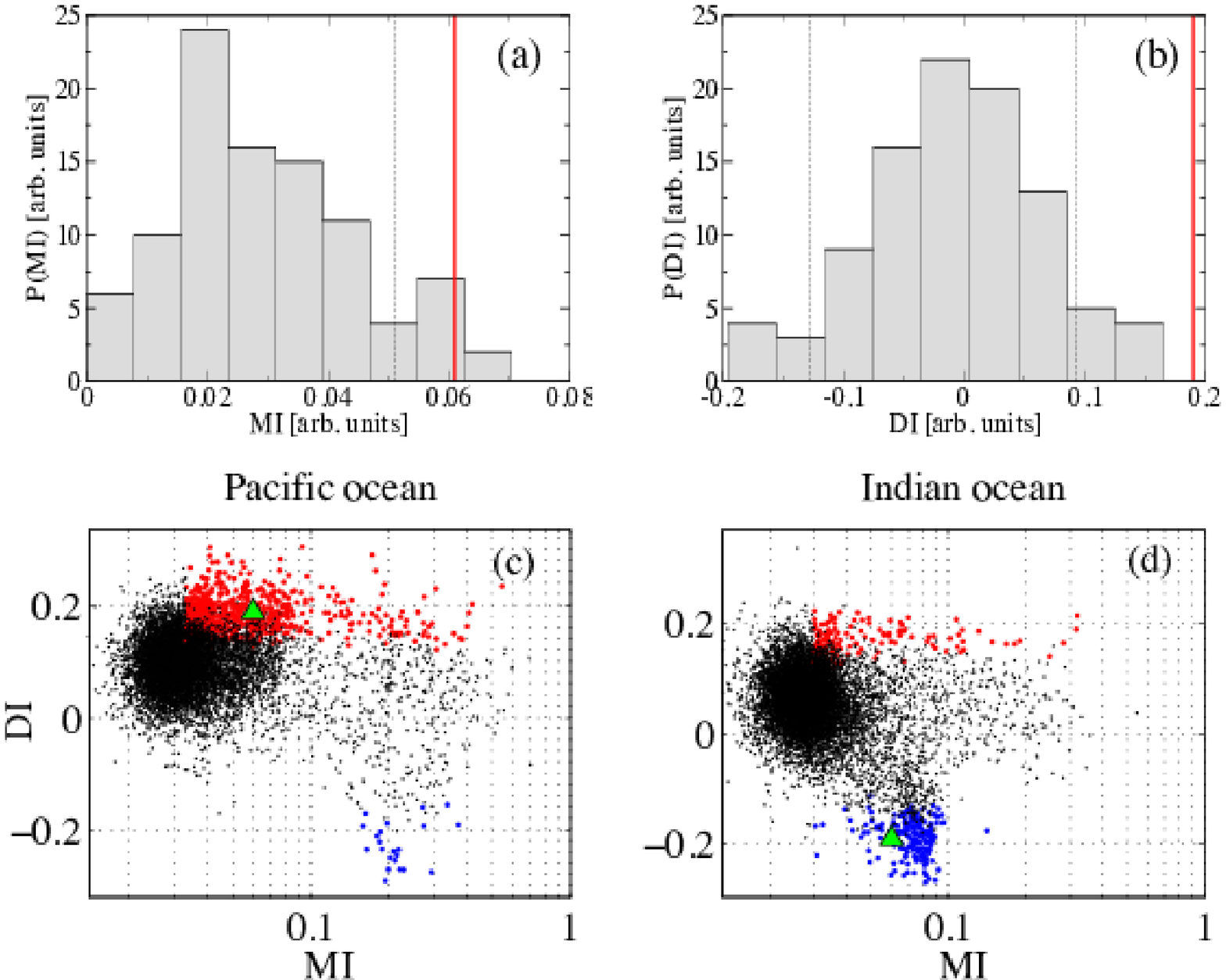}
\caption{(Color Online). (a), (b) Histograms of the $\mbox{MI}$ and $\mbox{DI}$ values ($\tau=1$ month) computed from $100$ BS surrogates, the dashed vertical lines indicate the significance thresholds and the solid lines (red online) indicate the $\mbox{MI}$ and $\mbox{DI}$ values computed from the original data. (c), (d) Plot of $\mbox{DI}$ ($\tau=1$ month) \textit{vs.} $\mbox{MI}$ for all the links of the two nodes in the Pacific and Indian oceans considered in Fig. \ref{fig:construction}. Incoming links are indicated in blue and outgoing links, in red. Significant links are plotted in red or blue dots; the black dots indicate the disregarded links (either because $\mbox{MI}$ is not statistically significant, or because $\mbox{MI}$ is significant but $\mbox{DI}$ is not). The triangles indicate the ($\mbox{MI}$, $\mbox{DI}$) values of the particular link between the two nodes.} \label{fig:scatter}
\end{center}
\end{figure}

To address the significance of $\mbox{DI}$ values, 100 surrogates were generated using the bootstrap (BS) algorithm \cite{Mudelsee2010}. {The  null hypothesis considered for $\mbox{MI}$ is that the processes are independent from each other. The null hypothesis for $\mbox{DI}$ is that there is no preferential direction of the interaction.}

 The BS algorithm randomly resamples with replacement from the original datasets using blocks of data of approximately the size of the autocorrelation time of the time series, and then computes the estimators ($\mbox{MI}$ and  $\mbox{DI}$) from the resampled data. {\color{black} Doing so, both the statistics (histogram) and the power spectrum of the original time series are approximately preserved. In this way, two different empirical distributions (one-tailed for $\mbox{MI}$ and two-tailed for $\mbox{DI}$), were obtained for each link, and significance thresholds were extracted from them.}

{Afterwards, for each link, the $\mbox{MI}$ value was calculated from the original datasets and a significance test was applied. A MI value is considered significant if: $\mbox{MI}>\Theta_{MI}$, where $\Theta_{MI}$ is the threshold derived from the bootstrap MI distribution.
In addition---for the $\mbox{MI}$ significant links only---the $\mbox{DI}$ value computed from the original dataset was compared to the $\mbox{DI}$ thresholds $\Theta^+_{DI}$ and $\Theta^-_{DI}$. The $\mbox{DI}$ value was considered significant if $\mbox{DI}>\Theta^+_{DI}$ or if $\mbox{DI}<\Theta^-_{DI}$. This two-step significance test was performed with  the scope of assessing firstly the presence of a link, and afterwards its directionality.}

A graphical explanation of the full procedure is shown in Fig. \ref{fig:construction} where, for two nodes (one over the Pacific and one over the Indian ocean) the unfiltered $\mbox{DI}$ maps are displayed in panels (a,d),  the significant $\mbox{MI}$ values are displayed in panels (b,e) and finally the $\mbox{DI}$ significant  values for  the $\mbox{MI}$ significant links are displayed in panels (c,f). The $\mbox{DI}$ maps show positive $\mbox{DI}$ values  in red, which mean outgoing links, while the incoming links are shown in blue.

The significance thresholds, $\Theta_{MI}$, $\Theta^+_{DI}$ and $\Theta^-_{DI}$, extracted from the BS surrogates, were computed as 
$\Theta_{MI}=\mu_{MI} + 3 \times \sigma_{MI}$, $\Theta^+_{DI}=\mu_{DI} + 3 \times \sigma_{DI}$, and $\Theta^-_{DI}=\mu_{DI} - 3 \times \sigma_{DI}$, 
with $\mu$ and $\sigma$ being the mean and standard deviation of the corresponding $\mbox{MI}$ and $\mbox{DI}$ bootstrap distributions. 
{\color{black} The process is illustrated in Fig. \ref{fig:scatter}, panels (a) and (b). It can be noticed that these distributions are bell-shaped, which justifies the use of classical statistics. Also notice from panel \ref{fig:scatter}(b) that for a sufficiently large number of BS surrogates, the distribution tends to be centered around a rather small $\mbox{DI}$ value, which motivates the null hypothesis of no preferential direction of the interaction. }

The more time and memory consuming 99nth percentile test was also used, and yielded similar results.

{{In order to visualize the results of the significance criterion employed, in Fig. \ref{fig:scatter} panels (c) and (d) display---as an example---the $\mbox{DI}$ \textit{vs} $\mbox{MI}$ values of all the links of the two nodes, one located in the central Pacific and the other in the Indian ocean (the nodes are the same as in Fig. \ref{fig:construction}). The black dots indicate the disregarded links (either because $\mbox{MI}$ is not significant, or because $\mbox{MI}$ is significant but $\mbox{DI}$ is not). Significant links are plotted in red (outgoing links, $\mbox{DI} >\Theta^+_{DI}$) and in blue (incoming links, $\mbox{DI} <\Theta^-_{DI}$).}}

{{In Fig. \ref{fig:scatter}, panels (c) and (d) also show that high $\mbox{MI}$ values do not imply high $\mbox{DI}$ values.
In Fig. \ref{fig:scatter} (d) one can notice that most of the blue dots are located in a narrow range of $\mbox{MI}$ values while they are distributed in terms of $\mbox{DI}$ values. An inspection of panel (f) in Fig. \ref{fig:construction} shows that the blue links come to the node in the Indian Ocean from a well-defined region in the central Pacific Ocean. On the other hand, one can notice in Fig. \ref{fig:scatter} (d) that the few red dots are more distributed in the $\mbox{MI}$, $\mbox{DI}$ plane, and in Fig. \ref{fig:construction} (f) the red outgoing links connect the node in the Indian Ocean to various regions on Earth.}}

%{The second row of figure \ref{fig:scatter} indicates the typical distribution of surrogate data for a single link, for  \ref{fig:scatter} (c) $\mbox{MI}$, and  \ref{fig:scatter} (d) $\mbox{DI}$. Observe that the distributions are bell-shaped. In this example, the link considered is the one between the location of the panels \ref{fig:scatter} (a,b), where  a green triangle marks the position of this link in both panels.  As can be seen, both MI and DI are significative as the measures calculated with the original data---marked with a red line---are located outside the thresholds---marked with red lines---of the surrogate data.}

\begin{figure}[htb]
\begin{center}
\includegraphics[width=0.23\textwidth]{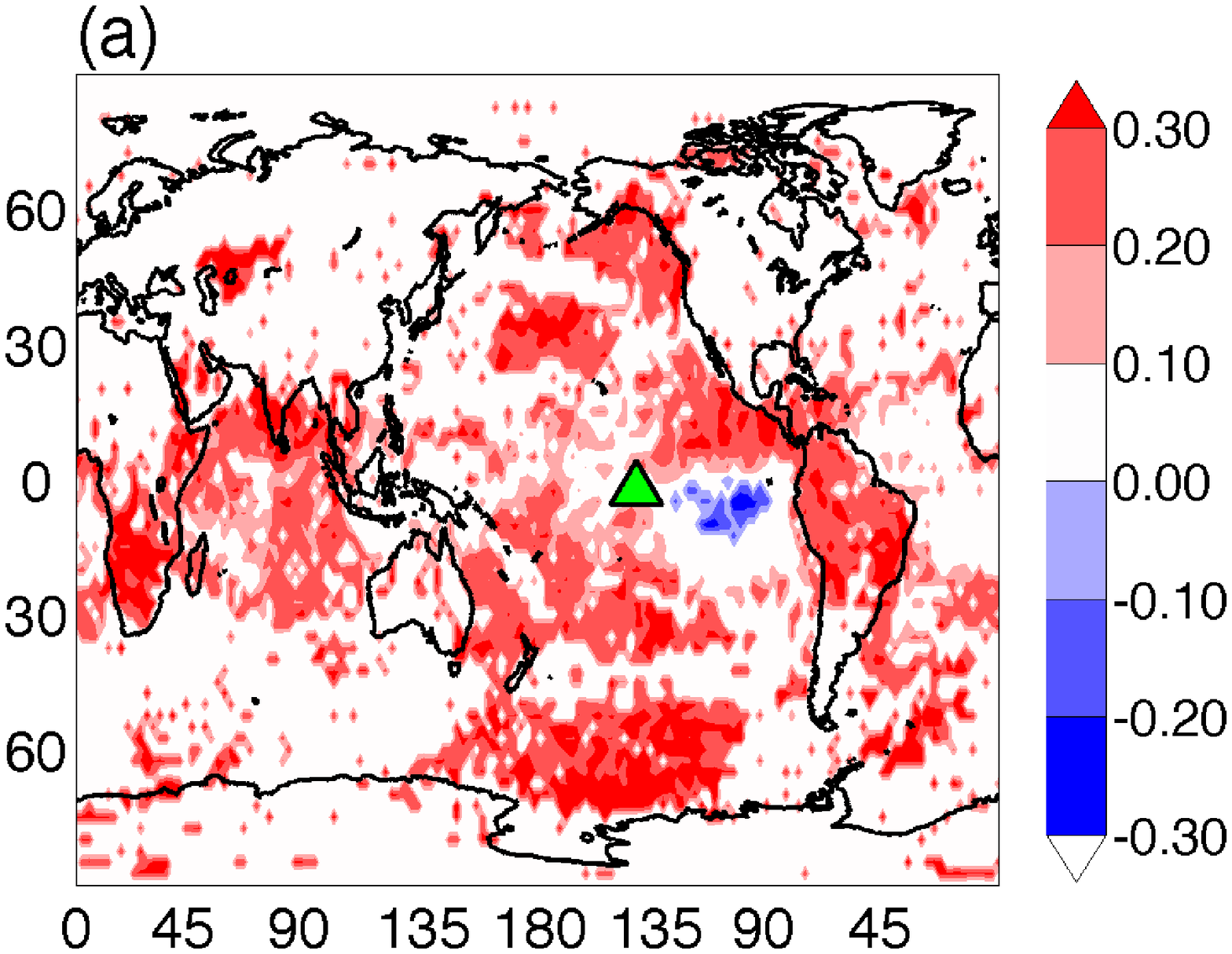}
\includegraphics[width=0.23\textwidth]{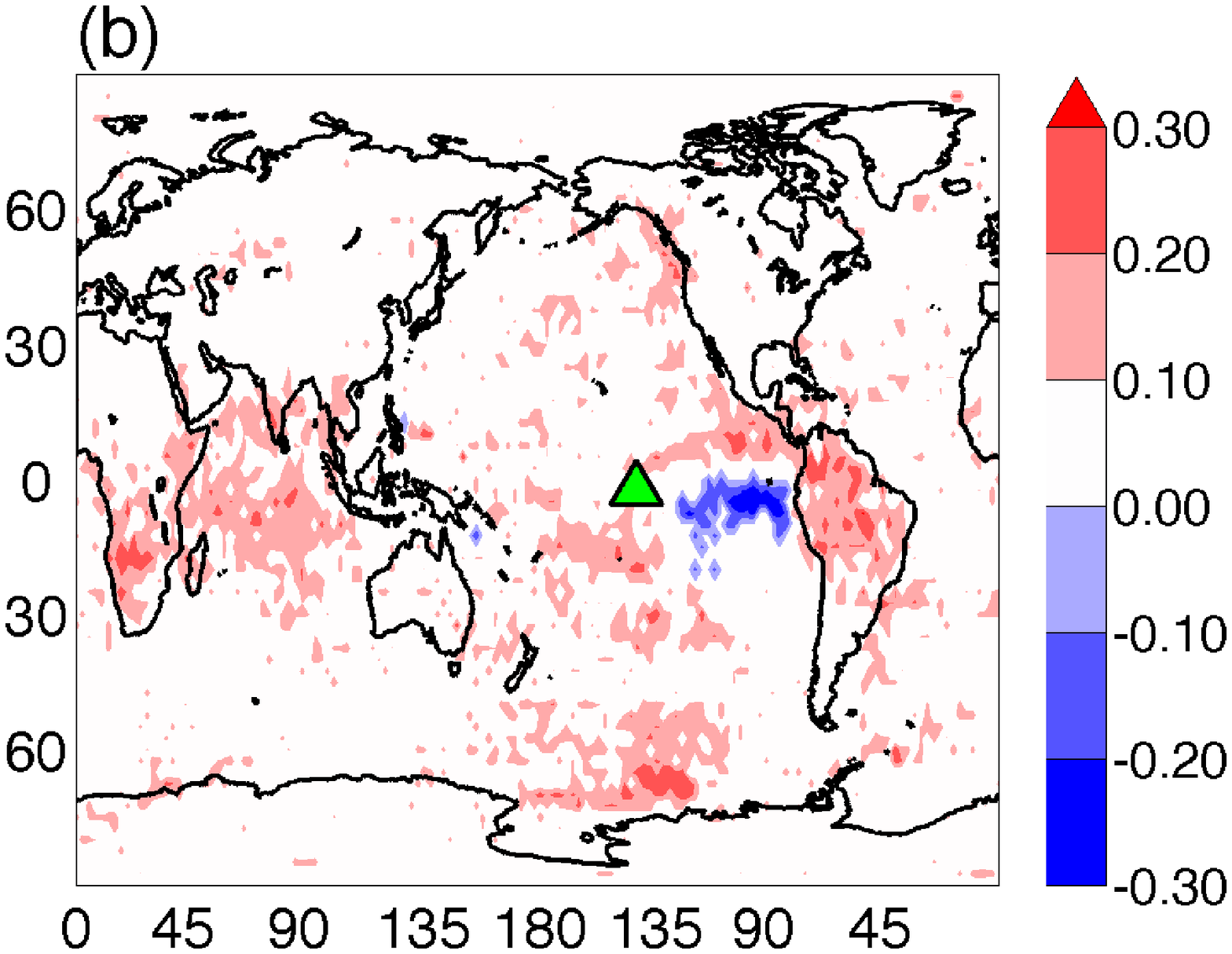}
\includegraphics[width=0.23\textwidth]{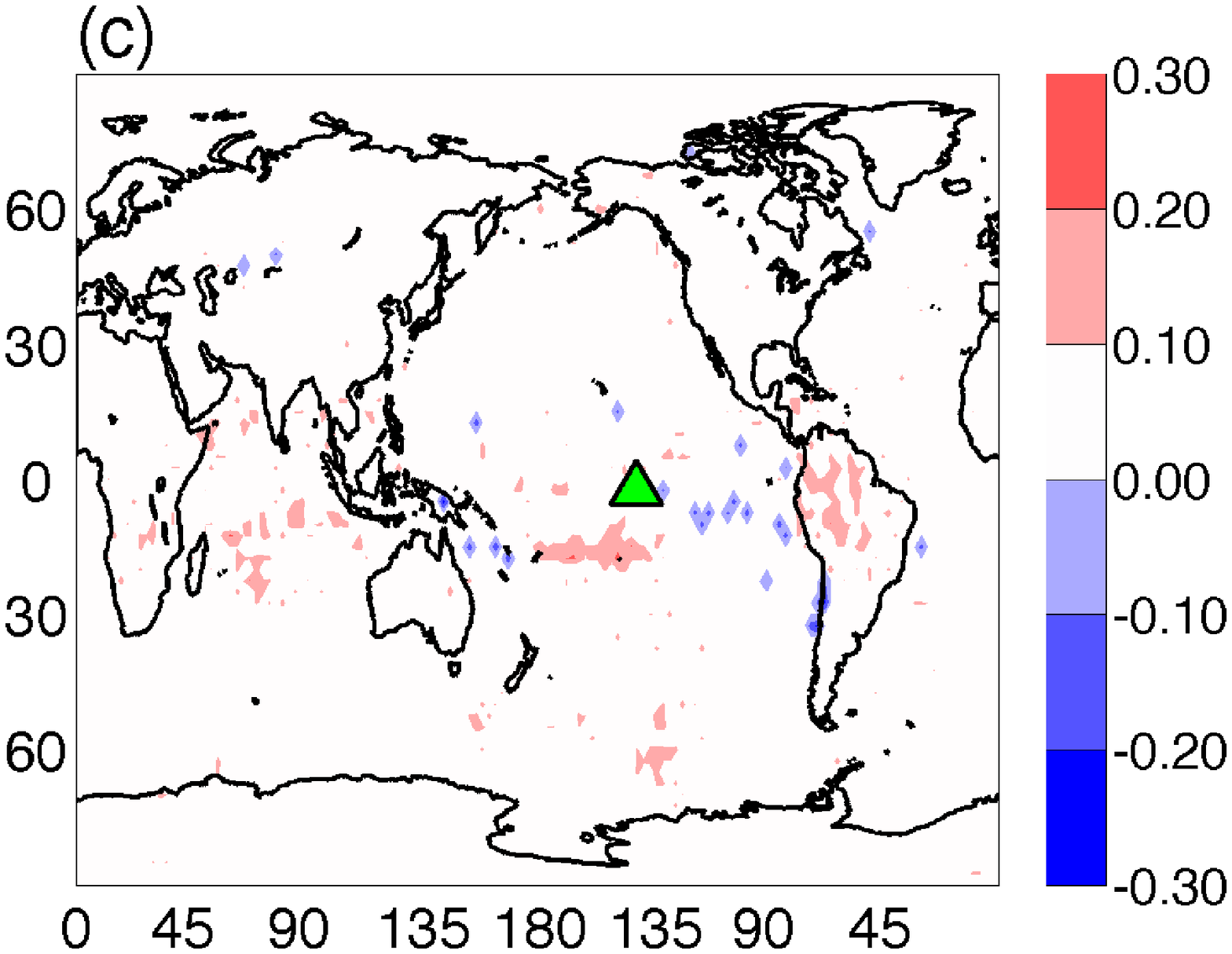}
\includegraphics[width=0.23\textwidth]{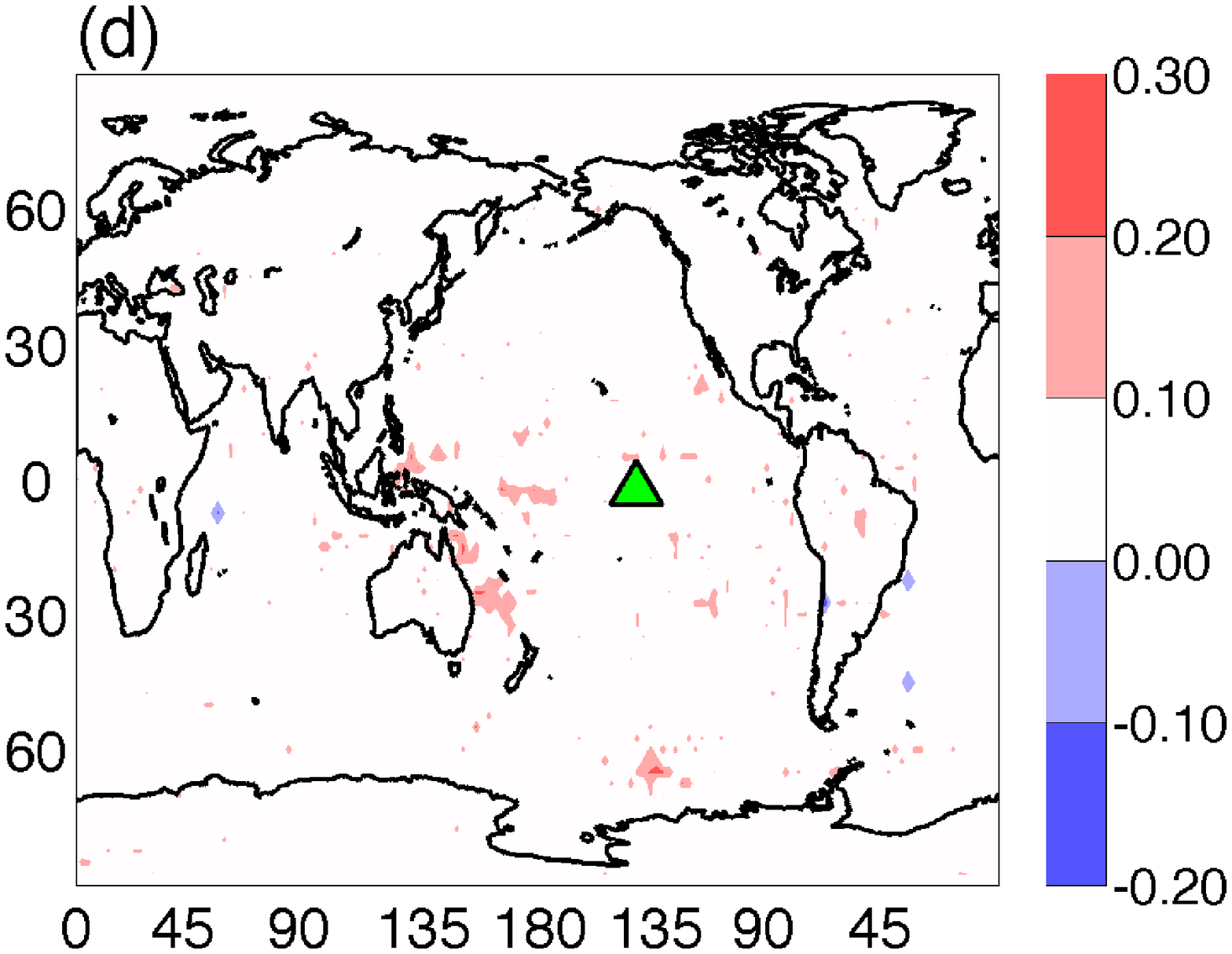}
\caption{(Color Online).
Effect of $\tau$ on tropical areas using monthly-averaged data.  In this case a point in central pacific (the same point used in the top row of Fig. \ref{fig:construction} and in \cite{Deza2013} is considered.  The values of $\tau$ are: (a) 1 month, (b) 3 months, (c) 6 months, (d) 12 months. Notice the decorrelation of the time series for large $\tau$. Incoming links are in blue while outgoing links are in red.
} \label{fig:tauMonthlyTropics}
\end{center}
\end{figure}

\begin{figure}[htb]
\begin{center}
\includegraphics[width=0.23\textwidth]{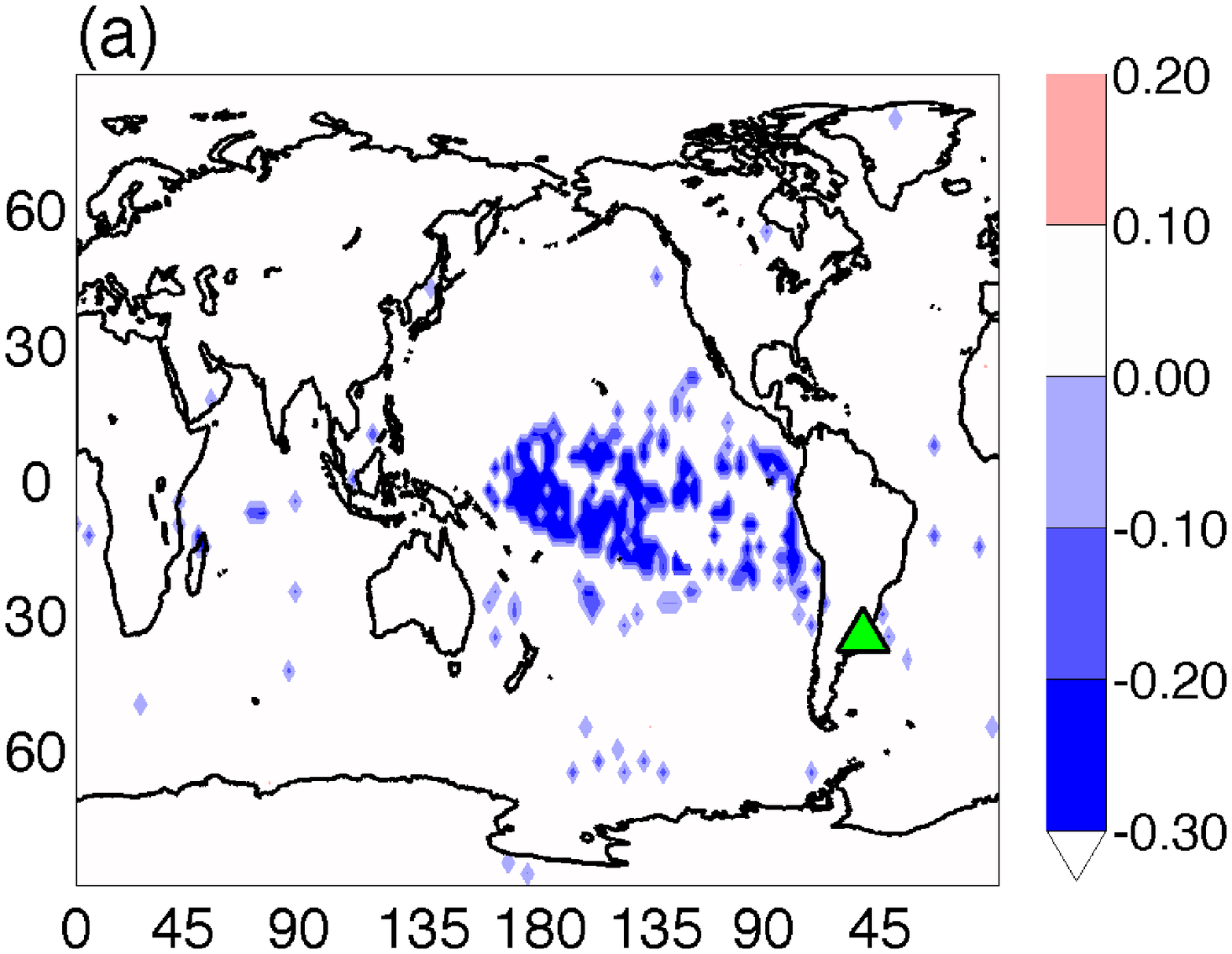}
\includegraphics[width=0.23\textwidth]{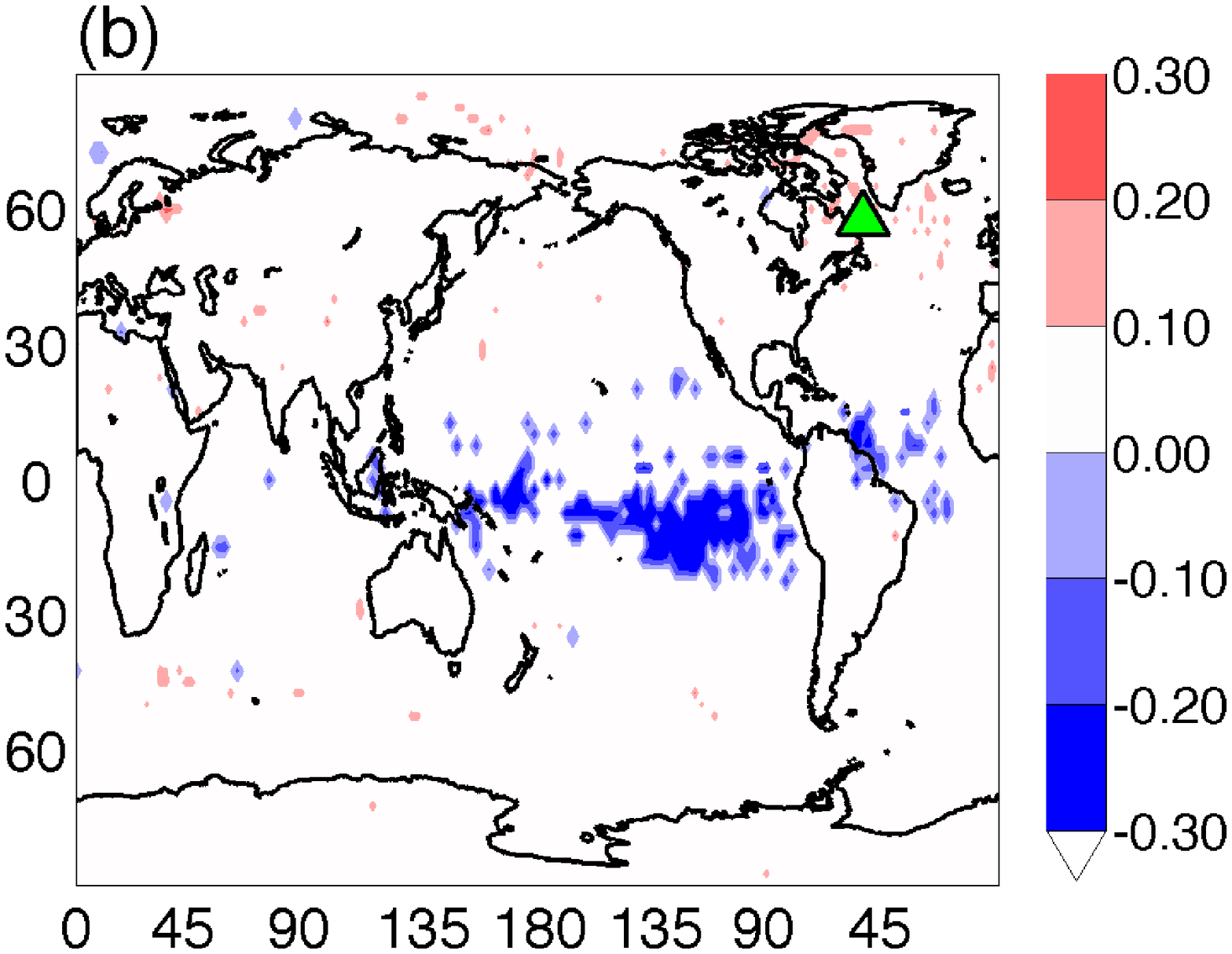}
\includegraphics[width=0.23\textwidth]{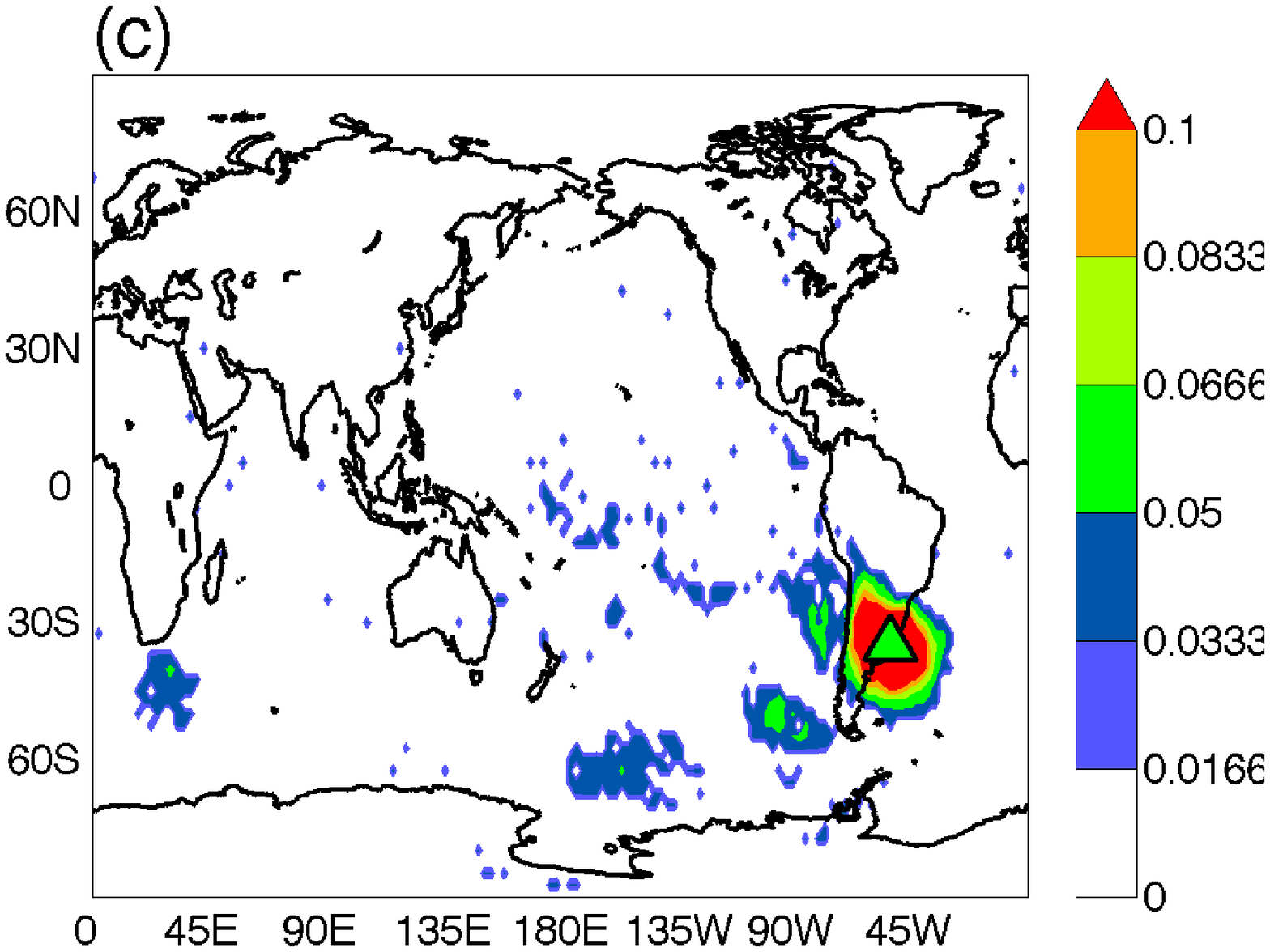}
\includegraphics[width=0.23\textwidth]{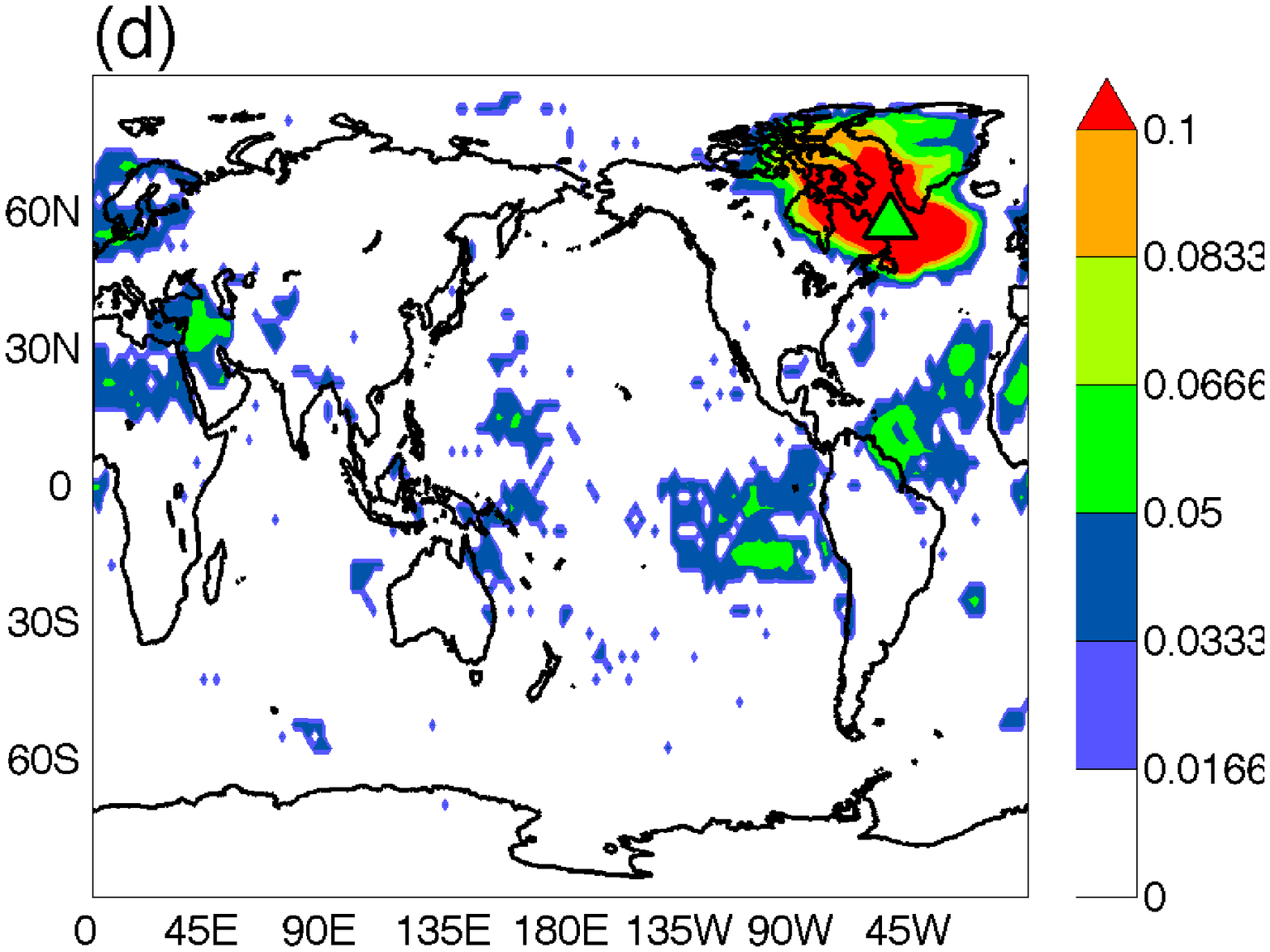}
\caption{(Color Online).
{(a), (b) Directionality of the significant links ($\tau=1$ month) of two nodes in the extratropics indicated with triangles: (a) in southern South America (de la Plata basin) and (b) in Labrador Sea. As the decorrelation time on the extratropics is very fast, no clear structures are seen. In order to show that it is not a problem of statistics, panels (c) and (d) display the significant $\mbox{MI}$ values.}
}\label{fig:tauMonthlyExtraTropics}
\end{center}
\end{figure}

\subsection{Analysis of monthly-averaged SAT anomalies}

\subsubsection{Influence of the parameter $\tau$} 

As stated in the methodology the correct choice of the value of $\tau$ is necessary for obtaining consistent results.  As $\tau$ can be only an integer, using monthly averaged data, its minimum value will be of one month.  In the tropical areas the  influence of the ocean on the surface air temperature is a dominant characteristic. Moreover, because of the large heat capacity of water and the ocean's dynamics, the sea surface temperature (SST) anomalies vary in the scale of months. Calculating  $\mbox{DI}$ for a point in the central pacific (NINO3 area) for different values of $\tau$ yields the results shown in Fig.  \ref{fig:tauMonthlyTropics}. The point considered is the same as in Fig.  \ref{fig:construction} (a-c) and moreover, the panel \ref{fig:construction} (c) is the same as panel \ref{fig:tauMonthlyTropics} (a).

For $\tau=1$ Fig.\ref{fig:tauMonthlyTropics} (a) shows the central Pacific influenced by (in blue) the eastern Pacific and influencing (in red), presumably through atmospheric teleconnections, the global tropics and the extratropical Pacific ocean. However, as $\tau$ grows the number of significant connections decreases, suggesting that the time-scale of decorrelation of the SAT is less than 6 months. This is consistent with the persistence time scale of 3 to 6 months of observed sea surface temperature.

The extratropical atmosphere shows larger internal variability than the tropics and the impact of the extratropical SST on the atmosphere is much more limited than in the tropics. Thus, the variability of extratropical SAT is dominated by synoptic atmospheric dynamics and has time scales of a few days. Longer persistence time scales might appear in the extratropics if the region is influenced by tropical SST. This motivates the use of a small value of $\tau$ when considering extra-tropical variability.

In Fig.  \ref{fig:tauMonthlyExtraTropics} the $\mbox{DI}$ and $\mbox{MI}$ maps for two points in the extratropics are shown.  Panels \ref{fig:tauMonthlyExtraTropics} (a,c) show links related to a point in southeastern South America, while panels \ref{fig:tauMonthlyExtraTropics} (b,d) show links related to a point in the Labrador sea, whose characteristics are linked to the North Atlantic Oscillation (NAO)\cite{Deza2014}. The top panels show $\mbox{DI}$ for $\tau=1$ month while the bottom panels show $\mbox{MI}$.

Consistent with the previous description, the extratropical SAT show only some incoming links from the tropical region for $\tau=1$ month. The point over the Labrador Sea seems to show also some outgoing links to the northeast, although there is no clear structure. {\color{black}As shown in Sec. \ref{daily}, a better identification of link directionality is obtained by using daily data.}

\subsubsection{Comparison between monthly and daily datasets}
\begin{figure}[htb]
\begin{center}
\includegraphics[width=0.23\textwidth]{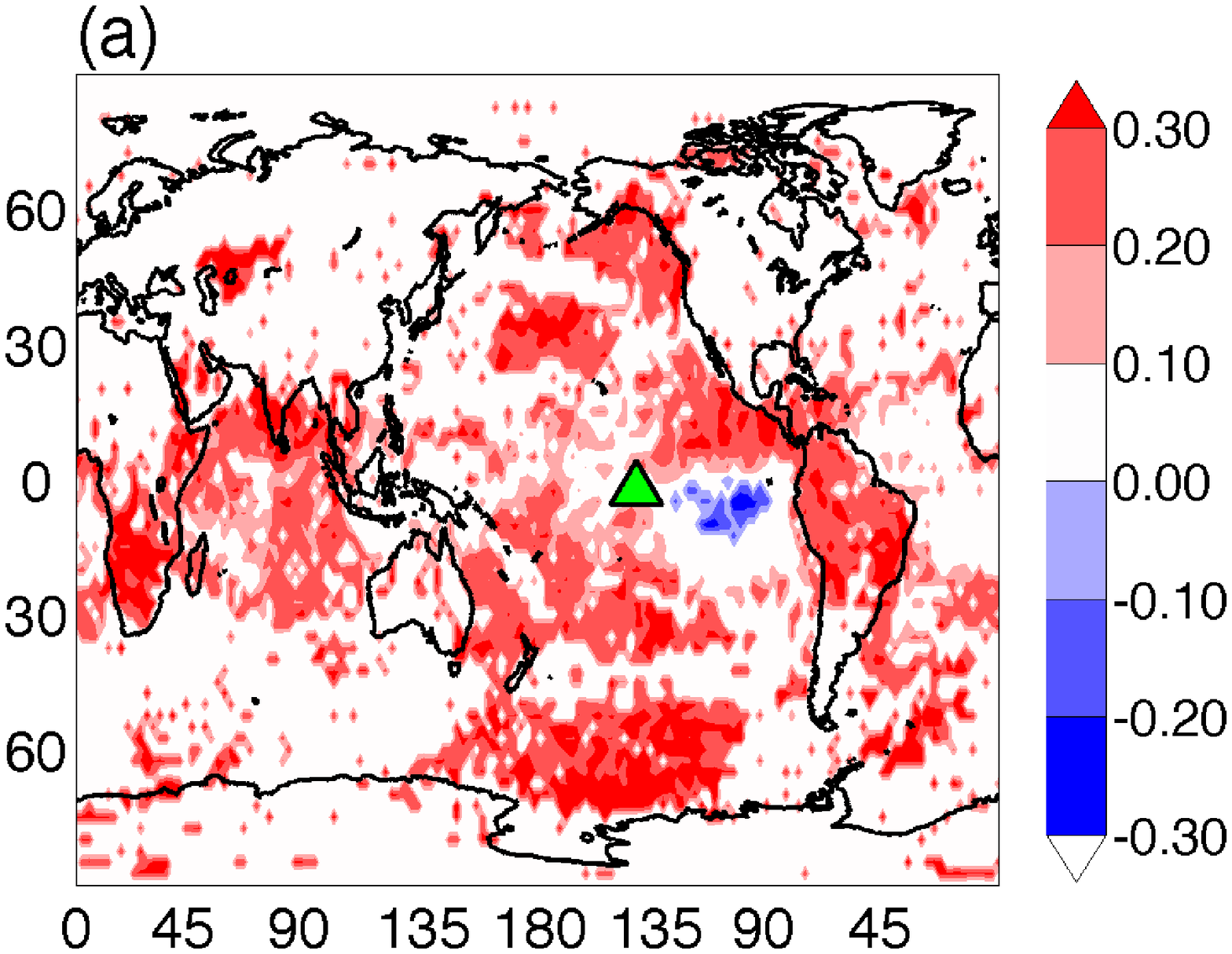}
\includegraphics[width=0.23\textwidth]{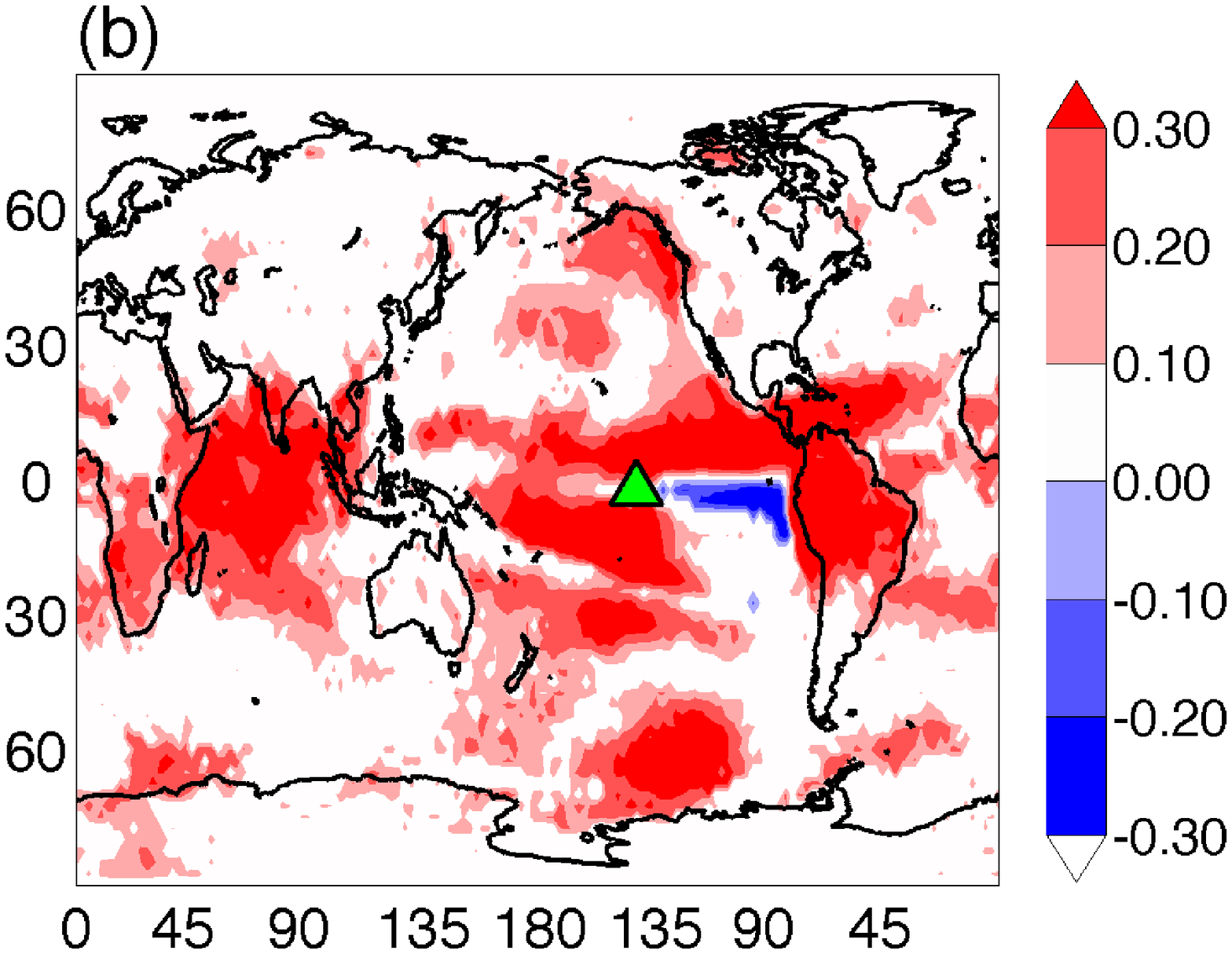}
\includegraphics[width=0.23\textwidth]{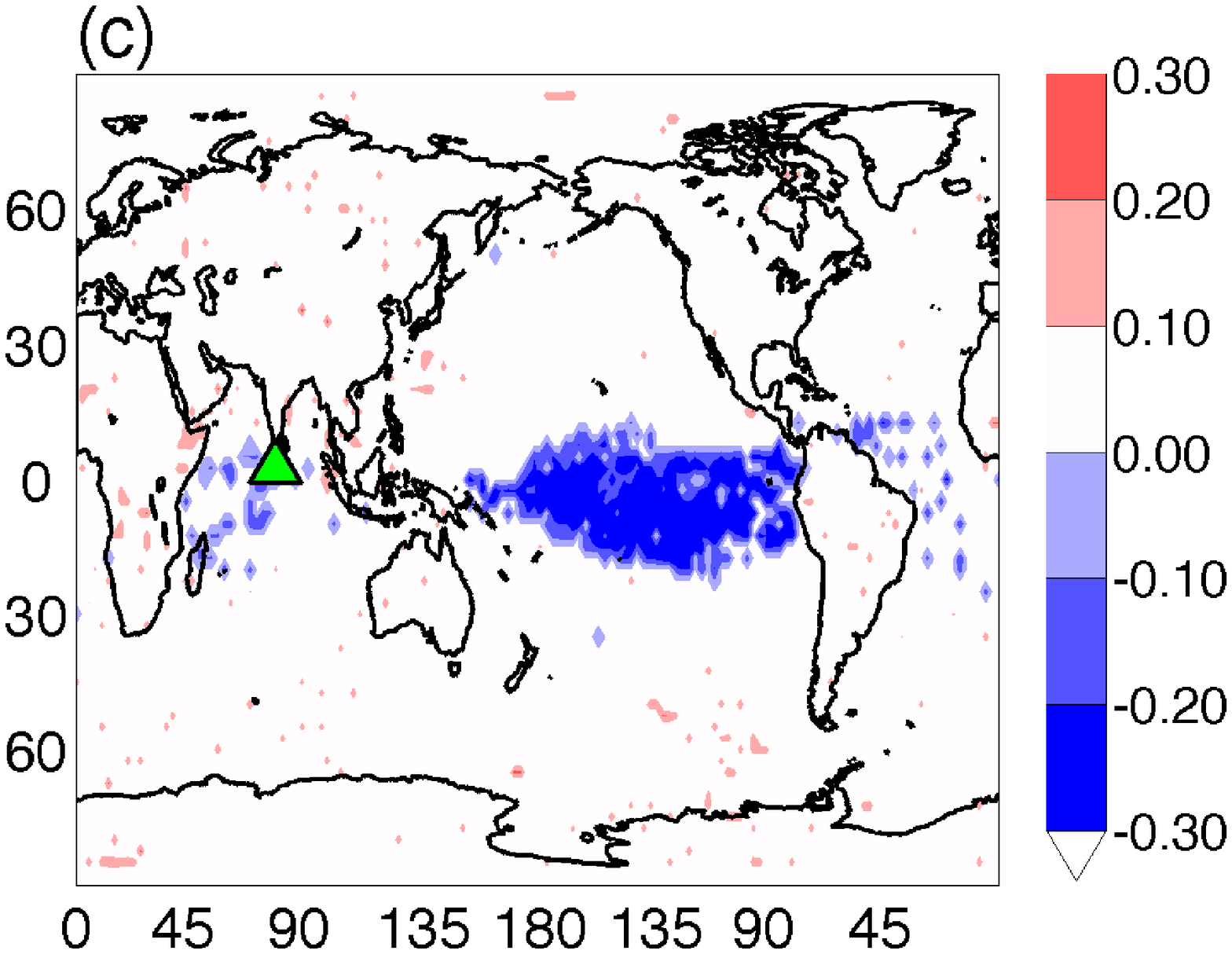}
\includegraphics[width=0.23\textwidth]{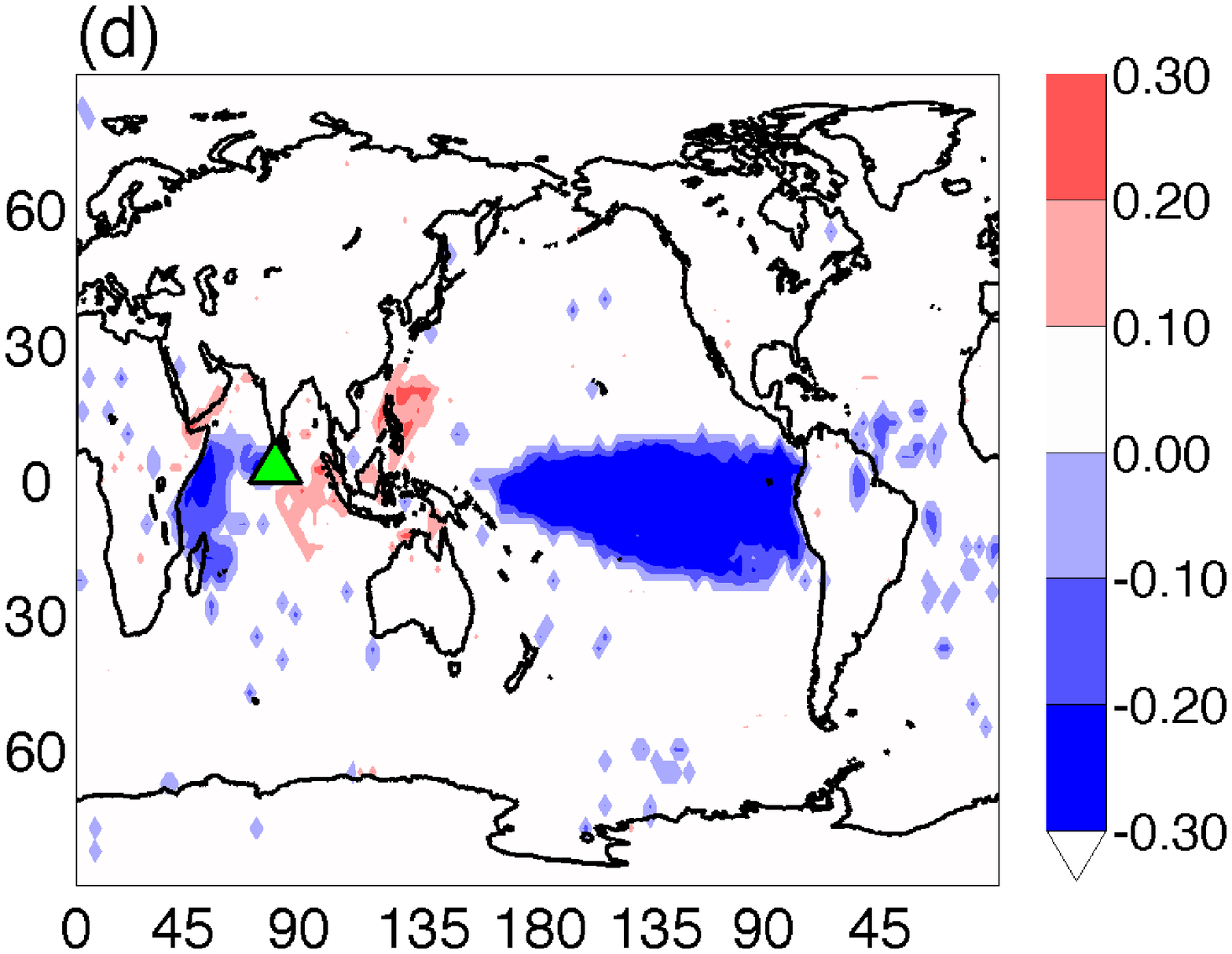}
\caption{(Color Online).
{\color{black}Comparison of monthly-averaged and daily-averaged datasets. The nodes considered are as in Fig.  \ref{fig:construction}.  Panels (a) and (c) display the $\mbox{DI}$ measure ($\tau=1$ month) calculated from monthly data ---same as Figs.  \ref{fig:construction} (c) and (f).  Panels (b) and (d) display the $\mbox{DI}$ measure ($\tau=30$ days) calculated from daily data.  Results are consistent and the resolution using daily data is better.}
} \label{fig:comparisonMonthDay}
\end{center}
\end{figure}

In order to obtain more temporal resolution daily data has been used.  Figure \ref{fig:comparisonMonthDay} shows a comparison between $\mbox{DI}$ for monthly data  --panels \ref{fig:comparisonMonthDay} (a) and  \ref{fig:comparisonMonthDay} (c)---and for daily data---panels \ref{fig:comparisonMonthDay} (b) and  \ref{fig:comparisonMonthDay} (d).  Corresponding to the same points considered in Fig. \ref{fig:construction}. In order to adequately compare the datasets, $\tau$ was chosen equal to one month in the monthly data and and 30 days in the daily data.

The maps using monthly and daily data show similar features and no inconsistencies are found. Furthermore, the map constructed using daily data captures much better the local and remote dependencies and directionality of the links. Areas with significant links are better defined and some regions that are known to be influenced by equatorial Pacific SST, like the tropical north Atlantic \cite{Chang2000}, clearly appear using daily data, but only very roughly using monthly data. Thus, the increase in temporal resolution improves the representation of the links related to tropical regions.

\subsection{\label{daily} Analysis of daily-averaged SAT anomalies}

{\color{black}
\subsubsection{Influence of the parameter $\tau$} \label{sec:depOnTau}

\begin{figure}[htb]
\begin{center}
\includegraphics[width=0.23\textwidth]{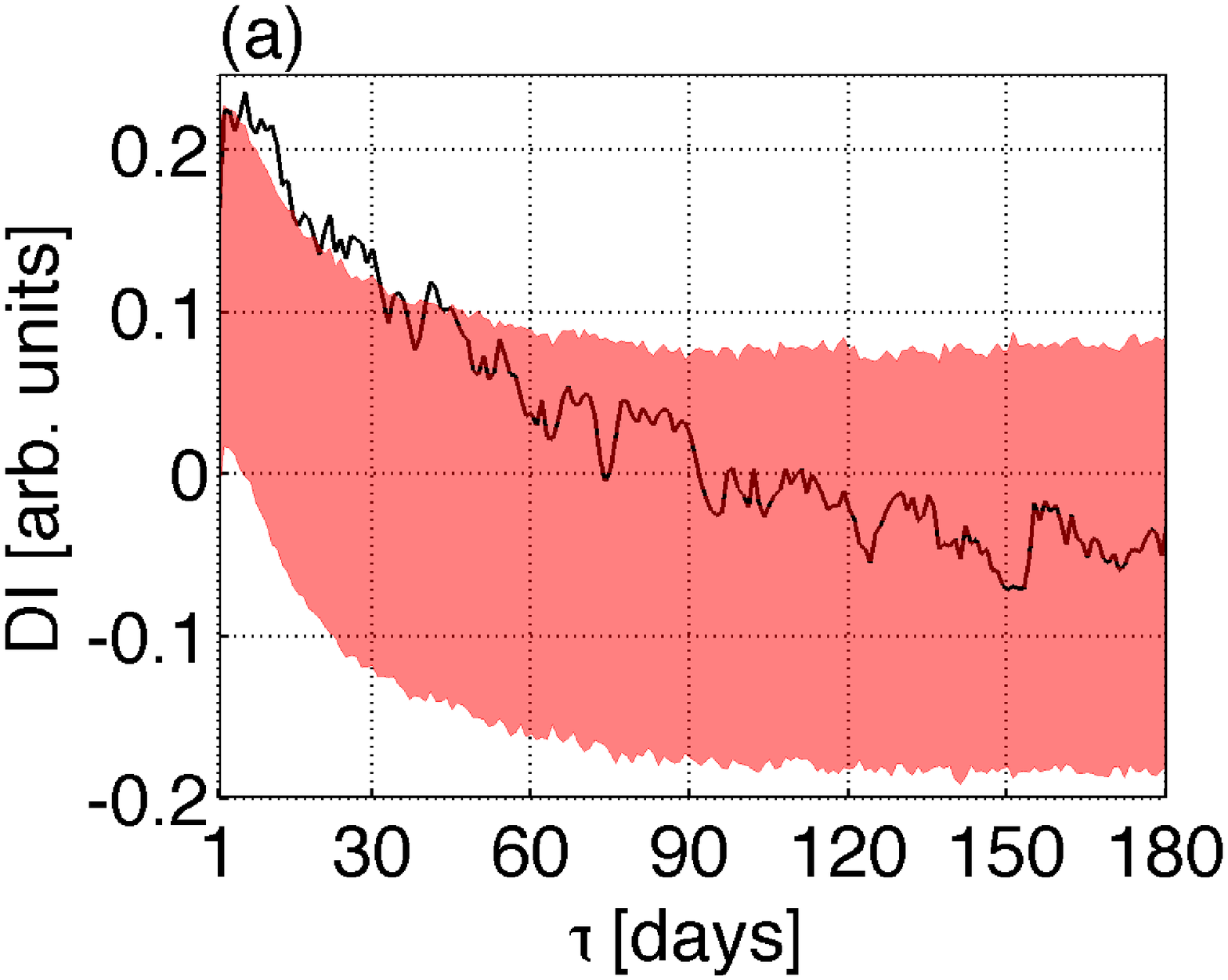}
\includegraphics[width=0.23\textwidth]{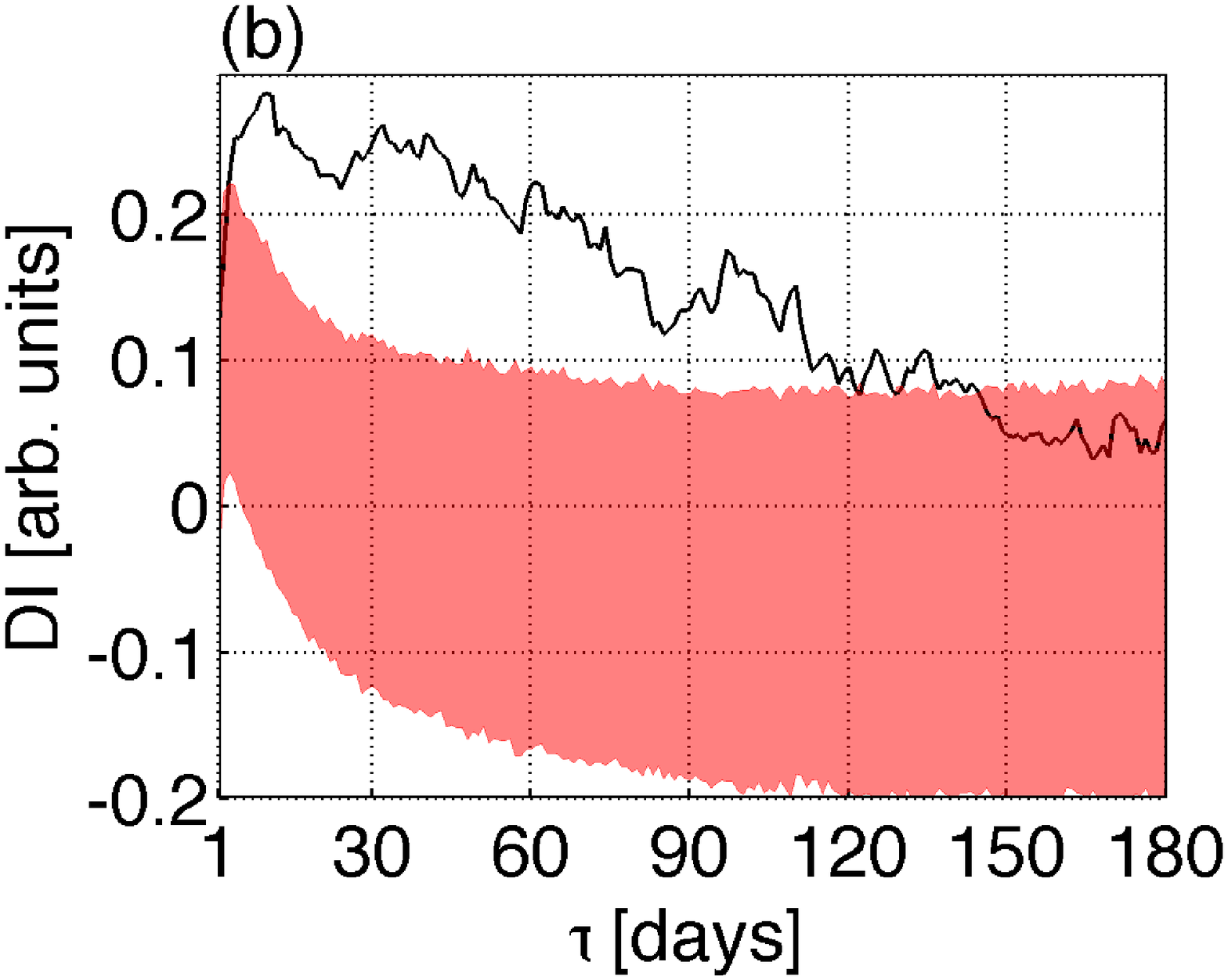}
\includegraphics[width=0.23\textwidth]{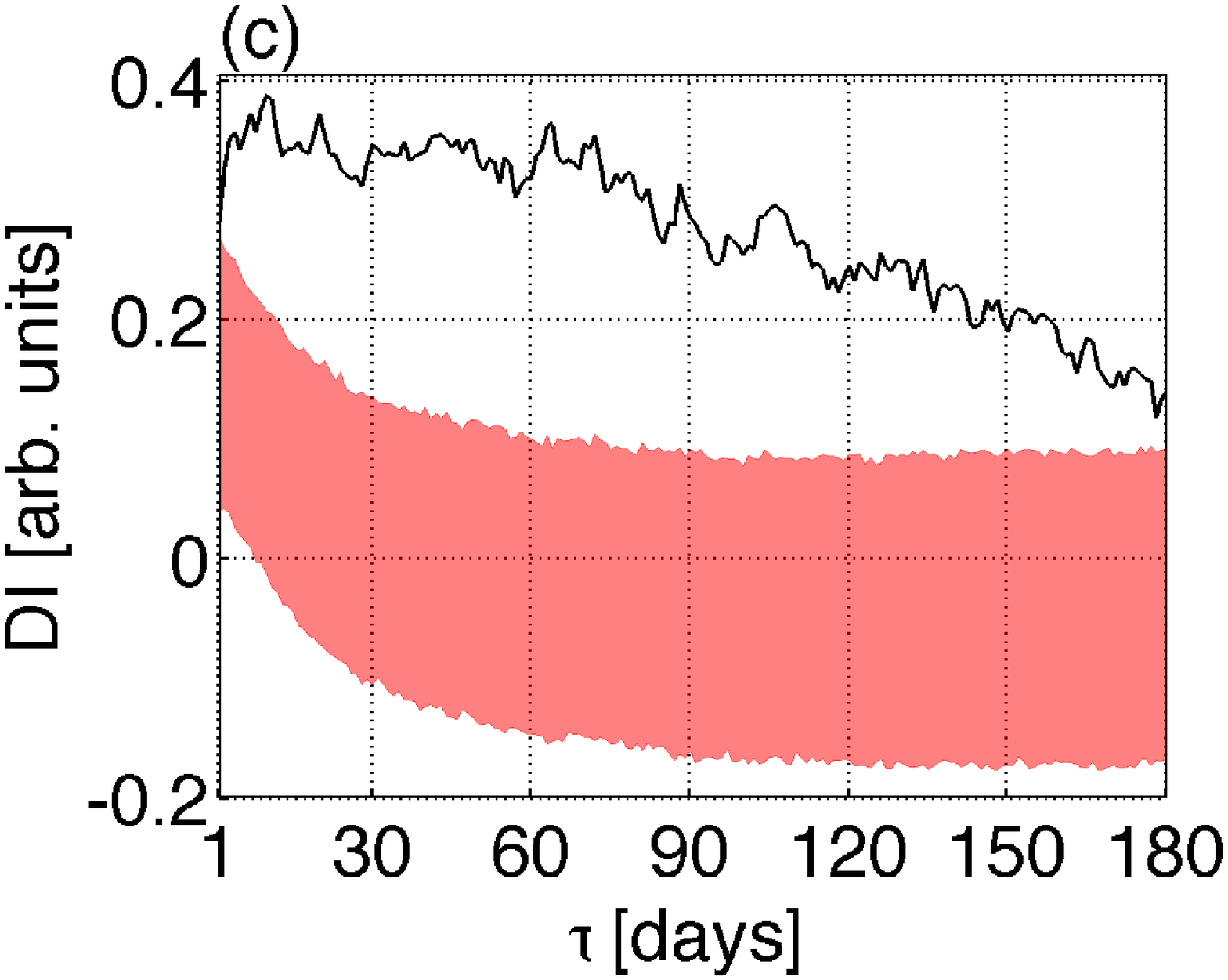}
\includegraphics[width=0.23\textwidth]{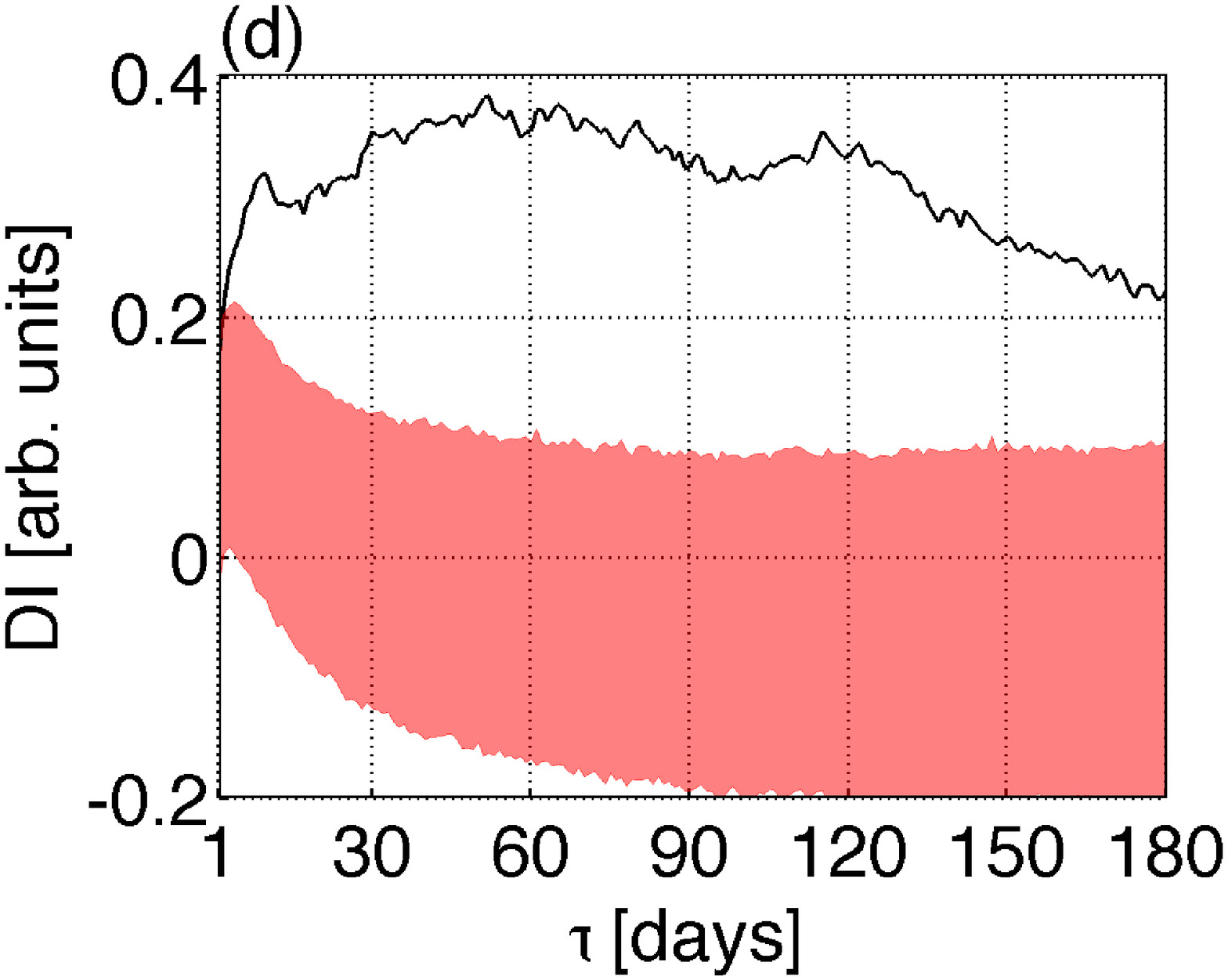}
\caption{(Color Online).
Influence of the parameter $\tau$ when $\mbox{DI}$ is computed from daily-averaged data. The $\mbox{DI}$ value is plotted \textit{vs.} $\tau$ for the four links that connect a node in the  Pacific ocean with (a) a node in Labrador sea, (b) a node in tropical north Atlantic (in the red area in Fig. \ref{fig:comparisonMonthDay} between the Caribbean and Africa), (c) a node in southern Pacific (the red spot in Fig. \ref{fig:comparisonMonthDay} near Antarctica) and (d) a node in the Indian ocean. In order to filter the noise the $\mbox{DI}$ value was averaged over second neighbors in the grid. The shaded area (red online) indicates the $\mbox{DI}$ value computed from 100 BS surrogates, as explained in section \ref{statistAnalysis}.
}\label{fig:DIvstau}
\end{center}
\end{figure}

To analyze the influence of the parameter $\tau$ in the $\mbox{DI}$ measure when it is computed from daily data,  four links connecting the node in the Pacific ocean are considered in Figs. 1, 5 with:

(a) a node in Labrador sea, 

(b) a node in tropical north Atlantic (in the red area in Fig. \ref{fig:comparisonMonthDay} between the Caribbean and Africa), 

(c) a node in southern Pacific (the red spot in Fig. \ref{fig:comparisonMonthDay} near Antarctica) and 

(d) a node in the Indian ocean.

Figure \ref{fig:DIvstau} displays, for each link, the $\mbox{DI}$ value vs $\tau$. 

In panel (a), the directionality index that characterizes the influence of the equatorial Pacific to the Labrador Sea shows a small increase for small $\tau$ that persists up to about $\tau=10$ days. This is the typical time scale associated with the setup of the atmospheric anomalies forced by anomalies in tropical convection. Afterwards, the $\mbox{DI}$ decreases exponentially-like, becoming non-significant at $\tau=30$ days---as the $\mbox{DI}$ value enters the shaded area. In the tropical north Atlantic---Fig. \ref{fig:DIvstau} (b)---the $\mbox{DI}$ to the equatorial Pacific shows similar values as for the Labrador Sea case, for up to $\tau=10$ days. For larger values of $\tau$, however, the $\mbox{DI}$ has significant values (approximately constant) up to about $\tau=60$ days. Afterwards, it decreases becoming non-significant for $\tau$ larger than $4$ months. The difference in behavior between the tropical north Atlantic and the Labrador Sea might be because the remote forcing from ENSO induces a clear regional response in the surface temperatures of the tropical Atlantic \cite{Chang2000}, which will add persistence to the remote signal. On the other hand, in the Labrador Sea, the ocean does not respond and the large atmospheric variability obscures the signal from the equatorial Pacific.

A similar behavior to the one seen in the tropical Atlantic is also found in the Southern ocean---Fig. \ref{fig:DIvstau} (c). There is a fast time scale for small values of $\tau$, but in this case the influence of the equatorial Pacific persists for $\tau$ up to $80$ days, with $\mbox{DI}$ values significantly larger than in the Atlantic. The ENSO influence over the south Pacific is one of the most robust signals in the extratropics, consequence of atmospheric teleconections associated with the Pacific South American pattern \cite{Mo1998}. The time scale of about 3 months seen in the $\mbox{DI}$ is likely associated with the time it takes to the surface ocean to respond to anomalous atmospheric fluxes and to the seasonal dependence of the atmospheric teleconnection pattern on the mean state of the extratropical atmosphere.

The behavior of the influence of the equatorial Pacific onto the Indian ocean---Fig. \ref{fig:DIvstau} (d)---also shows a fast time scale of a few days, but in this case the largest value of the $\mbox{DI}$ is seen for a $\tau$ of about 60 days. Also, there are large values of $\mbox{DI}$ for $\tau$ values of more than 4 months. This suggests that the Indian ocean responds to the incoming ENSO signal in a time scale of about 2 months, through thermodynamic and dynamic coupling \cite{Annamalai2003a,Wang2013c}.

Thus, the $\mbox{DI}$ plots as a function of $\tau$ for different ocean regions suggest that there is one fast time scale of about 10 days associated to the setup of the atmospheric teleconections associated with changes in tropical convection. Moreover, in some regions the $\mbox{DI}$ shows a second longer time scale of about 2 or 3 months associated with the response of the local ocean to the circulation anomalies forced from the tropics. Finally, longer time scales related to oceanic inertia also affect the $\mbox{DI}$ value but longer datasets are needed in order to perform a robust estimation of their effects.} %there is a time scale of decay of $\mbox{DI}$ for larger values of $\tau$ that will depend on the oceanic inertia.}

\begin{figure}[htb]
\begin{center}
\includegraphics[width=0.23\textwidth]{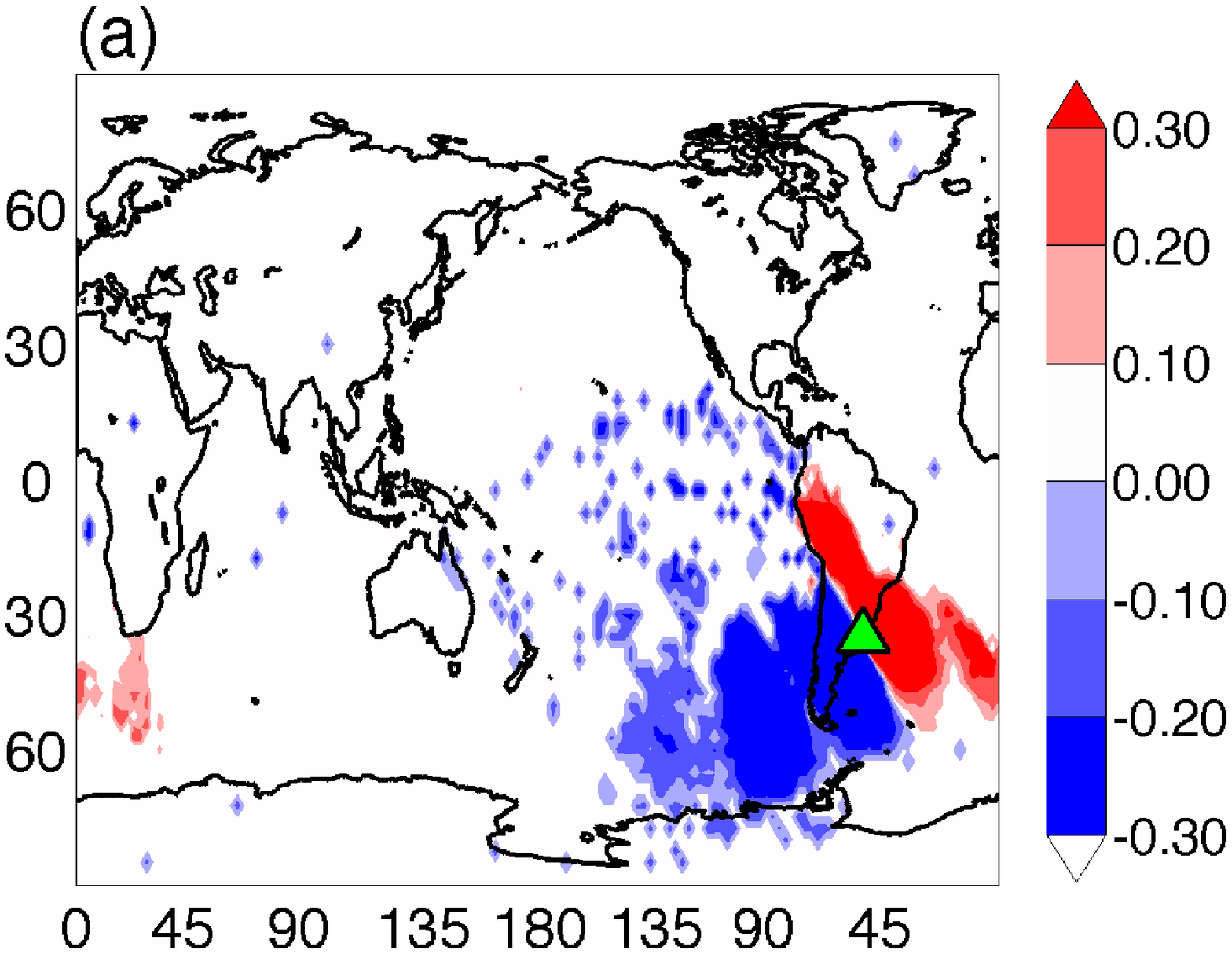}
\includegraphics[width=0.23\textwidth]{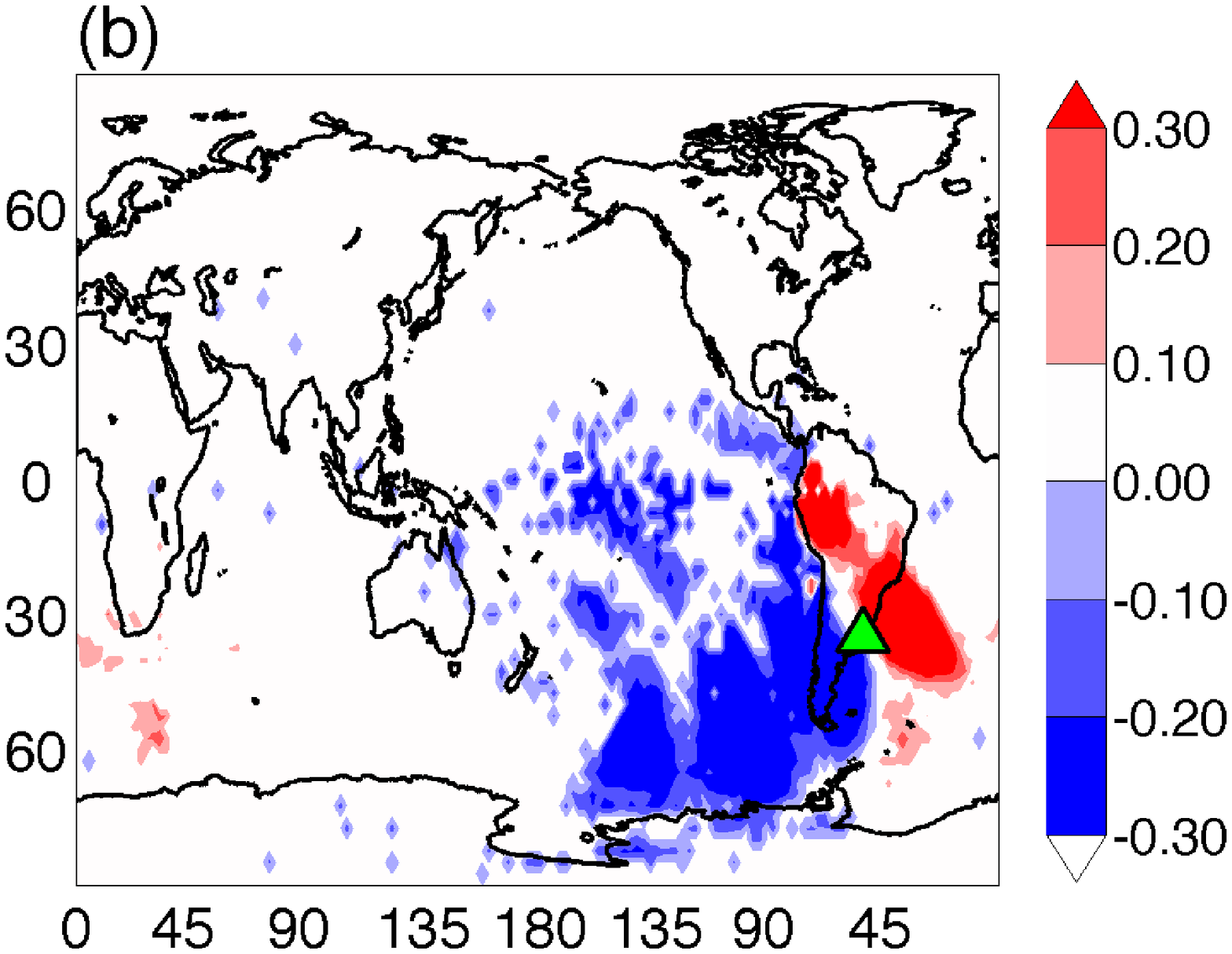}
\includegraphics[width=0.23\textwidth]{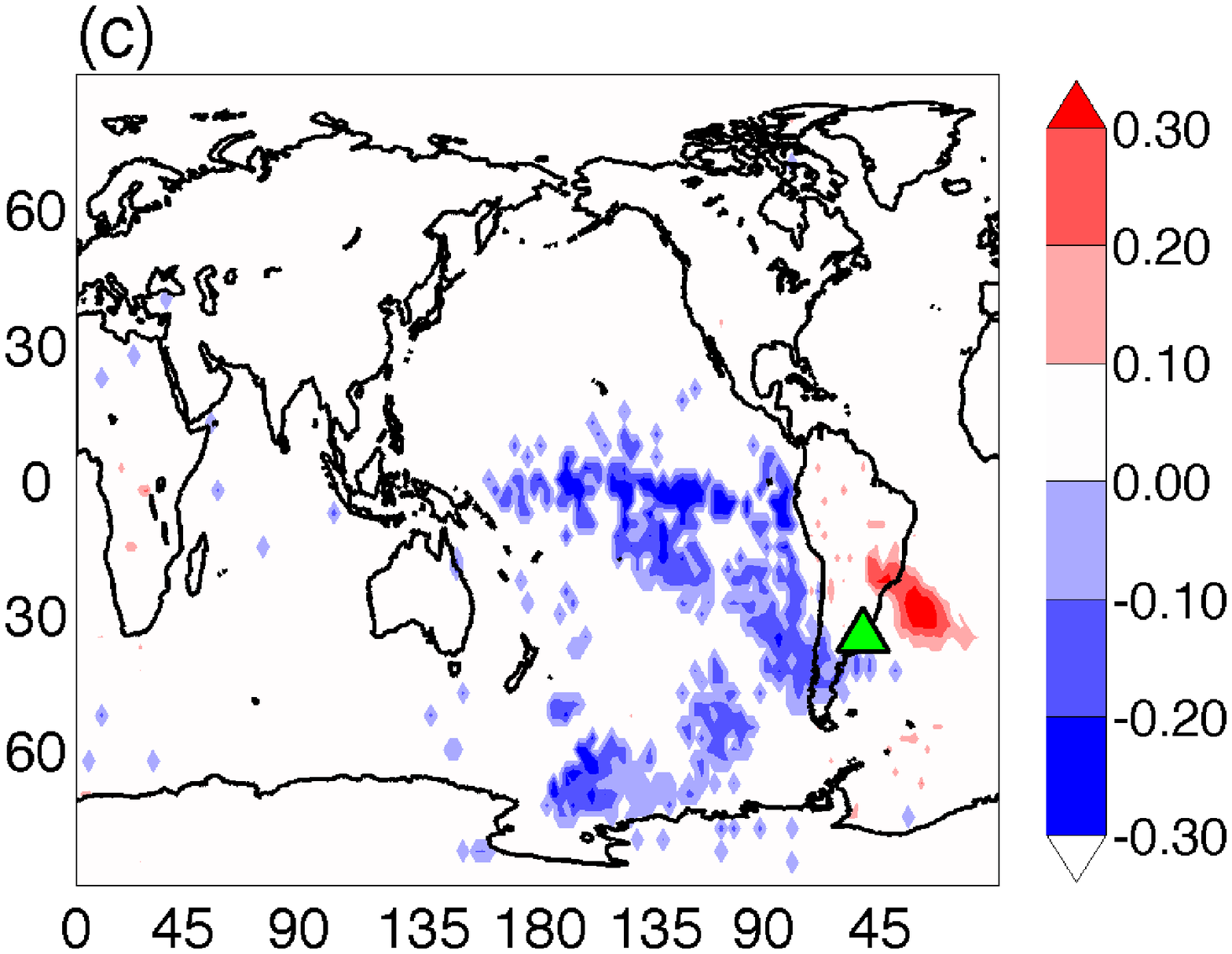}
\includegraphics[width=0.23\textwidth]{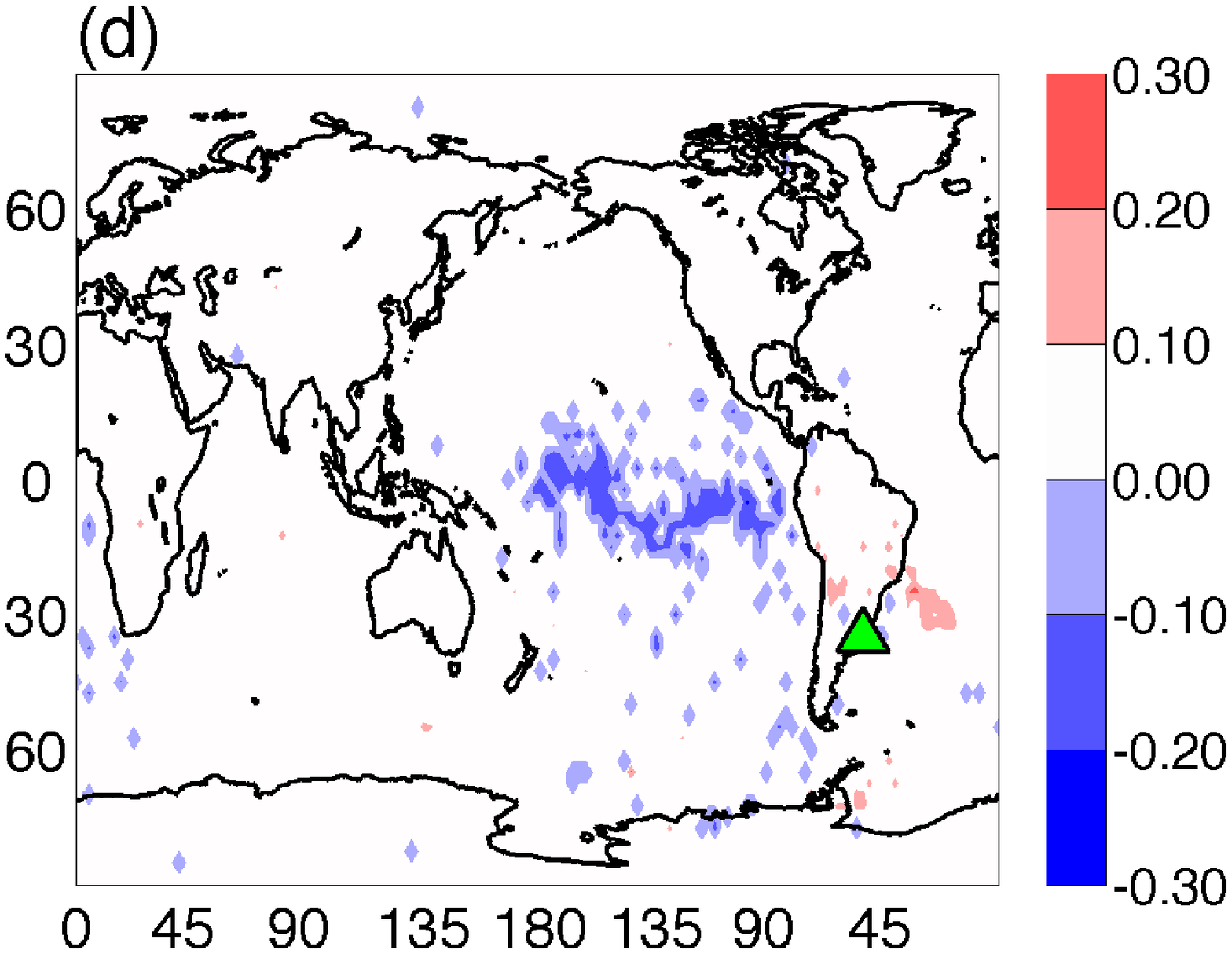}
\caption{(Color Online). {Effect of $\tau$ in the southern extratropics, when $\mbox{DI}$ is computed from daily data. The same node in Fig. \ref{fig:tauMonthlyExtraTropics} (a) is considered, and the values of $\tau$ are: (a) $1$ day, (b) $3$ days, (c) $7$ days, and (d) $30$ days.  Small $\tau$ capture wave trains propagating from west to east, while for $\tau$ values longer than a week the decorrelation is higher and only the influence from the Pacific ocean persists. }} \label{fig:uruguay}
\end{center}
\end{figure}

\begin{figure}[htb]
\begin{center}
\includegraphics[width=0.23\textwidth]{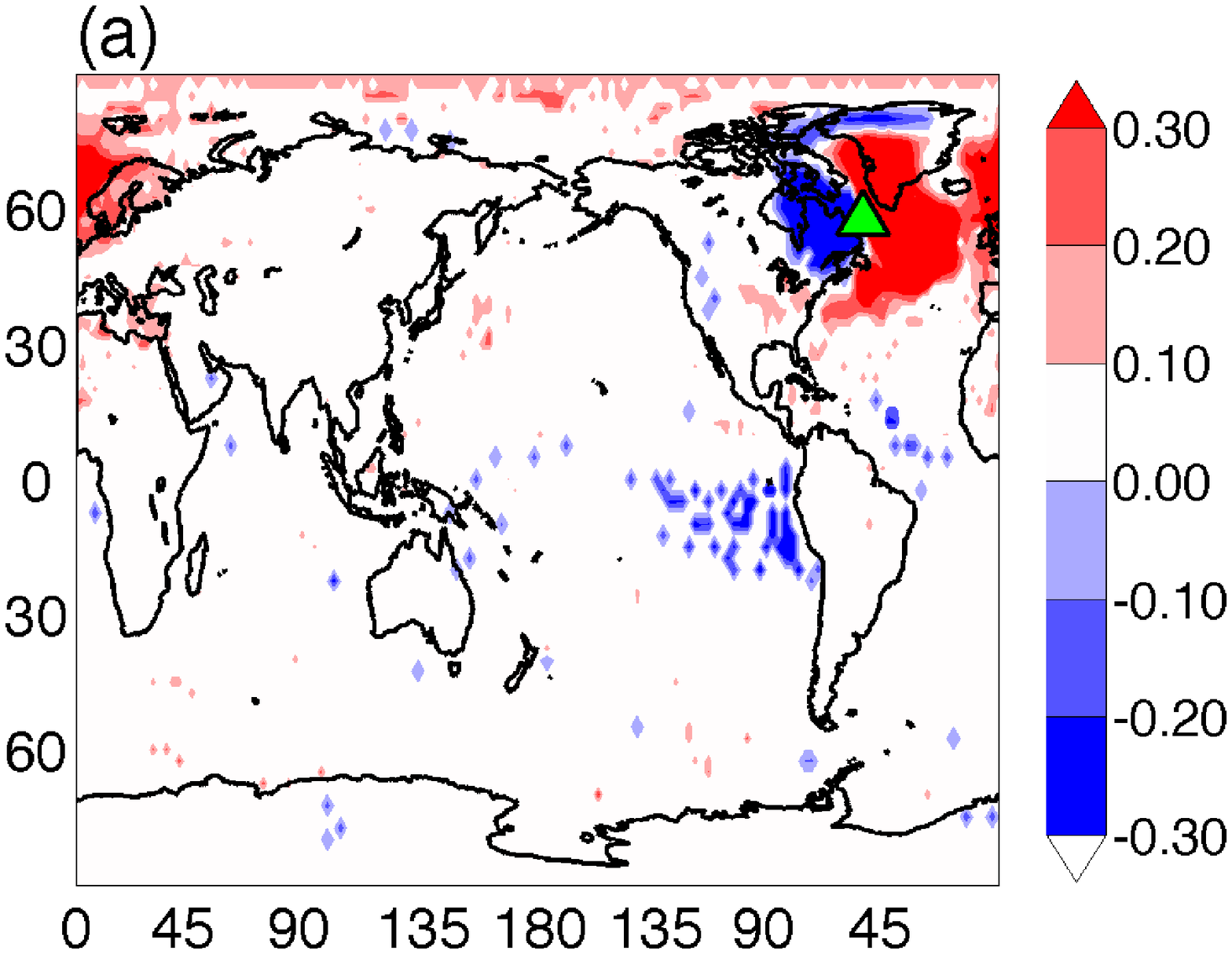}
\includegraphics[width=0.23\textwidth]{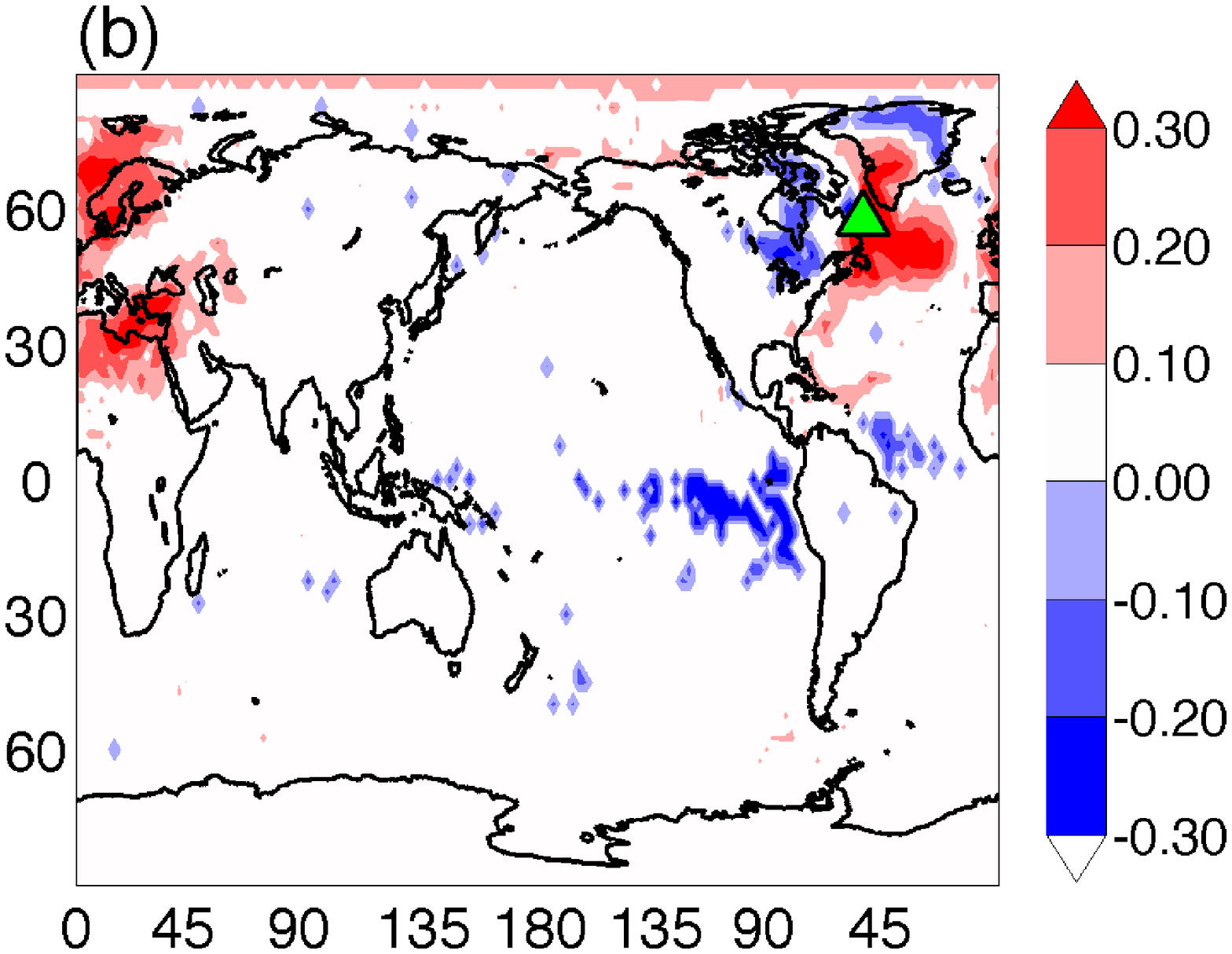}
\includegraphics[width=0.23\textwidth]{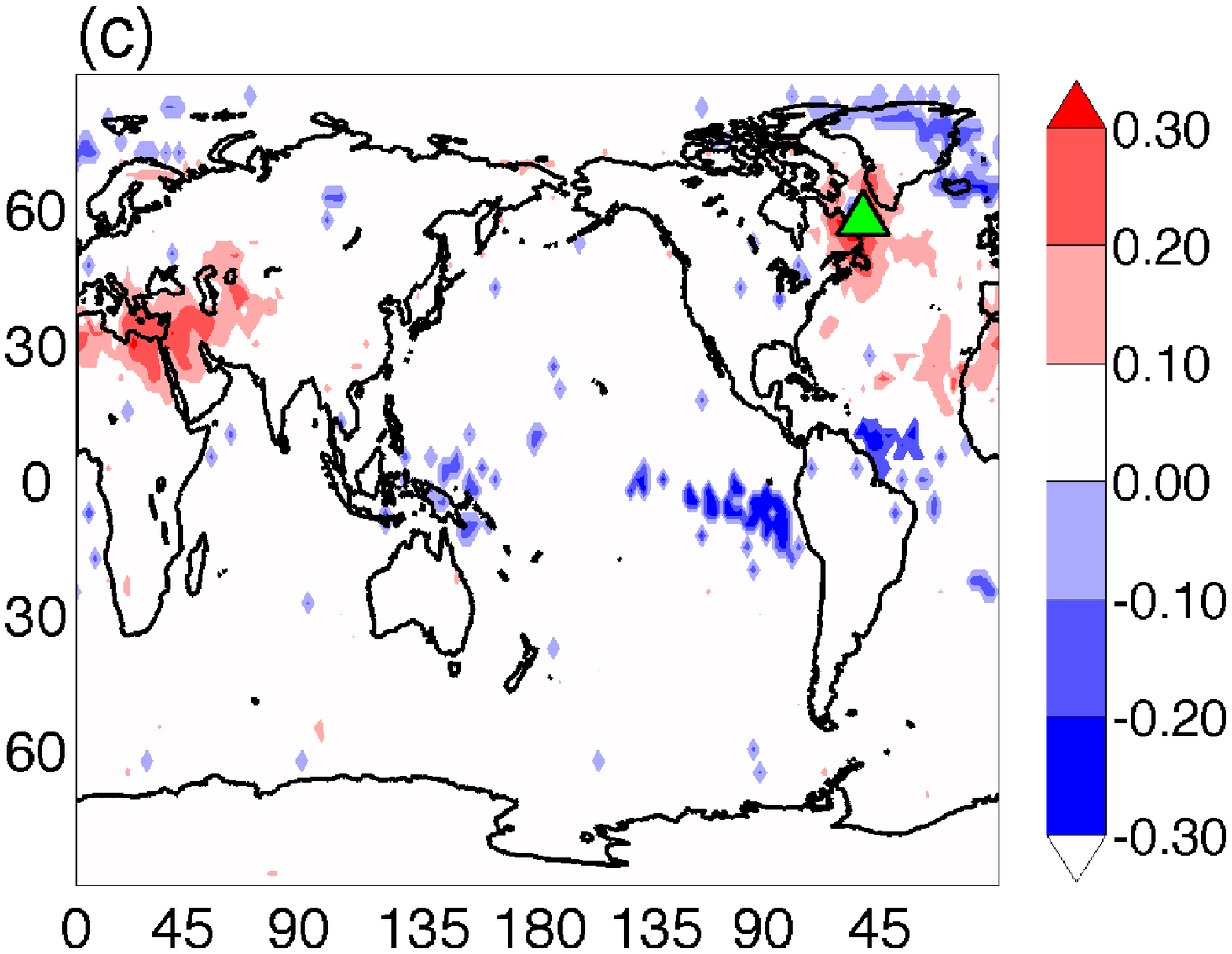}
\includegraphics[width=0.23\textwidth]{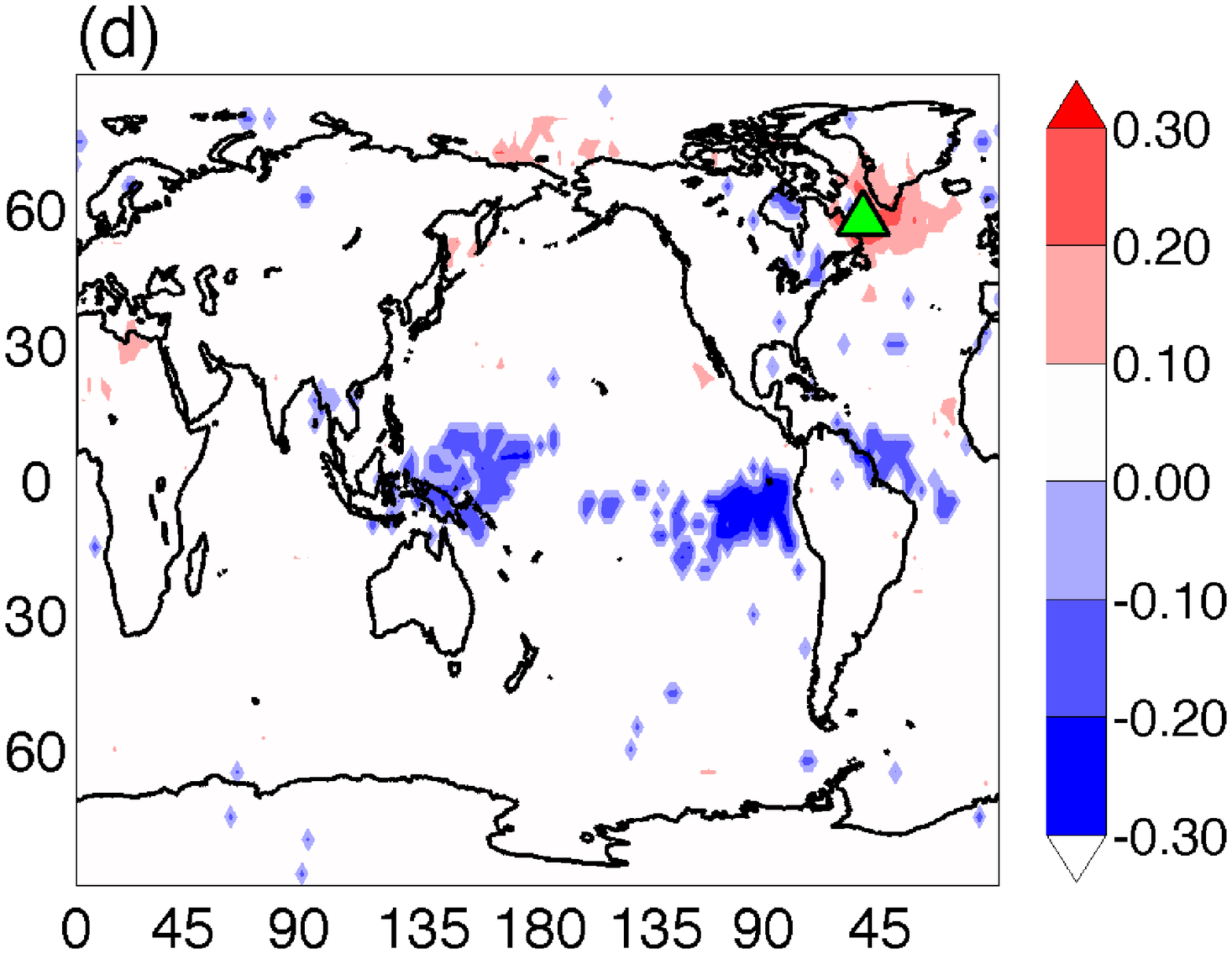}
\caption{(Color Online). {As in Fig \ref{fig:uruguay} but for a node in the northern extratropics ---the Labrador Sea, as in Fig.  \ref{fig:tauMonthlyExtraTropics} (b). The values of $\tau$ are as in Fig \ref{fig:uruguay}.  The Labrador sea area is related to a source of atmospheric variability of the north Atlantic ocean that affects Europe.  As in Fig. \ref{fig:uruguay}, smaller values of $\tau$ can capture wave trains propagating from west to east, over Europe, while values of $\tau$ longer than a week loose this effect and only the influence from the Pacific ocean persists. }
} \label{fig:labrador}
\end{center}
\end{figure}

\subsubsection{Influence of $\tau$ in the extra-tropics} \label{sec:depOnTau}

The improvement in characterizing directionality using daily data is even larger in the extratropics. As  mentioned above, the extratropical SAT is strongly dependent on synoptic scale perturbations (a time scale of a few days). Thus, the use of daily data should allow to uncover these relationships and investigate the direction of the links as the lag increases.
{In order to do so, the point in southeastern South America shown in Fig. \ref{fig:tauMonthlyExtraTropics} is considered and the directionality network for several values of $\tau$ ranging from a day to one month (Fig. \ref{fig:uruguay}) is constructed. For synoptic time scales of a few days the methodology uncovers the existence of a wave train connected to southeastern South America propagating with a southwest-northeast direction.} Moreover, there is a clear separation line between regions with incoming and outgoing links. This configuration is characteristic of the progression of a front through the reference point and does not imply that the SAT over the reference point influences the region to the northeast but it only happens to be in the path of the perturbation. As the lag time increases, the extratropical wave train associated with synoptic time scales fades and only the points in the tropics remain, consistent with an influence of the equatorial Pacific on the region on longer time scales, perhaps related to ENSO.

A similar behavior is seen taking as reference point the SAT over the Labrador Sea (Fig. \ref{fig:labrador}). For small values of $\tau$ the progression of a front is clearly detected using this procedure: given the mean westerly winds at these latitudes, the front moves from west to east and is clearly marked as the boundary between the incoming and outgoing links. It is also seen for $\tau=3$ {suggesting} that in about three days the front reaches the Mediterranean region affecting temperatures there. Again, as $\tau$ increases mainly the tropical links remain. However, even for $\tau=30$ there is a well defined region of outgoing links that remain over the Labrador Sea, suggesting that the SAT in the region may have relatively long time scales of variability, perhaps related to the North Atlantic Oscillation. {\color{black}Another possibility is the influence of the local ocean that increases the persistence of atmospheric temperature anomalies through thermodynamic coupling, as shown by Barsugli \textit{et. al.} \cite{Barsugli1998}.}

\subsubsection{Influence of $\tau$ in the tropics} 

\begin{figure*}[htb]
\begin{center}
\includegraphics[width=0.32\textwidth]{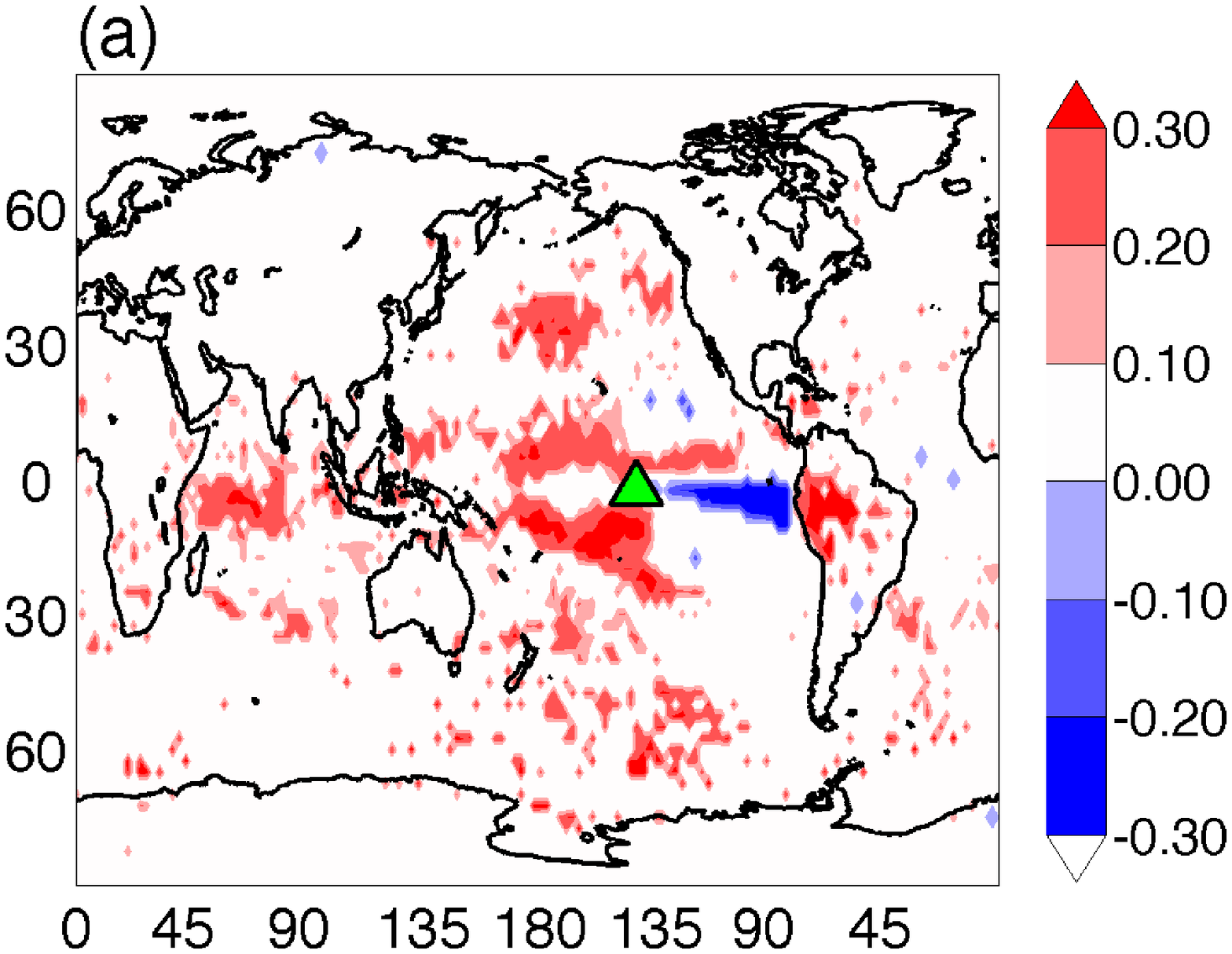}
\includegraphics[width=0.32\textwidth]{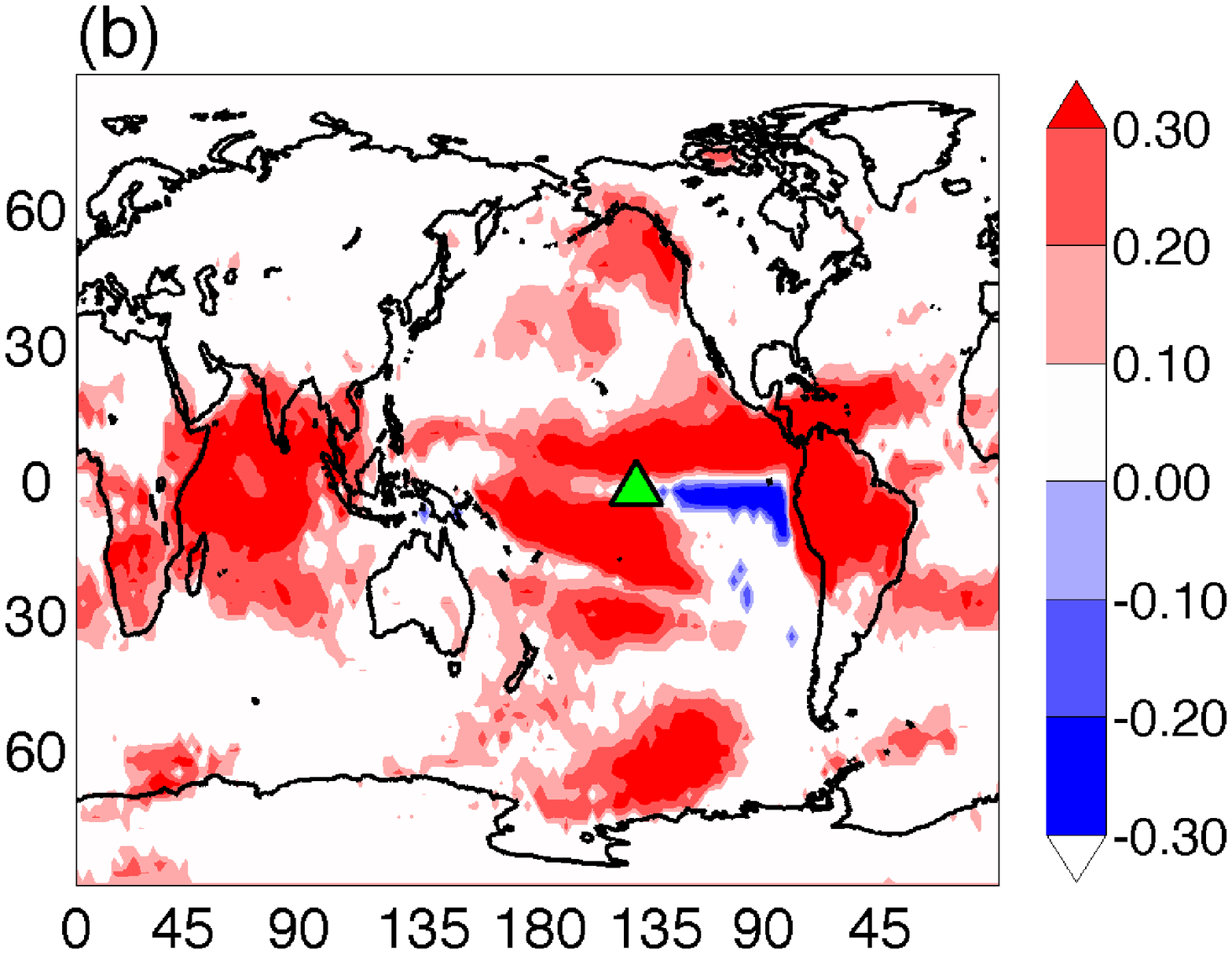}
\includegraphics[width=0.32\textwidth]{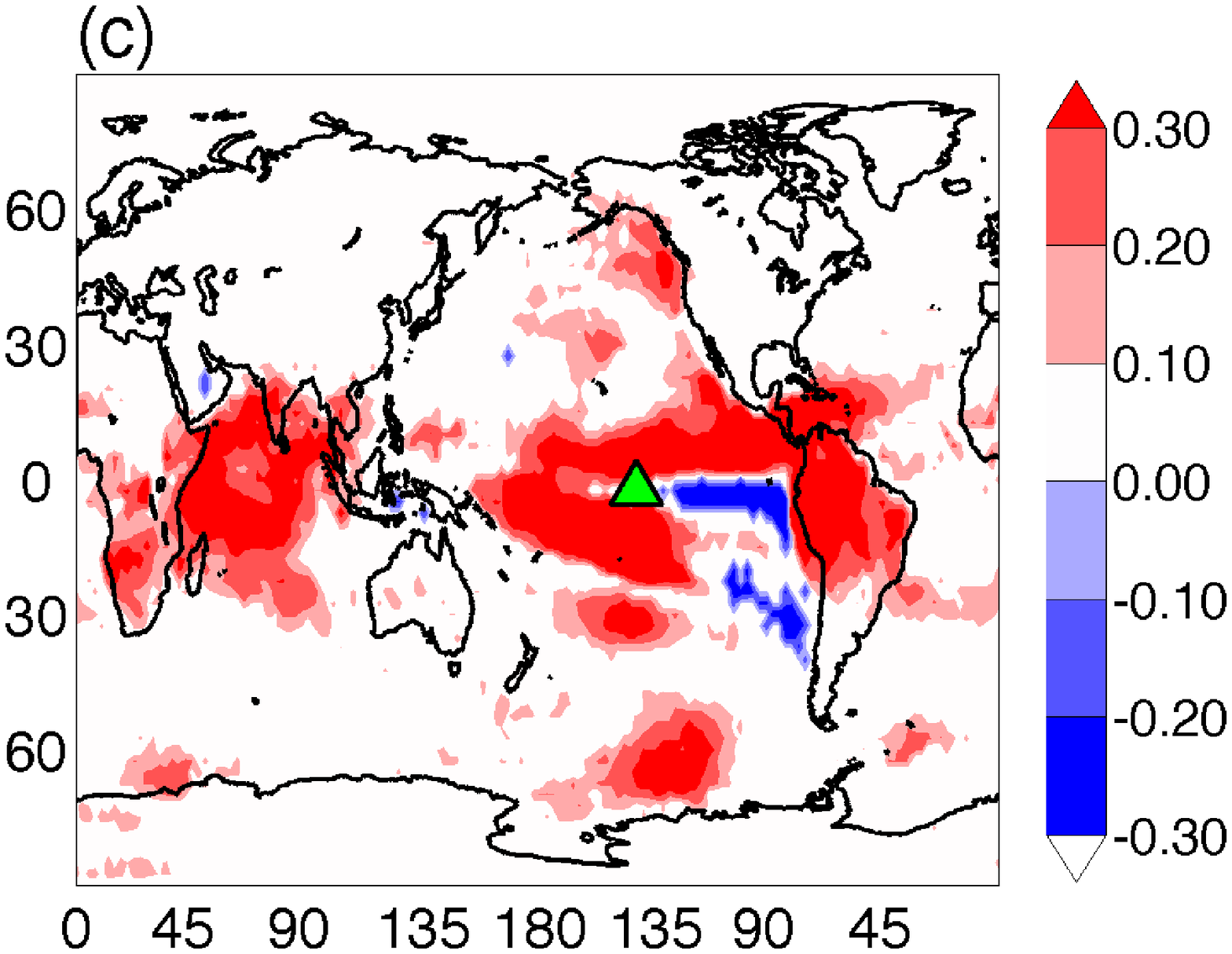}
\includegraphics[width=0.32\textwidth]{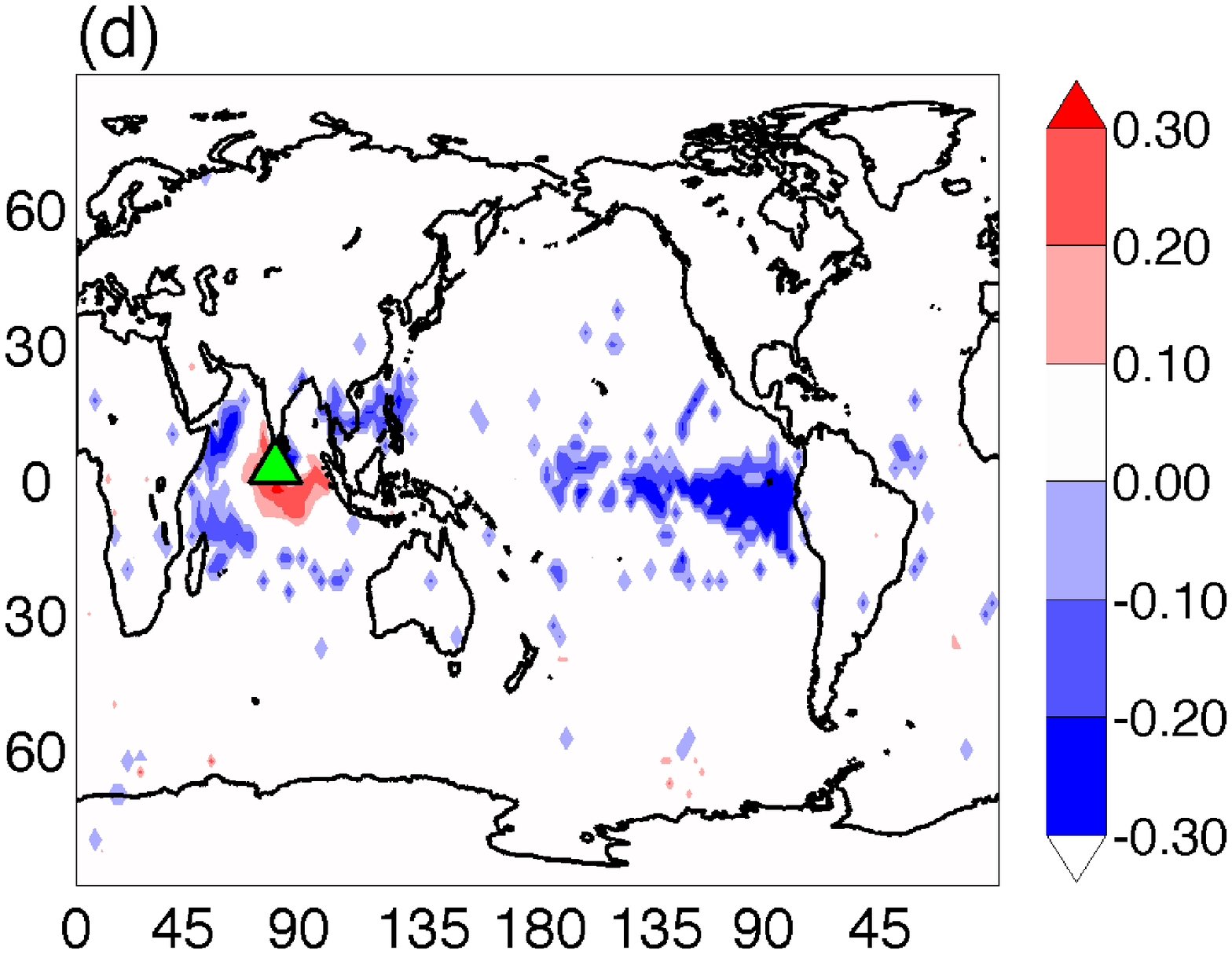}
\includegraphics[width=0.32\textwidth]{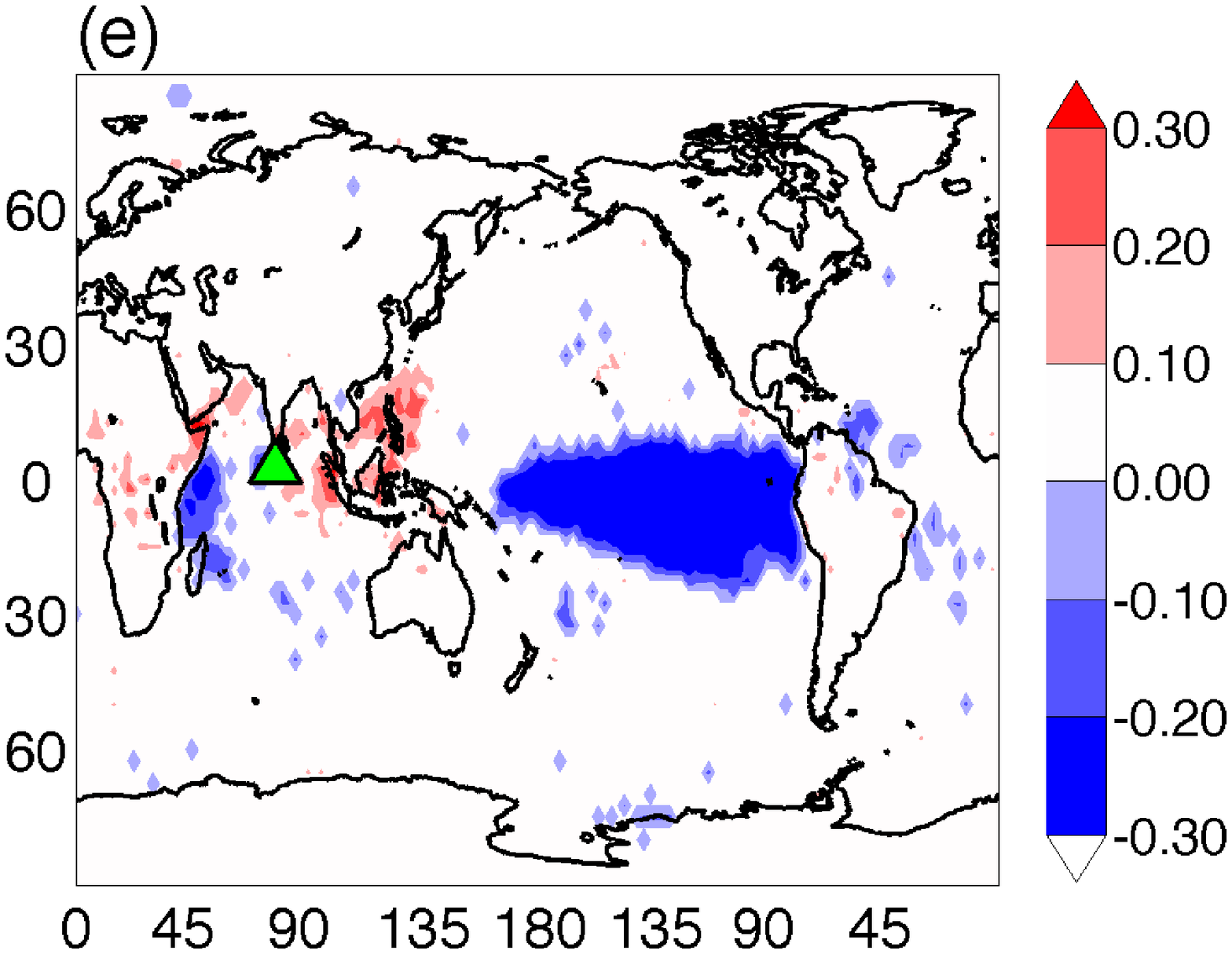}
\includegraphics[width=0.32\textwidth]{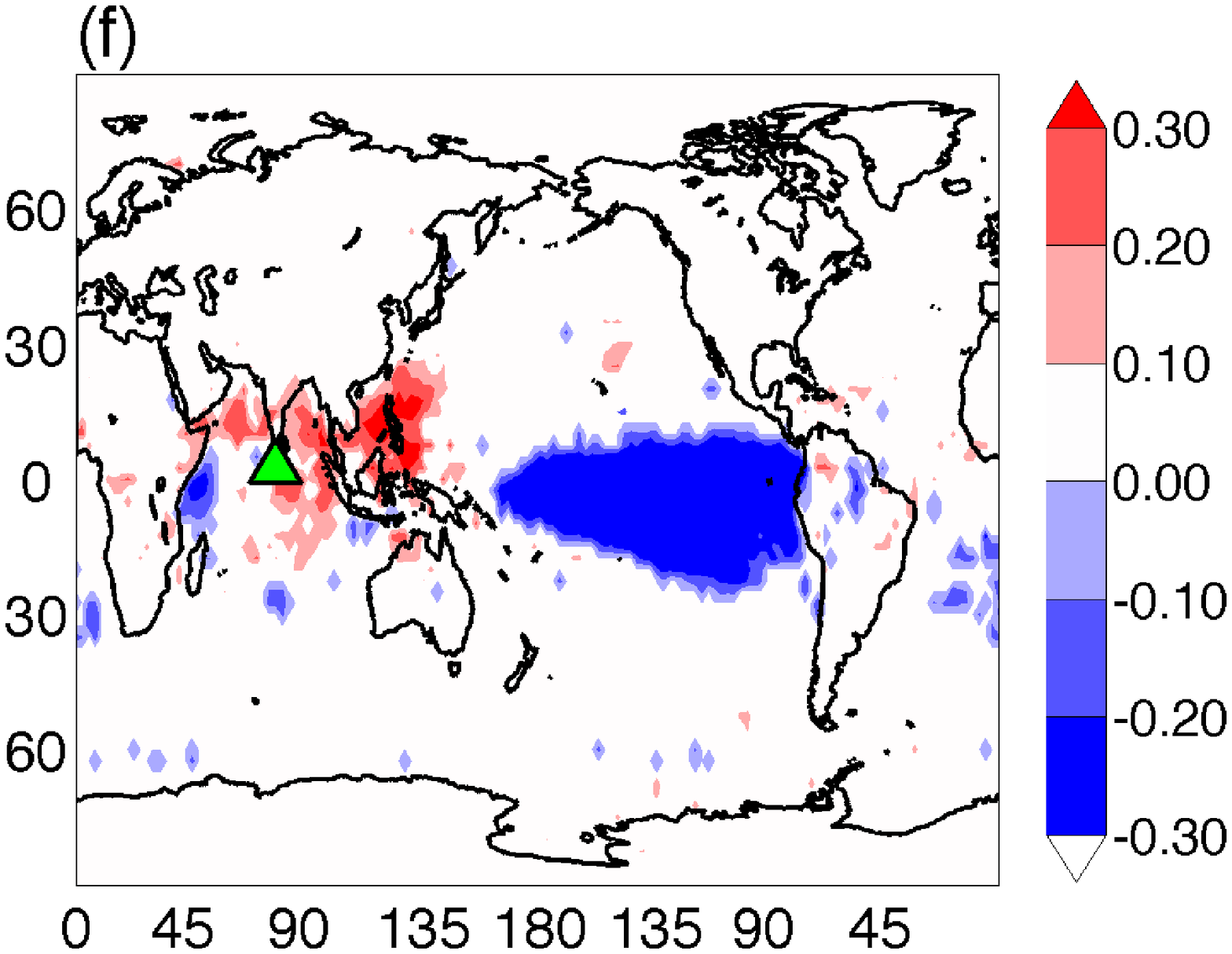}
\caption{(Multimedia view) (Color Online). Effect of $\tau$ using daily data for a node in the tropics. In panels (a-c) the node is in the Pacific ocean, and (d-f), in the Indian ocean. The  values of $\tau$ are: (a,d) $1$ day, (b,e) $45$ days, and (c,f) $90$ days. The maps for $\tau=30$ were shown in Fig.  \ref{fig:tauMonthlyTropics}.   In contrast to the extra-tropics, where the propagation of waves in the scale of days is dominant, and are clearly seen in the maps for time scales of a few days, in the tropics the variability is longer as the ocean adds  persistence to the nearly barotropic atmosphere. For higher values of $\tau$  only long lasting phenomena are observed, related to the strongest variability pattern in this area, which is ENSO. {In the supplementary information a movie showing the links to the node of the fist row, for $\tau=1$ to $\tau=180$ is presented \cite{movie2014}.} }
\label{fig:ninoIndianTau}
\end{center}
\end{figure*}

{Figure  \ref{fig:ninoIndianTau} presents the effect of $\tau$ on daily data in the tropics. Panel \ref{fig:ninoIndianTau} (a)
shows the point considered in the Pacific ocean for $\tau=1$ day. Contrarily to e.g. figure \ref{fig:uruguay}, where front propagation shows a big response in the tropics due to the existence of wave trains in the extratropics, in this case the connections are weak. The opposite case, for $\tau > 30$ days---Panels \ref{fig:ninoIndianTau} (b,c)---are consistent with in figure  \ref{fig:DIvstau} as the structures are robust for a wide range of $\tau$. Notice in Panel \ref{fig:ninoIndianTau} (c) the presence of a second blue area near the coast of Chile, This is in agreement with the easterly trades  advection over the Pacific ocean. 
{In order to complement the information provided in panels \ref{fig:ninoIndianTau} (a,b,c)  and also in Fig. \ref{fig:DIvstau}, in the supplementary information a movie was included \cite{movie2014} (Multimedia view), which presents the links to the node in the Pacific ocean, for $\tau=1$ day to $\tau=180$ days.}
Panels \ref{fig:ninoIndianTau} (d,e,f) show maps for a point in the Indian ocean, for values of $\tau=1$, $45$, and $90$ days respectively where similar results are obtained. }

\subsubsection{Directionality on the tropical Pacific Ocean}

\begin{figure}[htb]
\begin{center}
\includegraphics[width=0.23\textwidth]{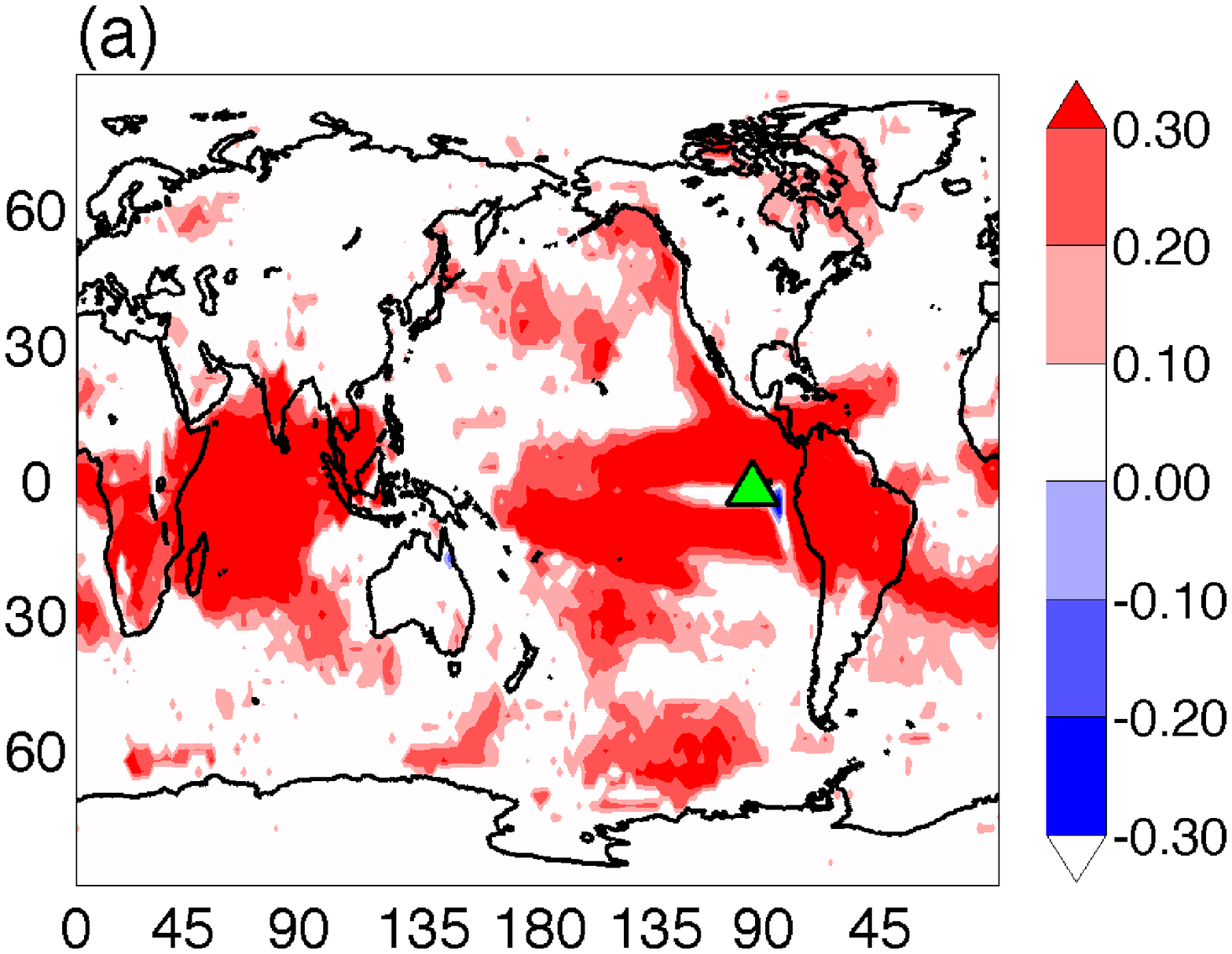}
\includegraphics[width=0.23\textwidth]{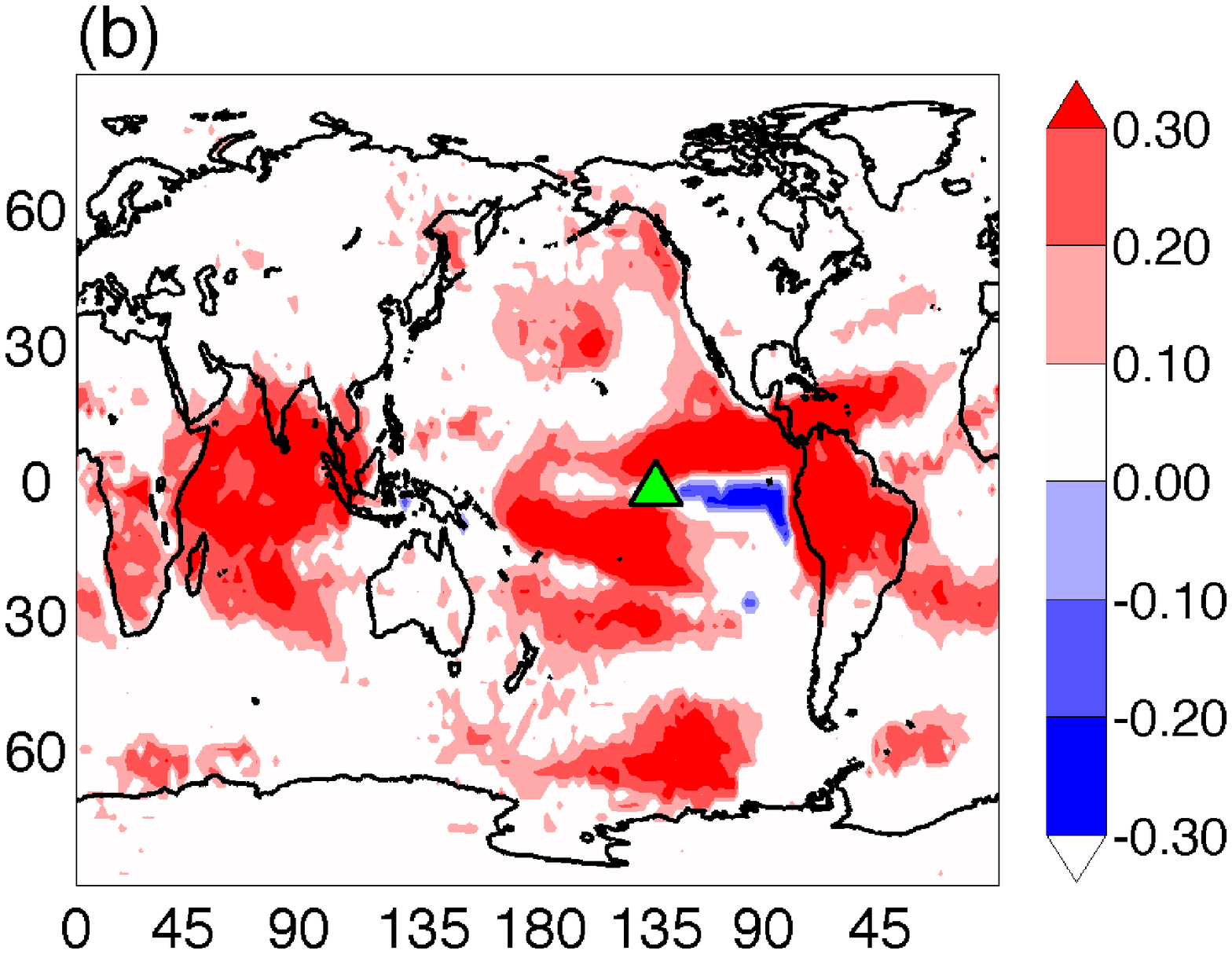}
\includegraphics[width=0.23\textwidth]{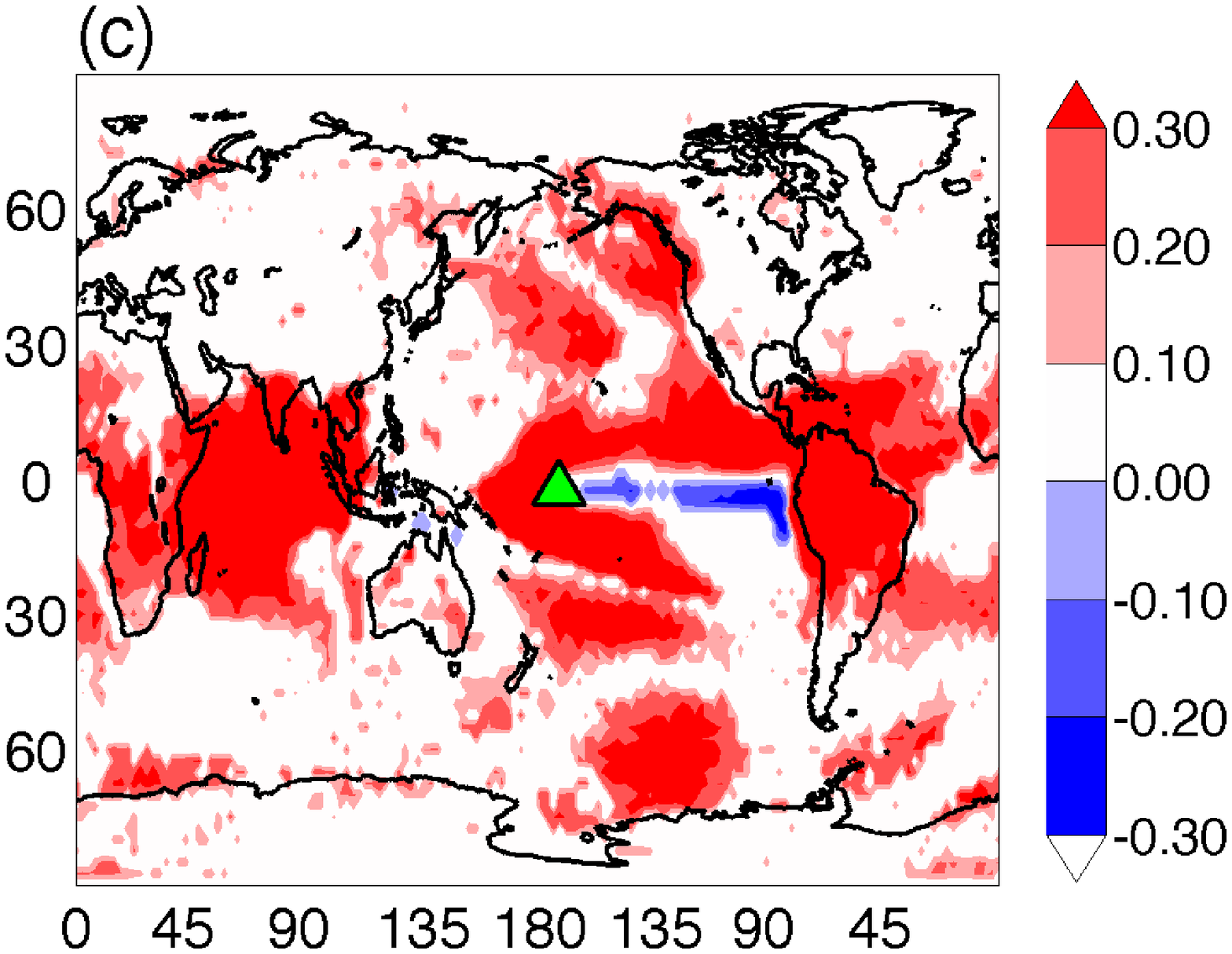}
\includegraphics[width=0.23\textwidth]{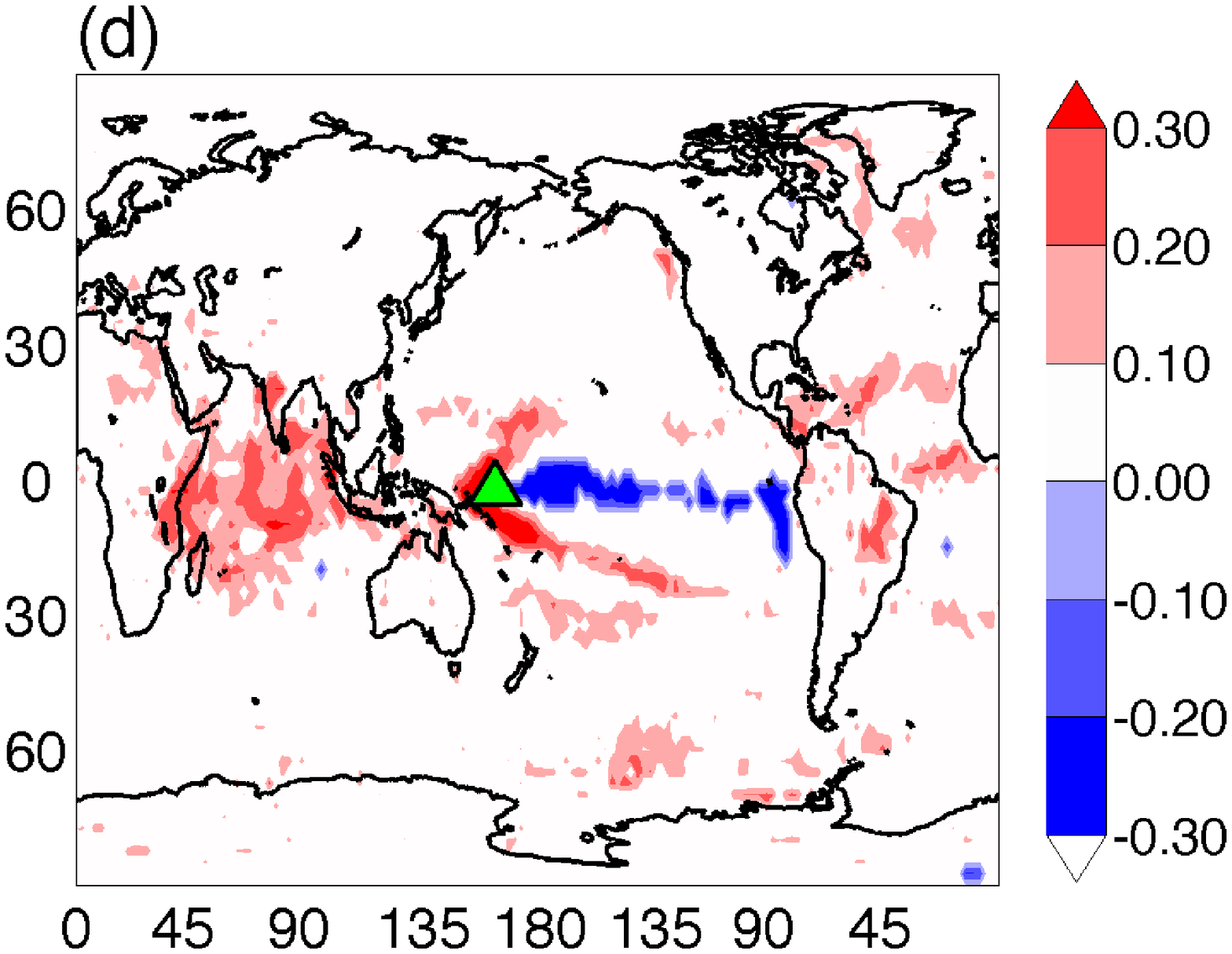}
\caption{(Color Online). {The zonal change of directionality over the equatorial pacific is shown.  In all cases $\tau=30$ days. From (a) near the south American coast, to (d) in the western Pacific ocean.  As seen in the maps, most of the points over central and eastern Pacific have an important effect over a large part of the world; especially over the tropical areas and the rest of the Pacific ocean---notice that in (a)-(c) the teleconnections remain basically the same, although the intensity varies. Incoming links are in blue while outgoing are in red.} }
\label{fig:trades}
\end{center}
\end{figure}

{In Fig. \ref{fig:trades} the $\mbox{DI}$ for $\tau=30$ days has been plotted for points covering all the equatorial Pacific.  They begin at $95.0^\circ$ W --panel \ref{fig:trades} (a) -- and end at $125. 0^\circ$ E --panel \ref{fig:trades} (d).
Clearly, the influence the Pacific ocean exerts is almost global, over tropics and extratropics, in agreement with previous studies (e.g.  in \cite{Trenberth1998}). The $\mbox{DI}$ allows to show that ---even if there are feedbacks and the Pacific is affected by extratropical perturbations and other ocean basins (e.g. the tropical Atlantic)--- the influence is {\em effectively} from the Pacific to the rest of the world. Moreover, the maps show that the largest influence is done by the equatorial Pacific close to the dateline. This is clear in the extratropical atmosphere, as well as in the tropical north Atlantic. On the other hand, the connection to the Indian ocean and south Atlantic is not so sensitive of the point considered over the equatorial Pacific. As the reference point moves further west from the dateline the influence decreases substantially, only remaining weak connections to the tropical north Atlantic and Indian oceans. The methodology can thus be applied to find the best region to construct an index that describes the Pacific influence over the area where climate anomalies are studied.

Notice that all maps show a blue tongue of incoming links to the east of the point considered; it it is seen first in panel \ref{fig:trades} (a) and extends westward until covering the whole Pacific ocean in \ref{fig:trades} (d). This feature is related to the existence of the equatorial cold tongue and the fact that easterly trades blow over the equator thus advecting air from the east to the west of the point.}

\section{\label{sec:conclusions} Conclusions}

The climatological relevance of {\em directed} climate networks constructed via information-theoretic tools has been shown. The presence of significant links was inferred using mutual information while the direction of those links was inferred using the directionality index. Using monthly-averaged SAT data the results from our previous work \cite{Deza2013}  have been recovered  and  the analysis extended  to infer the net direction of propagation of the information. The inference method was tested against the value of the parameter $\tau$, that represents the time required for the information to travel from one node to another. It was found that by adequately tuning $\tau$ the network connectivity varied revealing the various time-scales of atmospheric processes; too short values of $\tau$ failed to capture several long-range links, while for too large values of $\tau$ (above the length of correlations in the data) the connectivity of the network decreased drastically.

In addition, when considering daily-averaged SAT data, the analysis revealed variability patterns consistent with known features of the global climate dynamics.
In the extratropics the long time average synoptic weather was correctly inferred:  as specific examples two geographical regions in different hemispheres---one node in de la Plata basin and another in the Labrador Sea---were considered, and the link direction revealing wave trains propagating from west to east, in both hemispheres was shown.

As a future work, it would be interesting to apply this methodology to analyze how the climate network changes in the different seasons. Another interesting issue is the role of permutation entropy. Computing the directionality index using the ordinal patterns approach \cite{Barreiro2011,Deza2013}, which would allow constructing networks revealing atmospheric processes with short time-scales (of days-months) or with long time-scales (few years). These studies are in progress and will be reported elsewhere.

\acknowledgments
The research leading to these results has received funding from the European
Community's Seventh Framework Programme FP7/2007-2011 under grant agreement n$\circ$ 289447  (ITN LINC). C.M. also acknowledges financial support of grant FIS2012-37655-C02-01 of the Spanish Ministerio de Ciencia e Innovaci\'on.

%\nocite{*}
%\bibliographystyle{}
%\bibliographystyle{}
\bibliography{DezaBarreiroMasoller_revised2}% Produces the bibliography via BibTeX.

\begin{thebibliography}{42}%
\makeatletter
\providecommand \@ifxundefined [1]{%
 \@ifx{#1\undefined}
}%
\providecommand \@ifnum [1]{%
 \ifnum #1\expandafter \@firstoftwo
 \else \expandafter \@secondoftwo
 \fi
}%
\providecommand \@ifx [1]{%
 \ifx #1\expandafter \@firstoftwo
 \else \expandafter \@secondoftwo
 \fi
}%
\providecommand \natexlab [1]{#1}%
\providecommand \enquote  [1]{``#1''}%
\providecommand \bibnamefont  [1]{#1}%
\providecommand \bibfnamefont [1]{#1}%
\providecommand \citenamefont [1]{#1}%
\providecommand \href@noop [0]{\@secondoftwo}%
\providecommand \href [0]{\begingroup \@sanitize@url \@href}%
\providecommand \@href[1]{\@@startlink{#1}\@@href}%
\providecommand \@@href[1]{\endgroup#1\@@endlink}%
\providecommand \@sanitize@url [0]{\catcode `\\12\catcode `\$12\catcode
  `\&12\catcode `\#12\catcode `\^12\catcode `\_12\catcode `\%12\relax}%
\providecommand \@@startlink[1]{}%
\providecommand \@@endlink[0]{}%
\providecommand \url  [0]{\begingroup\@sanitize@url \@url }%
\providecommand \@url [1]{\endgroup\@href {#1}{\urlprefix }}%
\providecommand \urlprefix  [0]{URL }%
\providecommand \Eprint [0]{\href }%
\providecommand \doibase [0]{http://dx.doi.org/}%
\providecommand \selectlanguage [0]{\@gobble}%
\providecommand \bibinfo  [0]{\@secondoftwo}%
\providecommand \bibfield  [0]{\@secondoftwo}%
\providecommand \translation [1]{[#1]}%
\providecommand \BibitemOpen [0]{}%
\providecommand \bibitemStop [0]{}%
\providecommand \bibitemNoStop [0]{.\EOS\space}%
\providecommand \EOS [0]{\spacefactor3000\relax}%
\providecommand \BibitemShut  [1]{\csname bibitem#1\endcsname}%
\let\auto@bib@innerbib\@empty
%</preamble>
\bibitem [{\citenamefont {Watts}\ and\ \citenamefont
  {Strogatz}(1998)}]{Watts1998}%
  \BibitemOpen
  \bibfield  {author} {\bibinfo {author} {\bibfnamefont {D.~J.}\ \bibnamefont
  {Watts}}\ and\ \bibinfo {author} {\bibfnamefont {S.~H.}\ \bibnamefont
  {Strogatz}},\ }\href {\doibase 10.1038/30918} {\bibfield  {journal} {\bibinfo
   {journal} {Nature}\ }\textbf {\bibinfo {volume} {393}},\ \bibinfo {pages}
  {440} (\bibinfo {year} {1998})}\BibitemShut {NoStop}%
\bibitem [{\citenamefont {Albert}\ and\ \citenamefont
  {Barab\'{a}si}(2002)}]{Albert2002}%
  \BibitemOpen
  \bibfield  {author} {\bibinfo {author} {\bibfnamefont {R.}~\bibnamefont
  {Albert}}\ and\ \bibinfo {author} {\bibfnamefont {A.-L.}\ \bibnamefont
  {Barab\'{a}si}},\ }\href {\doibase 10.1103/RevModPhys.74.47} {\bibfield
  {journal} {\bibinfo  {journal} {Rev. Mod. Phys.}\ }\textbf {\bibinfo {volume}
  {74}},\ \bibinfo {pages} {47} (\bibinfo {year} {2002})}\BibitemShut {NoStop}%
\bibitem [{\citenamefont {Boccaletti}\ \emph {et~al.}(2006)\citenamefont
  {Boccaletti}, \citenamefont {Latora}, \citenamefont {Moreno}, \citenamefont
  {Chavez},\ and\ \citenamefont {Hwang}}]{Boccaletti2006a}%
  \BibitemOpen
  \bibfield  {author} {\bibinfo {author} {\bibfnamefont {S.}~\bibnamefont
  {Boccaletti}}, \bibinfo {author} {\bibfnamefont {V.}~\bibnamefont {Latora}},
  \bibinfo {author} {\bibfnamefont {Y.}~\bibnamefont {Moreno}}, \bibinfo
  {author} {\bibfnamefont {M.}~\bibnamefont {Chavez}}, \ and\ \bibinfo {author}
  {\bibfnamefont {D.~U.}\ \bibnamefont {Hwang}},\ }\href {\doibase
  10.1016/j.physrep.2005.10.009} {\bibfield  {journal} {\bibinfo  {journal}
  {Phys. Rep.}\ }\textbf {\bibinfo {volume} {424}},\ \bibinfo {pages} {175}
  (\bibinfo {year} {2006})}\BibitemShut {NoStop}%
\bibitem [{\citenamefont {Arenas}\ \emph {et~al.}(2008)\citenamefont {Arenas},
  \citenamefont {D\'{\i}az-Guilera}, \citenamefont {Kurths}, \citenamefont
  {Moreno},\ and\ \citenamefont {Zhou}}]{Arenas2008}%
  \BibitemOpen
  \bibfield  {author} {\bibinfo {author} {\bibfnamefont {A.}~\bibnamefont
  {Arenas}}, \bibinfo {author} {\bibfnamefont {A.}~\bibnamefont
  {D\'{\i}az-Guilera}}, \bibinfo {author} {\bibfnamefont {J.}~\bibnamefont
  {Kurths}}, \bibinfo {author} {\bibfnamefont {Y.}~\bibnamefont {Moreno}}, \
  and\ \bibinfo {author} {\bibfnamefont {C.}~\bibnamefont {Zhou}},\ }\href
  {\doibase 10.1016/j.physrep.2008.09.002} {\bibfield  {journal} {\bibinfo
  {journal} {Phys. Rep.}\ }\textbf {\bibinfo {volume} {469}},\ \bibinfo {pages}
  {93} (\bibinfo {year} {2008})}\BibitemShut {NoStop}%
\bibitem [{\citenamefont {Colizza}\ \emph {et~al.}(2006)\citenamefont
  {Colizza}, \citenamefont {Barrat}, \citenamefont {Barth\'{e}lemy},\ and\
  \citenamefont {Vespignani}}]{Colizza2006}%
  \BibitemOpen
  \bibfield  {author} {\bibinfo {author} {\bibfnamefont {V.}~\bibnamefont
  {Colizza}}, \bibinfo {author} {\bibfnamefont {A.}~\bibnamefont {Barrat}},
  \bibinfo {author} {\bibfnamefont {M.}~\bibnamefont {Barth\'{e}lemy}}, \ and\
  \bibinfo {author} {\bibfnamefont {A.}~\bibnamefont {Vespignani}},\ }\href
  {\doibase 10.1073/pnas.0510525103} {\bibfield  {journal} {\bibinfo  {journal}
  {Proc. Natl. Acad. Sci. U. S. A.}\ }\textbf {\bibinfo {volume} {103}},\
  \bibinfo {pages} {2015} (\bibinfo {year} {2006})}\BibitemShut {NoStop}%
\bibitem [{\citenamefont {Barab\'{a}si}\ \emph {et~al.}(2002)\citenamefont
  {Barab\'{a}si}, \citenamefont {Jeong}, \citenamefont {N\'{e}da},
  \citenamefont {Ravasz}, \citenamefont {Schubert},\ and\ \citenamefont
  {Vicsek}}]{Barabasi2002}%
  \BibitemOpen
  \bibfield  {author} {\bibinfo {author} {\bibfnamefont {A.~L.}\ \bibnamefont
  {Barab\'{a}si}}, \bibinfo {author} {\bibfnamefont {H.}~\bibnamefont {Jeong}},
  \bibinfo {author} {\bibfnamefont {Z.}~\bibnamefont {N\'{e}da}}, \bibinfo
  {author} {\bibfnamefont {E.}~\bibnamefont {Ravasz}}, \bibinfo {author}
  {\bibfnamefont {A.}~\bibnamefont {Schubert}}, \ and\ \bibinfo {author}
  {\bibfnamefont {T.}~\bibnamefont {Vicsek}},\ }\href {\doibase
  10.1016/S0378-4371(02)00736-7} {\bibfield  {journal} {\bibinfo  {journal}
  {Phys. A Stat. Mech. its Appl.}\ }\textbf {\bibinfo {volume} {311}},\
  \bibinfo {pages} {590} (\bibinfo {year} {2002})}\BibitemShut {NoStop}%
\bibitem [{\citenamefont {Albert}\ \emph {et~al.}(1999)\citenamefont {Albert},
  \citenamefont {Jeong},\ and\ \citenamefont {Barab\'{a}si}}]{Albert1999}%
  \BibitemOpen
  \bibfield  {author} {\bibinfo {author} {\bibfnamefont {R.}~\bibnamefont
  {Albert}}, \bibinfo {author} {\bibfnamefont {H.}~\bibnamefont {Jeong}}, \
  and\ \bibinfo {author} {\bibfnamefont {A.~L.}\ \bibnamefont {Barab\'{a}si}},\
  }\href {\doibase 10.1038/43601} {\bibfield  {journal} {\bibinfo  {journal}
  {Nature}\ }\textbf {\bibinfo {volume} {401}},\ \bibinfo {pages} {398}
  (\bibinfo {year} {1999})}\BibitemShut {NoStop}%
\bibitem [{\citenamefont {Tsonis}\ and\ \citenamefont
  {Roebber}(2004)}]{Tsonis2004}%
  \BibitemOpen
  \bibfield  {author} {\bibinfo {author} {\bibfnamefont {A.~A.}\ \bibnamefont
  {Tsonis}}\ and\ \bibinfo {author} {\bibfnamefont {P.}~\bibnamefont
  {Roebber}},\ }\href {\doibase 10.1016/j.physa.2003.10.045} {\bibfield
  {journal} {\bibinfo  {journal} {Phys. A Stat. Mech. its Appl.}\ }\textbf
  {\bibinfo {volume} {333}},\ \bibinfo {pages} {497} (\bibinfo {year}
  {2004})}\BibitemShut {NoStop}%
\bibitem [{\citenamefont {Yamasaki}\ \emph
  {et~al.}(2008{\natexlab{a}})\citenamefont {Yamasaki}, \citenamefont {Mackin},
  \citenamefont {Ohshiro}, \citenamefont {Matsushita},\ and\ \citenamefont
  {Nunohiro}}]{Yamasaki2008a}%
  \BibitemOpen
  \bibfield  {author} {\bibinfo {author} {\bibfnamefont {K.}~\bibnamefont
  {Yamasaki}}, \bibinfo {author} {\bibfnamefont {K.~J.}\ \bibnamefont
  {Mackin}}, \bibinfo {author} {\bibfnamefont {M.}~\bibnamefont {Ohshiro}},
  \bibinfo {author} {\bibfnamefont {K.}~\bibnamefont {Matsushita}}, \ and\
  \bibinfo {author} {\bibfnamefont {E.}~\bibnamefont {Nunohiro}},\ }\href
  {\doibase 10.1007/s10015-007-0492-2} {\bibfield  {journal} {\bibinfo
  {journal} {Artif. Life Robot.}\ }\textbf {\bibinfo {volume} {12}},\ \bibinfo
  {pages} {122} (\bibinfo {year} {2008}{\natexlab{a}})}\BibitemShut {NoStop}%
\bibitem [{\citenamefont {Steinhaeuser}\ \emph {et~al.}(2009)\citenamefont
  {Steinhaeuser}, \citenamefont {Chawla},\ and\ \citenamefont
  {Ganguly}}]{Steinhaeuser2009}%
  \BibitemOpen
  \bibfield  {author} {\bibinfo {author} {\bibfnamefont {K.}~\bibnamefont
  {Steinhaeuser}}, \bibinfo {author} {\bibfnamefont {N.~V.}\ \bibnamefont
  {Chawla}}, \ and\ \bibinfo {author} {\bibfnamefont {A.~R.}\ \bibnamefont
  {Ganguly}},\ }\href@noop {} {\bibfield  {journal} {\bibinfo  {journal} {ACM
  SIGKDD Explor. Newsl.}\ }\textbf {\bibinfo {volume} {12}},\ \bibinfo {pages}
  {25} (\bibinfo {year} {2009})}\BibitemShut {NoStop}%
\bibitem [{\citenamefont {Zerenner}\ \emph {et~al.}(2014)\citenamefont
  {Zerenner}, \citenamefont {Friederichs}, \citenamefont {Lehnertz},\ and\
  \citenamefont {Hense}}]{Zerenner2014}%
  \BibitemOpen
  \bibfield  {author} {\bibinfo {author} {\bibfnamefont {T.}~\bibnamefont
  {Zerenner}}, \bibinfo {author} {\bibfnamefont {P.}~\bibnamefont
  {Friederichs}}, \bibinfo {author} {\bibfnamefont {K.}~\bibnamefont
  {Lehnertz}}, \ and\ \bibinfo {author} {\bibfnamefont {A.}~\bibnamefont
  {Hense}},\ }\href {\doibase 10.1063/1.4870402} {\bibfield  {journal}
  {\bibinfo  {journal} {Chaos}\ }\textbf {\bibinfo {volume} {24}},\ \bibinfo
  {pages} {023103} (\bibinfo {year} {2014})}\BibitemShut {NoStop}%
\bibitem [{\citenamefont {Bialonski}\ \emph {et~al.}(2010)\citenamefont
  {Bialonski}, \citenamefont {Horstmann},\ and\ \citenamefont
  {Lehnertz}}]{Bialonski2010}%
  \BibitemOpen
  \bibfield  {author} {\bibinfo {author} {\bibfnamefont {S.}~\bibnamefont
  {Bialonski}}, \bibinfo {author} {\bibfnamefont {M.-T.}\ \bibnamefont
  {Horstmann}}, \ and\ \bibinfo {author} {\bibfnamefont {K.}~\bibnamefont
  {Lehnertz}},\ }\href {\doibase 10.1063/1.3360561} {\bibfield  {journal}
  {\bibinfo  {journal} {Chaos}\ }\textbf {\bibinfo {volume} {20}},\ \bibinfo
  {pages} {013134} (\bibinfo {year} {2010})},\ \Eprint
  {http://arxiv.org/abs/arXiv:1105.2257v1} {arXiv:arXiv:1105.2257v1}
  \BibitemShut {NoStop}%
\bibitem [{\citenamefont {Donges}\ \emph
  {et~al.}(2009{\natexlab{a}})\citenamefont {Donges}, \citenamefont {Zou},
  \citenamefont {Marwan},\ and\ \citenamefont {Kurths}}]{Donges2009}%
  \BibitemOpen
  \bibfield  {author} {\bibinfo {author} {\bibfnamefont {J.~F.}\ \bibnamefont
  {Donges}}, \bibinfo {author} {\bibfnamefont {Y.}~\bibnamefont {Zou}},
  \bibinfo {author} {\bibfnamefont {N.}~\bibnamefont {Marwan}}, \ and\ \bibinfo
  {author} {\bibfnamefont {J.}~\bibnamefont {Kurths}},\ }\href {\doibase
  10.1140/epjst/e2009-01098-2} {\bibfield  {journal} {\bibinfo  {journal} {Eur.
  Phys. J. Spec. Top.}\ }\textbf {\bibinfo {volume} {174}},\ \bibinfo {pages}
  {157} (\bibinfo {year} {2009}{\natexlab{a}})}\BibitemShut {NoStop}%
\bibitem [{\citenamefont {Donges}\ \emph
  {et~al.}(2009{\natexlab{b}})\citenamefont {Donges}, \citenamefont {Zou},
  \citenamefont {Marwan},\ and\ \citenamefont {Kurths}}]{Donges2009a}%
  \BibitemOpen
  \bibfield  {author} {\bibinfo {author} {\bibfnamefont {J.~F.}\ \bibnamefont
  {Donges}}, \bibinfo {author} {\bibfnamefont {Y.}~\bibnamefont {Zou}},
  \bibinfo {author} {\bibfnamefont {N.}~\bibnamefont {Marwan}}, \ and\ \bibinfo
  {author} {\bibfnamefont {J.}~\bibnamefont {Kurths}},\ }\href {\doibase
  10.1209/0295-5075/87/48007} {\bibfield  {journal} {\bibinfo  {journal} {EPL}\
  }\textbf {\bibinfo {volume} {87}},\ \bibinfo {pages} {48007} (\bibinfo {year}
  {2009}{\natexlab{b}})}\BibitemShut {NoStop}%
\bibitem [{\citenamefont {Donges}\ \emph {et~al.}(2011)\citenamefont {Donges},
  \citenamefont {Schultz}, \citenamefont {Marwan}, \citenamefont {Zou},\ and\
  \citenamefont {Kurths}}]{Donges2011}%
  \BibitemOpen
  \bibfield  {author} {\bibinfo {author} {\bibfnamefont {J.~F.}\ \bibnamefont
  {Donges}}, \bibinfo {author} {\bibfnamefont {H.~C.~H.}\ \bibnamefont
  {Schultz}}, \bibinfo {author} {\bibfnamefont {N.}~\bibnamefont {Marwan}},
  \bibinfo {author} {\bibfnamefont {Y.}~\bibnamefont {Zou}}, \ and\ \bibinfo
  {author} {\bibfnamefont {J.}~\bibnamefont {Kurths}},\ }\href {\doibase
  10.1140/epjb/e2011-10795-8} {\bibfield  {journal} {\bibinfo  {journal} {Eur.
  Phys. J. B}\ }\textbf {\bibinfo {volume} {84}},\ \bibinfo {pages} {635}
  (\bibinfo {year} {2011})}\BibitemShut {NoStop}%
\bibitem [{\citenamefont {Tsonis}\ \emph {et~al.}(2008)\citenamefont {Tsonis},
  \citenamefont {Swanson},\ and\ \citenamefont {Wang}}]{Tsonis2008}%
  \BibitemOpen
  \bibfield  {author} {\bibinfo {author} {\bibfnamefont {A.~A.}\ \bibnamefont
  {Tsonis}}, \bibinfo {author} {\bibfnamefont {K.~L.}\ \bibnamefont {Swanson}},
  \ and\ \bibinfo {author} {\bibfnamefont {G.}~\bibnamefont {Wang}},\ }\href
  {\doibase 10.1175/2007JCLI1907.1} {\bibfield  {journal} {\bibinfo  {journal}
  {J. Clim.}\ }\textbf {\bibinfo {volume} {21}},\ \bibinfo {pages} {2990}
  (\bibinfo {year} {2008})}\BibitemShut {NoStop}%
\bibitem [{\citenamefont {Tantet}\ and\ \citenamefont
  {Dijkstra}(2014)}]{Tantet2014}%
  \BibitemOpen
  \bibfield  {author} {\bibinfo {author} {\bibfnamefont {A.}~\bibnamefont
  {Tantet}}\ and\ \bibinfo {author} {\bibfnamefont {H.~A.}\ \bibnamefont
  {Dijkstra}},\ }\href {\doibase 10.5194/esd-5-1-2014} {\bibfield  {journal}
  {\bibinfo  {journal} {Earth Syst. Dyn.}\ }\textbf {\bibinfo {volume} {5}},\
  \bibinfo {pages} {1} (\bibinfo {year} {2014})}\BibitemShut {NoStop}%
\bibitem [{\citenamefont {van~der Mheen}\ \emph {et~al.}(2013)\citenamefont
  {van~der Mheen}, \citenamefont {Dijkstra}, \citenamefont {Gozolchiani},
  \citenamefont {den Toom}, \citenamefont {Feng}, \citenamefont {Kurths},\ and\
  \citenamefont {Hernandez-Garcia}}]{VanderMheen2013}%
  \BibitemOpen
  \bibfield  {author} {\bibinfo {author} {\bibfnamefont {M.}~\bibnamefont
  {van~der Mheen}}, \bibinfo {author} {\bibfnamefont {H.~A.}\ \bibnamefont
  {Dijkstra}}, \bibinfo {author} {\bibfnamefont {A.}~\bibnamefont
  {Gozolchiani}}, \bibinfo {author} {\bibfnamefont {M.}~\bibnamefont {den
  Toom}}, \bibinfo {author} {\bibfnamefont {Q.}~\bibnamefont {Feng}}, \bibinfo
  {author} {\bibfnamefont {J.}~\bibnamefont {Kurths}}, \ and\ \bibinfo {author}
  {\bibfnamefont {E.}~\bibnamefont {Hernandez-Garcia}},\ }\href {\doibase
  10.1002/grl.50515} {\bibfield  {journal} {\bibinfo  {journal} {Geophys. Res.
  Lett.}\ }\textbf {\bibinfo {volume} {40}},\ \bibinfo {pages} {2714} (\bibinfo
  {year} {2013})}\BibitemShut {NoStop}%
\bibitem [{\citenamefont {Gozolchiani}\ \emph {et~al.}(2008)\citenamefont
  {Gozolchiani}, \citenamefont {Yamasaki}, \citenamefont {Gazit},\ and\
  \citenamefont {Havlin}}]{Gozolchiani2008}%
  \BibitemOpen
  \bibfield  {author} {\bibinfo {author} {\bibfnamefont {A.}~\bibnamefont
  {Gozolchiani}}, \bibinfo {author} {\bibfnamefont {K.}~\bibnamefont
  {Yamasaki}}, \bibinfo {author} {\bibfnamefont {O.}~\bibnamefont {Gazit}}, \
  and\ \bibinfo {author} {\bibfnamefont {S.}~\bibnamefont {Havlin}},\ }\href
  {http://stacks.iop.org/0295-5075/83/i=2/a=28005} {\bibfield  {journal}
  {\bibinfo  {journal} {EPL}\ }\textbf {\bibinfo {volume} {83}},\ \bibinfo
  {pages} {28005} (\bibinfo {year} {2008})}\BibitemShut {NoStop}%
\bibitem [{\citenamefont {Yamasaki}\ \emph
  {et~al.}(2008{\natexlab{b}})\citenamefont {Yamasaki}, \citenamefont
  {Gozolchiani},\ and\ \citenamefont {Havlin}}]{Yamasaki2008b}%
  \BibitemOpen
  \bibfield  {author} {\bibinfo {author} {\bibfnamefont {K.}~\bibnamefont
  {Yamasaki}}, \bibinfo {author} {\bibfnamefont {A.}~\bibnamefont
  {Gozolchiani}}, \ and\ \bibinfo {author} {\bibfnamefont {S.}~\bibnamefont
  {Havlin}},\ }\href {\doibase 10.1103/PhysRevLett.100.228501} {\bibfield
  {journal} {\bibinfo  {journal} {Phys. Rev. Lett.}\ }\textbf {\bibinfo
  {volume} {100}},\ \bibinfo {pages} {228501} (\bibinfo {year}
  {2008}{\natexlab{b}})}\BibitemShut {NoStop}%
\bibitem [{\citenamefont {Wang}\ \emph {et~al.}(2012)\citenamefont {Wang},
  \citenamefont {Yang}, \citenamefont {Zhou}, \citenamefont {Swanson},\ and\
  \citenamefont {Tsonis}}]{Wang2012}%
  \BibitemOpen
  \bibfield  {author} {\bibinfo {author} {\bibfnamefont {G.}~\bibnamefont
  {Wang}}, \bibinfo {author} {\bibfnamefont {P.}~\bibnamefont {Yang}}, \bibinfo
  {author} {\bibfnamefont {X.}~\bibnamefont {Zhou}}, \bibinfo {author}
  {\bibfnamefont {K.~L.}\ \bibnamefont {Swanson}}, \ and\ \bibinfo {author}
  {\bibfnamefont {A.~A.}\ \bibnamefont {Tsonis}},\ }\href {\doibase
  10.1029/2012GL052149} {\bibfield  {journal} {\bibinfo  {journal} {Geophys.
  Res. Lett.}\ }\textbf {\bibinfo {volume} {39}},\ \bibinfo {pages} {704}
  (\bibinfo {year} {2012})}\BibitemShut {NoStop}%
\bibitem [{\citenamefont {Mokhov}\ \emph {et~al.}(2011)\citenamefont {Mokhov},
  \citenamefont {Smirnov}, \citenamefont {Nakonechny}, \citenamefont
  {Kozlenko}, \citenamefont {Seleznev},\ and\ \citenamefont
  {Kurths}}]{Mokhov2011}%
  \BibitemOpen
  \bibfield  {author} {\bibinfo {author} {\bibfnamefont {I.~I.}\ \bibnamefont
  {Mokhov}}, \bibinfo {author} {\bibfnamefont {D.~a.}\ \bibnamefont {Smirnov}},
  \bibinfo {author} {\bibfnamefont {P.~I.}\ \bibnamefont {Nakonechny}},
  \bibinfo {author} {\bibfnamefont {S.~S.}\ \bibnamefont {Kozlenko}}, \bibinfo
  {author} {\bibfnamefont {E.~P.}\ \bibnamefont {Seleznev}}, \ and\ \bibinfo
  {author} {\bibfnamefont {J.}~\bibnamefont {Kurths}},\ }\href {\doibase
  10.1029/2010GL045932} {\bibfield  {journal} {\bibinfo  {journal} {Geophys.
  Res. Lett.}\ }\textbf {\bibinfo {volume} {38}},\ \bibinfo {pages} {L00F04}
  (\bibinfo {year} {2011})}\BibitemShut {NoStop}%
\bibitem [{\citenamefont {Hlavackovaschindler}\ \emph
  {et~al.}(2007)\citenamefont {Hlavackovaschindler}, \citenamefont {Palu\v{s}},
  \citenamefont {Vejmelka},\ and\ \citenamefont
  {Bhattacharya}}]{Hlavackovaschindler2007}%
  \BibitemOpen
  \bibfield  {author} {\bibinfo {author} {\bibfnamefont {K.}~\bibnamefont
  {Hlavackovaschindler}}, \bibinfo {author} {\bibfnamefont {M.}~\bibnamefont
  {Palu\v{s}}}, \bibinfo {author} {\bibfnamefont {M.}~\bibnamefont {Vejmelka}},
  \ and\ \bibinfo {author} {\bibfnamefont {J.}~\bibnamefont {Bhattacharya}},\
  }\href {\doibase 10.1016/j.physrep.2006.12.004} {\bibfield  {journal}
  {\bibinfo  {journal} {Phys. Rep.}\ }\textbf {\bibinfo {volume} {441}},\
  \bibinfo {pages} {1} (\bibinfo {year} {2007})}\BibitemShut {NoStop}%
\bibitem [{\citenamefont {Palu\v{s}}(2014)}]{Palus2014}%
  \BibitemOpen
  \bibfield  {author} {\bibinfo {author} {\bibfnamefont {M.}~\bibnamefont
  {Palu\v{s}}},\ }\href {\doibase 10.1103/PhysRevLett.112.078702} {\bibfield
  {journal} {\bibinfo  {journal} {Phys. Rev. Lett.}\ }\textbf {\bibinfo
  {volume} {112}},\ \bibinfo {pages} {078702} (\bibinfo {year}
  {2014})}\BibitemShut {NoStop}%
\bibitem [{\citenamefont {Hlinka}\ \emph {et~al.}(2013)\citenamefont {Hlinka},
  \citenamefont {Hartman}, \citenamefont {Vejmelka}, \citenamefont {Runge},
  \citenamefont {Marwan}, \citenamefont {Kurths},\ and\ \citenamefont
  {Palu\v{s}}}]{Hlinka2013}%
  \BibitemOpen
  \bibfield  {author} {\bibinfo {author} {\bibfnamefont {J.}~\bibnamefont
  {Hlinka}}, \bibinfo {author} {\bibfnamefont {D.}~\bibnamefont {Hartman}},
  \bibinfo {author} {\bibfnamefont {M.}~\bibnamefont {Vejmelka}}, \bibinfo
  {author} {\bibfnamefont {J.}~\bibnamefont {Runge}}, \bibinfo {author}
  {\bibfnamefont {N.}~\bibnamefont {Marwan}}, \bibinfo {author} {\bibfnamefont
  {J.}~\bibnamefont {Kurths}}, \ and\ \bibinfo {author} {\bibfnamefont
  {M.}~\bibnamefont {Palu\v{s}}},\ }\href {\doibase 10.3390/e15062023}
  {\bibfield  {journal} {\bibinfo  {journal} {Entropy}\ }\textbf {\bibinfo
  {volume} {15}},\ \bibinfo {pages} {2023} (\bibinfo {year}
  {2013})}\BibitemShut {NoStop}%
\bibitem [{\citenamefont {Bahraminasab}\ \emph {et~al.}(2008)\citenamefont
  {Bahraminasab}, \citenamefont {Ghasemi}, \citenamefont {Stefanovska},
  \citenamefont {McClintock},\ and\ \citenamefont {Kantz}}]{Bahraminasab2008}%
  \BibitemOpen
  \bibfield  {author} {\bibinfo {author} {\bibfnamefont {A.}~\bibnamefont
  {Bahraminasab}}, \bibinfo {author} {\bibfnamefont {F.}~\bibnamefont
  {Ghasemi}}, \bibinfo {author} {\bibfnamefont {A.}~\bibnamefont
  {Stefanovska}}, \bibinfo {author} {\bibfnamefont {P.}~\bibnamefont
  {McClintock}}, \ and\ \bibinfo {author} {\bibfnamefont {H.}~\bibnamefont
  {Kantz}},\ }\href {\doibase 10.1103/PhysRevLett.100.084101} {\bibfield
  {journal} {\bibinfo  {journal} {Phys. Rev. Lett.}\ }\textbf {\bibinfo
  {volume} {100}},\ \bibinfo {pages} {084101} (\bibinfo {year}
  {2008})}\BibitemShut {NoStop}%
\bibitem [{\citenamefont {Palu\v{s}}\ and\ \citenamefont
  {Stefanovska}(2003)}]{Palus2003}%
  \BibitemOpen
  \bibfield  {author} {\bibinfo {author} {\bibfnamefont {M.}~\bibnamefont
  {Palu\v{s}}}\ and\ \bibinfo {author} {\bibfnamefont {A.}~\bibnamefont
  {Stefanovska}},\ }\href {\doibase 10.1103/PhysRevE.67.055201} {\bibfield
  {journal} {\bibinfo  {journal} {Phys. Rev. E}\ }\textbf {\bibinfo {volume}
  {67}},\ \bibinfo {pages} {055201} (\bibinfo {year} {2003})}\BibitemShut
  {NoStop}%
\bibitem [{\citenamefont {Lehnertz}\ and\ \citenamefont
  {Dickten}(2014)}]{Lehnertz2014}%
  \BibitemOpen
  \bibfield  {author} {\bibinfo {author} {\bibfnamefont {K.}~\bibnamefont
  {Lehnertz}}\ and\ \bibinfo {author} {\bibfnamefont {H.}~\bibnamefont
  {Dickten}},\ }\href@noop {} {\bibfield  {journal} {\bibinfo  {journal} {Proc.
  R. Soc. A}\ }\textbf {\bibinfo {volume} {in press}} (\bibinfo {year}
  {2014})}\BibitemShut {NoStop}%
\bibitem [{\citenamefont {Kalnay}\ \emph {et~al.}(1996)\citenamefont {Kalnay}
  \emph {et~al.}}]{Kalnay1996}%
  \BibitemOpen
  \bibfield  {author} {\bibinfo {author} {\bibfnamefont {E.}~\bibnamefont
  {Kalnay}} \emph {et~al.},\ }\href {\doibase
  10.1175/1520-0477(1996)077<0437:TNYRP>2.0.CO;2} {\bibfield  {journal}
  {\bibinfo  {journal} {Bull. Am. Meteorol. Soc.}\ }\textbf {\bibinfo {volume}
  {77}},\ \bibinfo {pages} {437} (\bibinfo {year} {1996})}\BibitemShut
  {NoStop}%
\bibitem [{\citenamefont {Palu\v{s}}(2007)}]{Palus2007a}%
  \BibitemOpen
  \bibfield  {author} {\bibinfo {author} {\bibfnamefont {M.}~\bibnamefont
  {Palu\v{s}}},\ }\href
  {http://www.tandfonline.com/doi/abs/10.1080/00107510801959206} {\bibfield
  {journal} {\bibinfo  {journal} {Contemp. Phys.}\ }\textbf {\bibinfo {volume}
  {48}},\ \bibinfo {pages} {307} (\bibinfo {year} {2007})}\BibitemShut
  {NoStop}%
\bibitem [{\citenamefont {Amig\'{o}}(2010)}]{Amigo2010}%
  \BibitemOpen
  \bibfield  {author} {\bibinfo {author} {\bibfnamefont {J.}~\bibnamefont
  {Amig\'{o}}},\ }\href {\doibase 10.1007/978-3-642-04084-9} {\emph {\bibinfo
  {title} {{Permutation Complexity in Dynamical Systems}}}},\ Springer Series
  in Synergetics\ (\bibinfo  {publisher} {Springer Berlin Heidelberg},\
  \bibinfo {address} {Berlin, Heidelberg},\ \bibinfo {year} {2010})\ p.\
  \bibinfo {pages} {260}\BibitemShut {NoStop}%
\bibitem [{\citenamefont {Barreiro}\ \emph {et~al.}(2011)\citenamefont
  {Barreiro}, \citenamefont {Marti},\ and\ \citenamefont
  {Masoller}}]{Barreiro2011}%
  \BibitemOpen
  \bibfield  {author} {\bibinfo {author} {\bibfnamefont {M.}~\bibnamefont
  {Barreiro}}, \bibinfo {author} {\bibfnamefont {A.~C.}\ \bibnamefont {Marti}},
  \ and\ \bibinfo {author} {\bibfnamefont {C.}~\bibnamefont {Masoller}},\
  }\href {\doibase 10.1063/1.3545273} {\bibfield  {journal} {\bibinfo
  {journal} {Chaos}\ }\textbf {\bibinfo {volume} {21}},\ \bibinfo {pages}
  {013101} (\bibinfo {year} {2011})}\BibitemShut {NoStop}%
\bibitem [{\citenamefont {Deza}\ \emph {et~al.}(2013)\citenamefont {Deza},
  \citenamefont {Barreiro},\ and\ \citenamefont {Masoller}}]{Deza2013}%
  \BibitemOpen
  \bibfield  {author} {\bibinfo {author} {\bibfnamefont {J.~I.}\ \bibnamefont
  {Deza}}, \bibinfo {author} {\bibfnamefont {M.}~\bibnamefont {Barreiro}}, \
  and\ \bibinfo {author} {\bibfnamefont {C.}~\bibnamefont {Masoller}},\ }\href
  {\doibase 10.1140/epjst/e2013-01856-5} {\bibfield  {journal} {\bibinfo
  {journal} {Eur. Phys. J. Spec. Top.}\ }\textbf {\bibinfo {volume} {222}},\
  \bibinfo {pages} {511} (\bibinfo {year} {2013})}\BibitemShut {NoStop}%
\bibitem [{\citenamefont {Mudelsee}(2014)}]{Mudelsee2010}%
  \BibitemOpen
  \bibfield  {author} {\bibinfo {author} {\bibfnamefont {M.}~\bibnamefont
  {Mudelsee}},\ }\href {\doibase 10.1007/978-90-481-9482-7} {\emph {\bibinfo
  {title} {{Climate Time Series Analysis}}}},\ \bibinfo {series} {Atmospheric
  and Oceanographic Sciences Library}, Vol.~\bibinfo {volume} {42}\ (\bibinfo
  {publisher} {Springer Netherlands},\ \bibinfo {address} {Dordrecht},\
  \bibinfo {year} {2014})\BibitemShut {NoStop}%
\bibitem [{\citenamefont {Deza}\ \emph {et~al.}(2014)\citenamefont {Deza},
  \citenamefont {Masoller},\ and\ \citenamefont {Barreiro}}]{Deza2014}%
  \BibitemOpen
  \bibfield  {author} {\bibinfo {author} {\bibfnamefont {J.~I.}\ \bibnamefont
  {Deza}}, \bibinfo {author} {\bibfnamefont {C.}~\bibnamefont {Masoller}}, \
  and\ \bibinfo {author} {\bibfnamefont {M.}~\bibnamefont {Barreiro}},\ }\href
  {\doibase 10.5194/npg-21-617-2014} {\bibfield  {journal} {\bibinfo  {journal}
  {Nonlinear Process. Geophys.}\ }\textbf {\bibinfo {volume} {21}},\ \bibinfo
  {pages} {617} (\bibinfo {year} {2014})}\BibitemShut {NoStop}%
\bibitem [{\citenamefont {Chang}\ \emph {et~al.}(2000)\citenamefont {Chang},
  \citenamefont {Saravanan}, \citenamefont {Ji},\ and\ \citenamefont
  {Hegerl}}]{Chang2000}%
  \BibitemOpen
  \bibfield  {author} {\bibinfo {author} {\bibfnamefont {P.}~\bibnamefont
  {Chang}}, \bibinfo {author} {\bibfnamefont {R.}~\bibnamefont {Saravanan}},
  \bibinfo {author} {\bibfnamefont {L.}~\bibnamefont {Ji}}, \ and\ \bibinfo
  {author} {\bibfnamefont {G.}~\bibnamefont {Hegerl}},\ }\href
  {http://journals.ametsoc.org/doi/abs/10.1175/1520-0442(2000)013\%3C2195\%3ATEOLSS\%3E2.0.CO\%3B2}
  {\bibfield  {journal} {\bibinfo  {journal} {J. Clim.}\ }\textbf {\bibinfo
  {volume} {13}},\ \bibinfo {pages} {2195} (\bibinfo {year}
  {2000})}\BibitemShut {NoStop}%
\bibitem [{\citenamefont {Mo}\ and\ \citenamefont {Higgins}(1998)}]{Mo1998}%
  \BibitemOpen
  \bibfield  {author} {\bibinfo {author} {\bibfnamefont {K.~C.}\ \bibnamefont
  {Mo}}\ and\ \bibinfo {author} {\bibfnamefont {R.~W.}\ \bibnamefont
  {Higgins}},\ }\href {\doibase
  10.1175/1520-0493(1998)126<1581:TPSAMA>2.0.CO;2} {\bibfield  {journal}
  {\bibinfo  {journal} {Mon. Weather Rev.}\ }\textbf {\bibinfo {volume}
  {126}},\ \bibinfo {pages} {1581} (\bibinfo {year} {1998})}\BibitemShut
  {NoStop}%
\bibitem [{\citenamefont {Annamalai}\ \emph {et~al.}(2003)\citenamefont
  {Annamalai}, \citenamefont {Murtugudde}, \citenamefont {Potemra},
  \citenamefont {Xie}, \citenamefont {Liu},\ and\ \citenamefont
  {Wang}}]{Annamalai2003a}%
  \BibitemOpen
  \bibfield  {author} {\bibinfo {author} {\bibfnamefont {H.}~\bibnamefont
  {Annamalai}}, \bibinfo {author} {\bibfnamefont {R.}~\bibnamefont
  {Murtugudde}}, \bibinfo {author} {\bibfnamefont {J.}~\bibnamefont {Potemra}},
  \bibinfo {author} {\bibfnamefont {S.}~\bibnamefont {Xie}}, \bibinfo {author}
  {\bibfnamefont {P.}~\bibnamefont {Liu}}, \ and\ \bibinfo {author}
  {\bibfnamefont {B.}~\bibnamefont {Wang}},\ }\href {\doibase
  10.1016/S0967-0645(03)00058-4} {\bibfield  {journal} {\bibinfo  {journal}
  {Deep Sea Res. Part II Top. Stud. Oceanogr.}\ }\textbf {\bibinfo {volume}
  {50}},\ \bibinfo {pages} {2305} (\bibinfo {year} {2003})}\BibitemShut
  {NoStop}%
\bibitem [{\citenamefont {Wang}\ and\ \citenamefont {Wang}(2013)}]{Wang2013c}%
  \BibitemOpen
  \bibfield  {author} {\bibinfo {author} {\bibfnamefont {X.}~\bibnamefont
  {Wang}}\ and\ \bibinfo {author} {\bibfnamefont {C.}~\bibnamefont {Wang}},\
  }\href {\doibase 10.1007/s00382-013-1711-2} {\bibfield  {journal} {\bibinfo
  {journal} {Clim. Dyn.}\ }\textbf {\bibinfo {volume} {42}},\ \bibinfo {pages}
  {991} (\bibinfo {year} {2013})}\BibitemShut {NoStop}%
\bibitem [{\citenamefont {Barsugli}\ and\ \citenamefont
  {Battisti}(1998)}]{Barsugli1998}%
  \BibitemOpen
  \bibfield  {author} {\bibinfo {author} {\bibfnamefont {J.}~\bibnamefont
  {Barsugli}}\ and\ \bibinfo {author} {\bibfnamefont {D.}~\bibnamefont
  {Battisti}},\ }\href
  {http://journals.ametsoc.org/doi/abs/10.1175/1520-0469(1998)055\%3C0477\%3ATBEOAO\%3E2.0.CO\%3B2}
  {\bibfield  {journal} {\bibinfo  {journal} {J. Atmos. Sci.}\ }\textbf
  {\bibinfo {volume} {55}},\ \bibinfo {pages} {477} (\bibinfo {year}
  {1998})}\BibitemShut {NoStop}%
\bibitem [{mov()}]{movie2014}%
  \BibitemOpen
  \href@noop {} {\enquote {\bibinfo {title} {{See supplemental material at
  http://youtu.be/alcormjKIbM for a movie showing an DI maps for increasing
  $\tau$.} year = {2013}},}\ }\BibitemShut {NoStop}%
\bibitem [{\citenamefont {Trenberth}\ \emph {et~al.}(1998)\citenamefont
  {Trenberth}, \citenamefont {Branstator}, \citenamefont {Karoly},
  \citenamefont {Kumar}, \citenamefont {Lau},\ and\ \citenamefont
  {Ropelewsky}}]{Trenberth1998}%
  \BibitemOpen
  \bibfield  {author} {\bibinfo {author} {\bibfnamefont {K.~E.}\ \bibnamefont
  {Trenberth}}, \bibinfo {author} {\bibfnamefont {G.~W.}\ \bibnamefont
  {Branstator}}, \bibinfo {author} {\bibfnamefont {D.}~\bibnamefont {Karoly}},
  \bibinfo {author} {\bibfnamefont {A.}~\bibnamefont {Kumar}}, \bibinfo
  {author} {\bibfnamefont {N.-C.}\ \bibnamefont {Lau}}, \ and\ \bibinfo
  {author} {\bibfnamefont {C.}~\bibnamefont {Ropelewsky}},\ }\href@noop {}
  {\bibfield  {journal} {\bibinfo  {journal} {J. Geophys. Res.}\ }\textbf
  {\bibinfo {volume} {103}},\ \bibinfo {pages} {14291} (\bibinfo {year}
  {1998})}\BibitemShut {NoStop}%
\end{thebibliography}%

\end{document}